\documentclass[sigconf]{acmart}
\AtBeginDocument{%
  \providecommand\BibTeX{{%
    \normalfont B\kern-0.5em{\scshape i\kern-0.25em b}\kern-0.8em\TeX}}}

\setcopyright{acmcopyright}
\setcopyright{rightsretained}
\acmConference[CHI '24]{Proceedings of the 2024 CHI Conference on Human Factors in Computing Systems}{May 11--16, 2024}{Honolulu, Hawaii}
\acmBooktitle{Proceedings of the 2024 CHI Conference on Human Factors in Computing Systems (CHI '24), May 11--16, 2024, Honolulu, Hawaii}
\copyrightyear{2024}
\acmYear{2024}
\acmDOI{10.1145/3613904.3642625}

\usepackage{dirtytalk}
\usepackage{multirow}
\usepackage{xcolor}
\usepackage{cancel}
\usepackage{subcaption}
\usepackage{enumitem}
\usepackage{multirow}
\usepackage{color}
\usepackage{hyperref}

\newcommand{\tabincell}[2]{\begin{tabular}{@{}#1@{}}#2\end{tabular}}

\author{Zhuoyan Li}
\email{li4178@purdue.edu}
\affiliation{
  \institution{Purdue University}
  \city{West Lafayette}
  \state{Indiana}
  \country{USA}
  \postcode{47907}
}

\author{Chen Liang}
\email{chenliang@uconn.edu}
\affiliation{
  \institution{University of Connecticut}
  \city{Storrs}
  \state{Connecticut}
  \country{USA}
  \postcode{06268}
}

\author{Jing Peng}
\email{jing.peng@uconn.edu}
\affiliation{
  \institution{University of Connecticut}
  \city{Storrs}
  \state{Connecticut}
  \country{USA}
  \postcode{06268}
}

\author{Ming Yin}
\email{mingyin@purdue.edu}
\affiliation{
  \institution{Purdue University}
  \city{West Lafayette}
  \state{Indiana}
  \country{USA}
  \postcode{47907}
}

\begin{document}

\title[]{The Value, Benefits, and Concerns of Generative AI-Powered Assistance in Writing
}

\renewcommand{\shortauthors}{Li, et al.}
\newcommand{\squishlist}{
   \begin{list}{\small $\bullet$}
    { \setlength{\itemsep}{0pt}      \setlength{\parsep}{1pt}
      \setlength{\topsep}{1pt}       \setlength{\partopsep}{1pt}
     \setlength{\leftmargin}{1.2em} \setlength{\labelwidth}{1em}
      \setlength{\labelsep}{0.5em} } }
\newcommand{\squishend}{  \end{list}  }
\newcommand{\whosays}[3]{\begin{center}\textit{\say{#3}} (Subject\##1, Group\##2)\end{center}}
\begin{abstract}

Recent advances in generative AI technologies like large language models raise both excitement and concerns about the future of human-AI co-creation in writing. To unpack people’s attitude towards and experience with generative AI-powered writing assistants, in this paper, we conduct an experiment to understand whether and how much value people attach to AI assistance, and how the incorporation of AI assistance in writing workflows changes people’s writing perceptions and performance. Our results suggest that people are willing to forgo financial payments to receive writing assistance from AI, especially if AI can provide direct content generation assistance and the writing task is highly creative. Generative AI-powered assistance is found to offer benefits in increasing people’s productivity and confidence in writing. However, direct content generation assistance offered by AI also comes with risks, including decreasing people’s sense of accountability and diversity in writing. We conclude by discussing the implications of our findings.
\end{abstract}

\ccsdesc[500]{Human-centered computing~Empirical studies in HCI}
\ccsdesc[500]{Computing methodologies~Artificial intelligence}

\keywords{Human-AI co-creation, AI writing assistant, Large language model}


\maketitle

\section{Introduction}
Recent advances in generative artificial intelligence (AI) have opened up exciting opportunities for fostering synergistic human-AI collaborations in completing various tasks.  
For instance, large language models (LLMs) 
such as GPT-3~\cite{brown2020language} and GPT-4~\cite{Openaigpt4} have demonstrated impressive capabilities in language understanding and generation, highlighting the promise of incorporating LLM-powered assistance into humans' writing processes to enable ``human-AI co-writing'' and enhance human productivity ~\cite{doi:10.1126/science.adh2586} and creativity~\cite{ding2023mapping}. Compared to traditional writing assistive tools, LLMs can provide a wide range of highly flexible assistance beyond simple grammar and spelling corrections, from polishing text writing styles to generating completely new content from scratch to inspire human writers. Recent research has shown that integrating generative AI-powered assistance in the workflows of professional writing tasks improves the productivity of human in these tasks~\cite{doi:10.1126/science.adh2586}. As such, a growing line of research has emerged in designing more effective interfaces to facilitate the communication between humans and their generative AI-powered writing assistants, aiming to unleash the full potential of human-AI collaboration in writing~\cite{yang2022ai,yuan2022wordcraft,macneil2023prompt,chakrabarty2022help}.

The significant promise brought up by generative AI-powered writing assistants appears to suggest that people would naturally value their assistance and be willing to pay for them. However, there are still lingering concerns on the use of such AI assistance 
in writing tasks. For example, writers have reservations that a writing process led by the AI writing assistant engenders their control, autonomy, and ownership~\cite{biermann2022tool}, and there is a general fear that powerful automation technologies may completely replace human labor in the future~\cite{acemoglu2018race}.
With all these concerns, people may associate limited, if any, value with the assistance provided by AI, and they may be reluctant to pay for such assistance. So, given the pros and cons of generative AI-powered writing assistants, one may wonder whether people's attitudes towards these assistants are dominated by their benefits or risks, and therefore, whether people perceive some financial value from utilizing AI assistance in writing.  In particular, are people willing to forgo some monetary compensation to complete their writing jobs together with a generative AI-powered assistant rather than completing the jobs on their own? If so, how will the perceived financial value of AI assistance vary with the kind of writing assistance offered by the generative AI model and people's own characteristics? How much benefit do different kinds of AI 
assistance bring about on people's writing experience and  performance, and how does it match up with people's value on these assistance? 


To answer these questions, in this paper, we present a randomized human-subject experiment ($N=379$) 
assessing how much people value generative AI-powered writing assistance financially and how such assistance influences people's writing experience and performance.  
During our experiment, participants were asked to write a 200-250 word article---either an argumentative essay or a creative story---within 45 minutes.  
We designed three writing modes in our experiment by varying the presence and type of AI assistance during participants' writing processes: (a) \textbf{Independent writing} ({\em Independent}), where participants completed the writing job on their own without receiving any AI assistance; (b) \textbf{Writing with editing assistance} ({\em Human-primary}), where participants took the primary responsibility of drafting the article while they received text editing and polishing assistance provided by ChatGPT, a LLM-based chatbot; and (c) \textbf{Collaborative writing with AI} ({\em AI-primary}), where ChatGPT took the primary responsibility of drafting the initial version of the article, while participants provided feedback and instructions to ChatGPT for improving the draft and made the final call to merge the content generated by ChatGPT and themselves.  
To estimate how much financial value that people attach to different
kinds of writing assistance offered by LLMs, 
we created two experimental treatments by having participants make a forced binary choice between one job offer that pays them a fixed amount of \$3 to complete the writing job on their own (i.e., in the ``{\em independent}'' mode), and another offer that pays them a variable amount of \$$x$ ($1.5\le x\le 4.5$) to complete the writing job with the AI assistance. The two experimental treatments differ on whether the AI assistance participants received in the second offer was editing assistance only (i.e., participants would be placed in the ``{\em human-primary}'' mode) or content generation assistance (i.e., participants would be placed in the ``{\em AI-primary}'' mode). Finally, after participants made the selection and completed the writing task in their selected writing modes, they were asked to fill out an exit survey to report their perceptions of their writing experience on various aspects, including their cognitive load, perceptions of the writing process (e.g., enjoyment) and writing outcome (e.g., uniqueness), perceived accountability, and confidence in completing future writing tasks. 


Our experimental results suggest that people attach a noticeable level of financial value to generative AI-powered assistance in completing their writing tasks. For example, in the ``{\em AI-primary}'' writing mode wherein ChatGPT can provide content generation assistance, we found that participants in our study were willing to forgo \$0.85 to receive its  assistance. To put it into context, this value 
amounts to 28.3\% of the \$3 payment that participants would have received should they choose to write independently. When translating this value to participants' hourly wage, we found that an average participant was willing to give up an hourly wage of \$1.71 in order to receive ChatGPT's content generation assistance. 
Our data further indicates that people's value of generative AI is particularly salient when AI can offer content generation assistance instead of just editing assistance, and when the writing task has a higher demand for creativity (e.g., write a creative story). 
In addition,
people's own characteristics also play a role in determining how much they value AI assistance. For instance, people with higher confidence in their own writing abilities generally attach lower value to AI assistance than people with lower confidence. 
Upon examining the impacts of AI assistance on participants' writing experience perceptions and writing performance, we 
found that both editing assistance and content generation assistance provided by LLMs bring about significant benefits on people's productivity (e.g., reduced writing time and improved grammar) and self-efficacy (e.g., increased writing confidence), especially on argumentative essay writing. However, when LLMs directly provide content generation assistance (as in the ``{\em AI-primary}'' mode), these benefits also come at some cost. This includes people's decreased satisfaction with their writing experience and writing outcome (e.g., decreased ability to express oneself through writing), their reluctance to take responsibility for potential issues of the writing (e.g., the writing contains misinformation), and a decline of diversity across writings generated by different individuals. In this sense, there appears to exist a degree of ``mismatch'' between the financial value (i.e., the ``price'') that people attach to different types of AI assistance in writing and the ``true value'' of these assistance.

Together, our study offers important experimental evidence regarding the value, benefits, and concerns of generative AI-powered assistance in human-AI co-writing. We conclude by discussing the design implications of our findings, and outline limitations and future work. 

\section{Related Work}\label{sec:related}
\subsection{Interaction between humans and writing assistants}
Research on understanding and designing interactions between humans and writing assistants has surged in recent years. Before the rise of LLMs, writing assistants primarily focus on enhancing human writing by offering features such as word-level suggestions~\cite{arnold2016suggesting,dunlop2012multidimensional,fowler2015effects,quinn2016cost}, sentence-level suggestions~\cite{chen2019gmail,kannan2016smart}, and 
novel text entry methods to optimize both speed and accuracy~\cite{bi2014both,vertanen2015velocitap}. However, some studies~\cite{banovic2019limits,buschek2018researchime,dalvi2016does,palin2019people} indicated that certain features, such as word suggestions, could occasionally clash with a user's typing habits and adversely affect their writing experience. 

With the advent of LLMs~\cite{brown2020language} and their unparalleled capabilities in language understanding and generation~\cite{Openaigpt4}, LLM-powered assistance is now more actively integrated into human writing workflows~\cite{doi:10.1126/science.adh2586}, potentially acting more like writing companion or coauthor than mere tools for word suggestions or corrections. 
As such, an increasing stream of research starts to investigate the range of assistance that LLMs could offer to human writers during the writing processes, 
such as tailored ideation based on current content~\cite{yuan2022wordcraft,dang2023choice} and advanced text revisions beyond traditional word/grammar correction check~\cite{yuan2022wordcraft}. This assistance could also tailor to diverse writing tasks including stories~\cite{yuan2022wordcraft,clark2018creative,lee2022coauthor,chung2022talebrush,singh2022hide,yang2022ai,zhou2023synthetic}, slogans~\cite{clark2018creative}, argumentative essays~\cite{lee2022coauthor}, emails~\cite{fu2023comparing,buschek2021impact}, and script writing~\cite{mirowski2023co}.

While early research highlights the advantages of integrating LLMs into the writing process, such as generating suggestions to unblock writers' creative thoughts, recent studies have begun to examine the potential negative impacts these AI writing assistants might have on humans during and post-writing  
process beyond those benefits.  
For instance,  one study reveals that the sentiment of text suggestions from LLMs can sway the sentiment of human-generated text~\cite{hohenstein2020ai}.  Additionally, opinionated LLM-powered writing assistants 
can not only change the 
opinions expressed in human-written content but also subtly shift the human writers' own perspectives and beliefs~\cite{jakesch2023co}. In the context of online platforms, it is also found that  AI-generated profiles on Airbnb may be viewed as less trustworthy~\cite{jakesch2019ai}. 

\subsection{Value estimation via ``willingness to pay''}
The conventional economic literature defines the maximum amount an individual would be willing to pay to acquire an amenity (e.g., the work flexibility, the work assistance) as their willingness to pay (WTP) for the amenity ~\cite{rosen1986theory, mas2017valuing}, which can be intuitively comprehended as the value of the amenity to them. Researchers in the human-computer interaction community have previously worked on quantifying the value of various amenities~\cite{holtz2022much,goldstein2014economic,yin2018running}.   
According to the existing literature, the prevailing best practice for quantifying individuals’ willingness to pay or accept for the amenity is through discrete choice experiments ~\cite{mas2017valuing, hedegaard2018price, liang2023hidden}.  Mas and Pallais ~\cite{ mas2017valuing} quantify the value of work flexibility offering each prospective worker two job options, differing in wage levels: one job offering flexible scheduling and the other an inflexible job. Hedegaard and Tyran ~\cite{hedegaard2018price} measure the value that workers attribute to working with individuals of the same ethnicity by inviting workers to choose between two job options: working alongside individuals of the same ethnicity and the other one with individuals from a different ethnicity. In the same vein, Liang et al. ~\cite{liang2023hidden} estimate gig workers’ willingness to accept a monitored job (i.e., the value they place on avoiding monitoring) by presenting them with a choice between a non-monitored job and an alternative job, the latter being subject to a randomly selected monitoring policy. Using similar methods, ~\citeauthor{holtz2022much}~\shortcite{holtz2022much} also estimate the value that freelancers associate with a positive review on online labor platforms. In this study, we will also quantify the value people attach to generative AI-powered writing assistance by estimating their willingness to pay for the assistance. 

\section{Study Design}

To understand how much people value AI assistance from LLMs (e.g., ChatGPT)
in completing a writing job, and how the AI assistance impacts both people's 
perceptions about their writing experience
and their final  performance in the writing job, we conduct a randomized human-subject experiment.  

\subsection{Writing Task}\label{sec:task}
Participants of our experiment were asked to write a 200--250 word article within 45 minutes. To examine how people's value of AI assistance in writing as well as the AI assistance's impacts on people's writing perceptions and performance
may vary with the nature of the writing task, 
we considered two kinds of writing jobs in our experiment:
\begin{itemize}
\item \textbf{Argumentative essay writing}: Participants were provided with a statement, and they were asked to write an essay to discuss the argument in the statement. In their essays, participants had the freedom to either support or oppose the argument in the statement. We considered three statements that were sampled from the pool of TOEFL writing exam topics: 
\begin{enumerate}
\item ``Some people think that if companies prohibit sending emails to staff on weekend or during other time out of office hours, staff's dissatisfactions with their companies will decrease. Others think this will not reduce the overall dissatisfactions among staff.''
\item ``Govenrment should put higher tax on junk food to reduce consumption.''
\item ``Nowadays it is easier to maintain good health than it was in the past.''
\end{enumerate} 

\item \textbf{Creative story writing}: Participants were given a prompt, and they were asked to write a story that includes the prompt. Drawing from the array of popular creative writing tasks featured on Reedsy’s Short Story Contest\footnote{ \url{https://blog.reedsy.com/creative-writing-prompts/terms/}.}, we curated three writing prompts:

\begin{enumerate}
    \item (Someone) ``realizes they're on the wrong path.''
    \item Including the line ``We're just too different.''
    \item Someone saying ``Let’s go for a walk.''
\end{enumerate}
\end{itemize}

We focused on argumentative essay and creative story writing in this experiment because these two types of writing tasks are associated with different structures and purposes of writing, and potentially require different kinds of capability in writing ~\cite{lee2022coauthor}. Effective argumentative essays have a clear and specific thesis statement, and are based on credible and even persuasive evidence. In contrast, creative stories are meant to express one's experience and feelings, and can be more imaginary and loosely-structured while placing high demand on creativity. To ensure that the argumentative essay and creative story writing tasks chosen for our experiment have a comparable and reasonable difficulty level, we conducted a pilot study to test the difficulty of different argumentative essay statements and creative story prompts.      
In this pilot study, participants were asked to complete the writing task on their own. 
For the final set of 3 statements and 3 prompts selected for our experiment, as detailed above, our pilot study results suggested that participants could successfully complete the essay/story writing job within the time limit, yielding articles of satisfactory quality.

\subsection{Experimental Treatments}
\label{sec:treatments}
In our experimental design, we evaluate three different writing modes, each reflecting varying degrees of assistance from and collaboration with a state-of-the-art LLM, i.e., ChatGPT. These modes encompass: (1) Independent writing without AI assistance (``{\em Independent}''), (2) Writing with editing assistance from ChatGPT (``{\em Human-Primary}''), and (3) Collaborative writing with ChatGPT (``{\em AI-Primary}'').

\begin{itemize}
    \item \textbf{Independent writing ({\em Independent})}: In this writing mode, participants needed to complete the writing job independently without any assistance from ChatGPT.
    
    \item \textbf{Writing with editing assistance from ChatGPT ({\em Human-Primary})}: In this writing mode, participants were asked to take the primary responsibility for writing the article, while ChatGPT would be provided to assist them in editing and polishing their written content. 
    Figure~\ref{fig:interface_human_primary} in the Appendix shows an example of the writing interface for participants in this mode. Specifically, participants could send any sentences or paragraphs in their writing to ChatGPT for polishing (via the ``Text Polishing Zone'' in Figure~\ref{fig:interface_human_primary}). To ensure that ChatGPT would only provide editing assistance to participants, we covertly crafted a prompt employing the OpenAI API by appending the following instructions ahead of the text that participants sought to polish:
    ``You should only edit or polish the texts I send to you. Please do not write any new content.'' We then sent this formulated prompt to ChatGPT, and subsequently displayed the polished text returned by ChatGPT to the participants (in the ``History Zone'' in Figure~\ref{fig:interface_human_primary}). Participants would be then granted the autonomy to decide whether and how to incorporate the polished text into their writing. 
        

    \item \textbf{Collaborative writing with ChatGPT ({\em AI-Primary}):} In this writing mode, ChatGPT would take the primary responsibility for writing the initial version of the article. Thereafter, participants could provide feedback to ChatGPT and engage in conversational interactions to steer the refinement of the article. Figure~\ref{fig:interface_ai_primary} in the Appendix shows an example of the writing interface for participants in this mode. Specifically, in accordance with the randomly assigned writing task, we initiated the process by directly sending a prompt aligned with the writing topic instruction (e.g., ``Write an article for the statement `Govenrment should put higher tax on junk food to reduce consumption.' '' or ``Write a story that includes someone saying `Let’s go for a walk.' ''). 
    This initial draft generated by ChatGPT was subsequently presented to participants (in the ``Interaction History Zone'' in Figure~\ref{fig:interface_ai_primary}). Participants were then given the opportunity to offer feedback to ChatGPT, guiding it in refining the initial draft. This feedback would be directly relayed to ChatGPT, and the revised draft returned by ChatGPT would again be shown to participants. Through several iterative cycles, participants could collaborate with ChatGPT to further refine the content. In the final stage, participants could freely incorporate any part of the ChatGPT-generated texts with their own composition.

    
\end{itemize}

Recall that a key objective of our study is to quantify 
how much financial value that people attach to
different kinds of writing 
assistance that generative AI-powered assistants could offer. This amount can be computed as 
the differences between the wage that is acceptable for people to complete a writing job independently and the wage that is acceptable for people to complete the writing job with various kinds of AI assistance of interests. 
Thus, following the classical methods in economics for estimating the ``willingness to pay/accept'' ~\cite{rosen1986theory, mas2017valuing,liang2023hidden}, we created two experimental treatments:
\begin{itemize}
    \item \textbf{Independent vs. Human-Primary}: In this treatment, after the topic of the writing job (i.e., the statement for writing an argumentative essay or the prompt for writing a creative story) was revealed to participants, we presented participants with two job offers: The first offer paid the participant \$3 to complete the writing job in the \textbf{\em independent} writing mode, 
    while the second offer paid the participant \$$x$ to complete the writing job in the \textbf{\em human-primary} mode, where $x$ was randomly sampled from the set $\{1.5, 2, 2.25, 2.5, 2.75, 3, 3.25, 3.5, \\ 3.75, 4, 4.5\}$. Participants were asked to make a selection between these two offers, and subsequently complete the writing job in accordance with the writing mode specified by the selected offer\footnote{We conducted a pilot test to validate the appropriateness of the lower and upper bounds, \$1.5 and \$4.5, respectively, for estimating participants’ willingness to pay for AI assistance. This consideration was made given that a majority of participants in our pilot study would 
    prefer not 
    to choose the job with AI assistance at the \$1.5 wage level, while conversely, most participants would select the job with AI assistance at the \$4.5 wage level.}.
    \item \textbf{Independent vs. AI-Primary}: In this treatment, after the topic of the writing job was revealed to participants, we presented participants with two job offers: The first offer compensated the participant with \$3 for completing the writing job in the 
    \textbf{\em independent} writing mode, 
    while the second offer compensated the participant with \$$x$ for completing the writing job 
    in the \textbf{\em AI-primary} mode, where $x$ was again randomly sampled from the set of values as that in the previous treatment. Participants were asked to make a selection between these two offers, and would then complete the writing job in the writing mode specified by the offer that they chose. 
\end{itemize}

Note that to ensure participants could make an informed selection between the two job offers, in both treatments, participants would be initially directed to watch a 2-minute video that introduces to them the type of writing assistance that they could receive from ChatGPT and acquaints them with the writing interface they would use upon selecting the job offer with AI assistance
(i.e., the offers associated with \textbf{\em human-primary} or \textbf{\em AI-primary} writing modes). Participants could only make their job offer selection after finish watching this video. 

\subsection{Experimental Procedure}

Our study was opened only to U.S. workers whose primary language is English on Prolific, and each worker was only allowed to participate in our study once. Each participant went through a few stages in our study, as detailed below.

\paragraph{\textbf{Background assessment.}}
Upon arrival of the study, participants were first asked to fill out a questionnaire to report their demographic information (e.g., gender, age, education). We then asked  participants to indicate how confident they were in completing six types of writing tasks, including creative writing (e.g., stories, novels), writing argumentative essays, writing emails or letters, writing product or book reviews, writing business reports or proposals, and writing blogs. For each type of writing tasks, participants reported their confidence on a 5-point Likert scale from 1 (very low) to 5 (very high). 

\paragraph{\textbf{Treatment assignment and writing mode selection.}} Subsequently, participants were randomized into one of the two treatments, ``{\em independent vs. human-primary}'' or ``{\em independent vs. AI-primary}''. They were then presented with their writing task, which could be either writing an argumentative essay or writing a creative story, and the statement/prompt used for the writing task was also selected at random from the candidate pool. 
Next, depending on the experimental treatment the participant was assigned, they would be presented with the corresponding two job offers, and the random payment value \$$x$ ($1.5\le x\le 4.5$) of the offer that provided AI assistance to participants would be realized from its 
set of candidate values. 
Participants were told that their final payment from the study would consist of three parts: (1) a {\em base payment} of \$2; (2) the {\em writing job payment} as specified in the job offer that they would choose; and (3) an (optional) {\em performance-based payment} of \$2. We informed participants that their submitted articles would be sent to other crowd workers for review. If the average rating of their article would rank within the top 10\% of the articles written for the same topic, they would receive the performance-based payment\footnote{Upon the completion of this study, we indeed recruited additional crowd workers from Prolific to evaluate the submitted articles. Their ratings were then used to determine whether each participant in this study would receive the performance-based payment.}.    
Once they were clear on the compensation structure, participants were required to watch an introductory video elucidating how they could potentially collaborate with ChatGPT through the designated interface to accomplish the writing task should they choose the AI-assisted writing mode (i.e., ``{\em human-primary}'' or ``{\em AI-primary}''). 
With all this information, participants then selected their preferred job offer. 

\paragraph{\textbf{Main writing task.}} After the job offer was selected, participants proceeded to the main writing task. They were asked to complete this writing task using the writing mode specified in their chosen offer, and they had a maximum duration of 45 minutes for finishing the writing task. Note that the time required for ChatGPT to respond to participants’ prompts would not be included in the allocated time limit. 

\paragraph{\textbf{Exit survey.}} After completing the main writing task, participants were asked to complete an exit survey. In this survey, participants were again asked to indicate their confidence in 
completing the same six types of writing tasks (e.g., creative writing, argumentative essay, emails/letters, etc.) as we surveyed at the beginning of the study, should they have the chance in the future to complete those tasks {\em in the same writing mode} as they had experienced in our study. 
We also asked a series of survey questions to guage participants' perceptions of their writing experience in our study. 
For example, 
the NASA Task Load Index (NASA TLX)~\cite{hart1988development} was used to measure the cognitive load that participants experienced during the writing task, including their mental demand, time pressure, amount of effort taken, and frustration level. 
To understand participants' perceptions of the overall writing processes, we presented the following statements to participants and asked them to rate how much they agreed with each statement on a 5-point Likert scale from 1 (strong disagree) to 5 (strongly agree): 
\begin{itemize}
\item \textbf{(Satisfaction)}: ``I am satisfied with the writing process.''
\item \textbf{(Enjoyment)}: ``I enjoy the writing process.''
\item \textbf{(Ease)}: ``I find it easy to complete the writing process.''
\item \textbf{(Ability of self-expression)}: ``I was able to express my creative goals during the writing process.'' 
\end{itemize}

Similarly, we asked participants to rate their agreement with the following statements, again on a 5-point Likert scale, to understand their perceptions of the final writing outcome:
\begin{itemize}
\item \textbf{(Quality)}: ``I am satisfied with the quality of the final article.''
\item \textbf{(Ownership)}: ``I feel ownership over the final article.''
\item \textbf{(Pride)}: ``I'm proud of the final article.''
\item \textbf{(Uniqueness)}: ``The article I submitted feels unique.''
\end{itemize}

In addition, to understand participants' accountability should their article be criticized for various issues during the evaluation process, 
we asked participants to rate their agreement in the following statements, on a 5-point Likert scale:
\begin{itemize}
\item \textbf{(Deceptive content)}: ``I'm willing to take the responsibility if my article is criticized for containing deceptive content (e.g., misinformation).''
\item \textbf{(Plagiarism)}: ``I'm willing to take the responsibility if my article is criticized for containing content that is highly similar to someone else's writing.''
\item \textbf{(Privacy invasion)}: ``I'm willing to take the responsibility if my article is criticized for containing content that invades someone else's privacy.''
\item \textbf{(Discrimination)}: ``I'm willing to take the responsibility if my article is criticized as exhibiting bias and discrimination.''
\end{itemize}

Finally, participants reported their familiarity with ChatGPT (1: very unfamiliar; 5: very familiar) and their frequency of using ChatGPT in their daily life or work (1: never; 5: very frequently--more than once a day). 

\paragraph{\textbf{Attention check.}} To filter out inattentive participants, we included two attention check questions in our study. The first attention check question was presented to participants right before they took the exit survey, and it asked the participant to select again which job offer they had previously chosen in the study. The second attention check question was included in the exit survey, where participants were required to select a randomly pre-specified option in the question. We considered only the data from participants who passed both attention check questions as valid data. 

\begin{figure*}[t]
\centering
\begin{subfigure}[b]{.33\textwidth}
\centering
\includegraphics[width=1.02\textwidth]{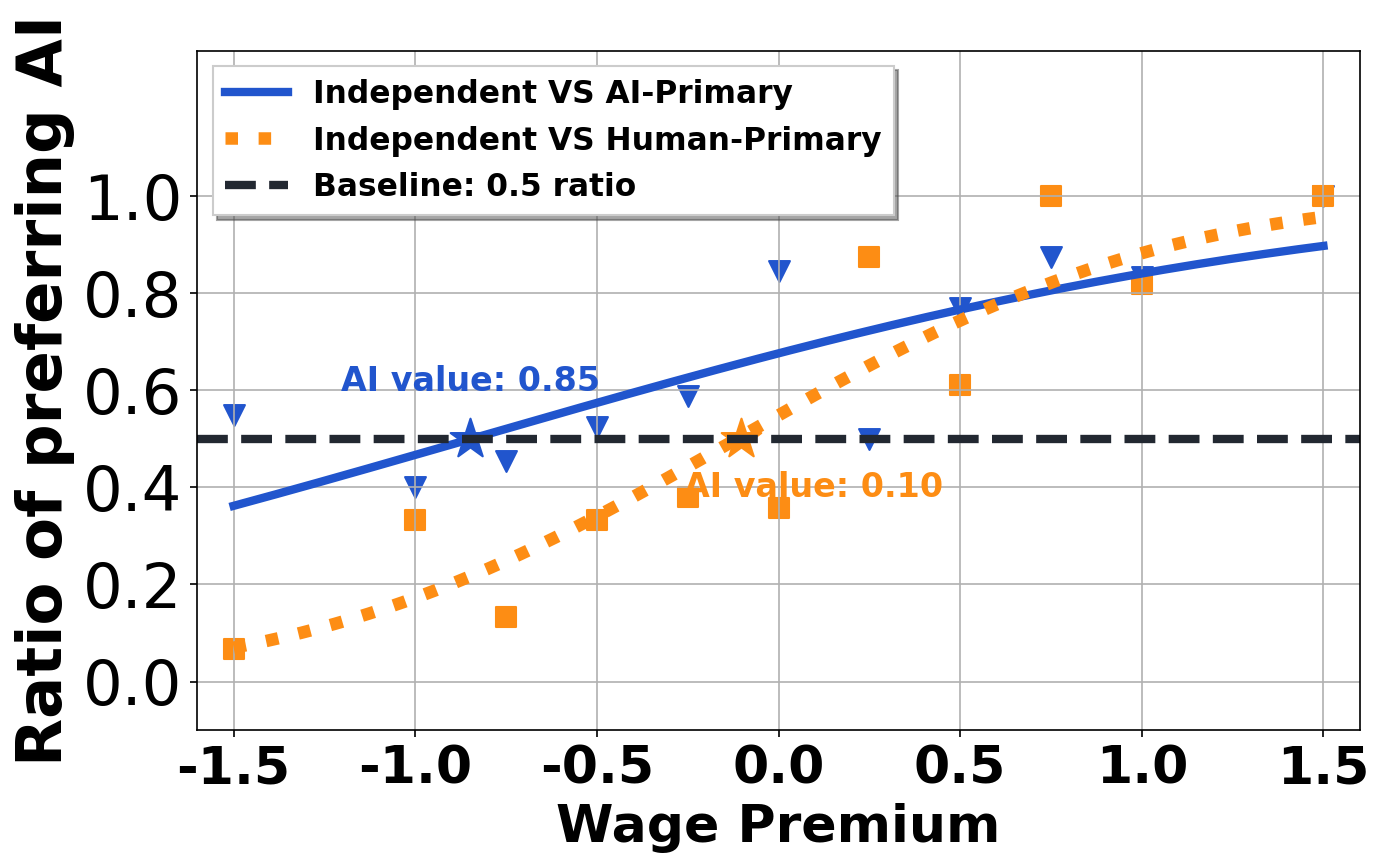}
\caption{Ratio of preferring AI: All data}
\label{wtp_all}

\end{subfigure}
\centering
\begin{subfigure}[b]{.33\textwidth}
\includegraphics[width=1.02\textwidth]{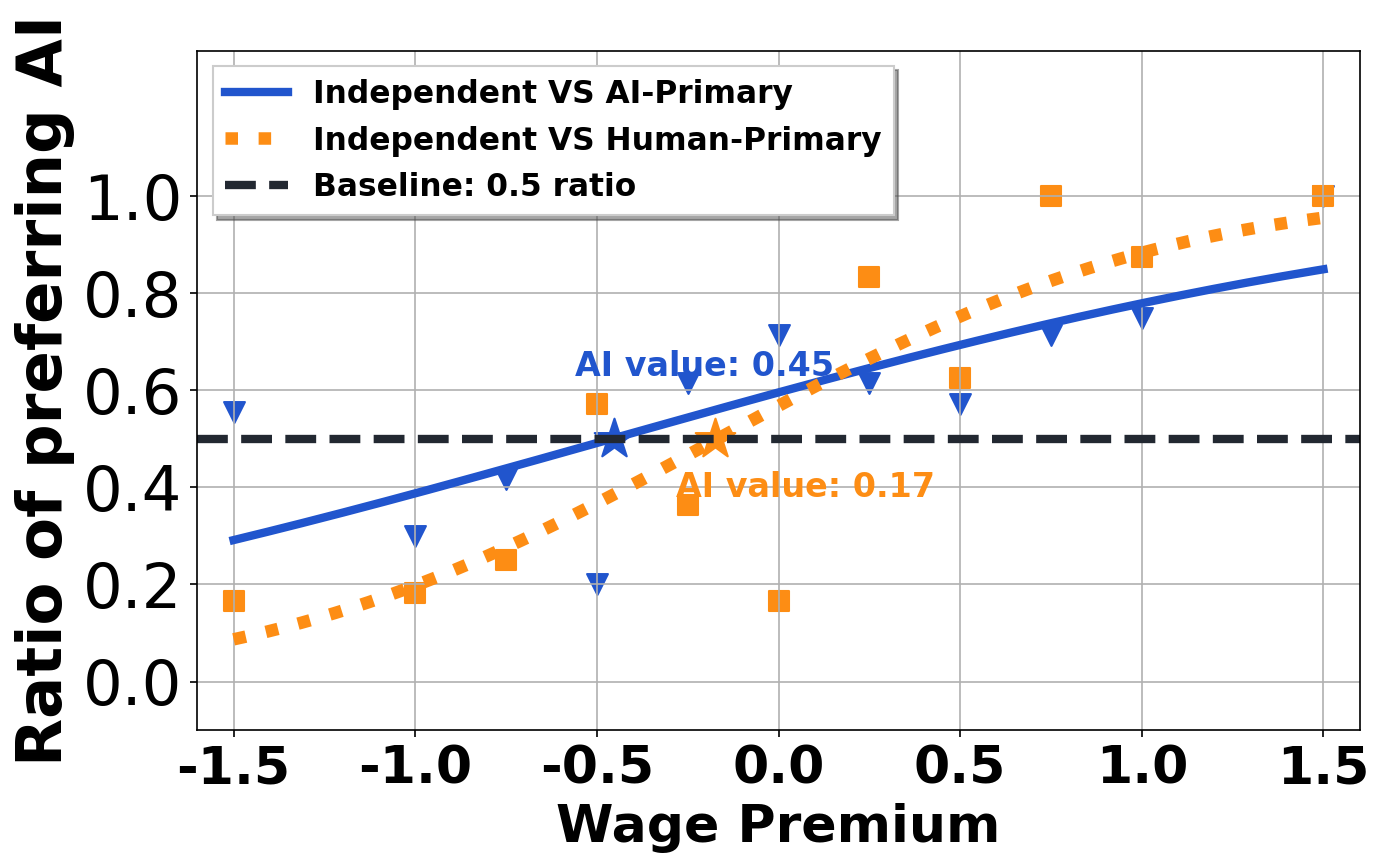}
\caption{Ratio of preferring AI: Argument only}
\label{wtp_statement}

\end{subfigure}
\centering
\begin{subfigure}[b]{.33\textwidth}
\includegraphics[width=1.02\textwidth]{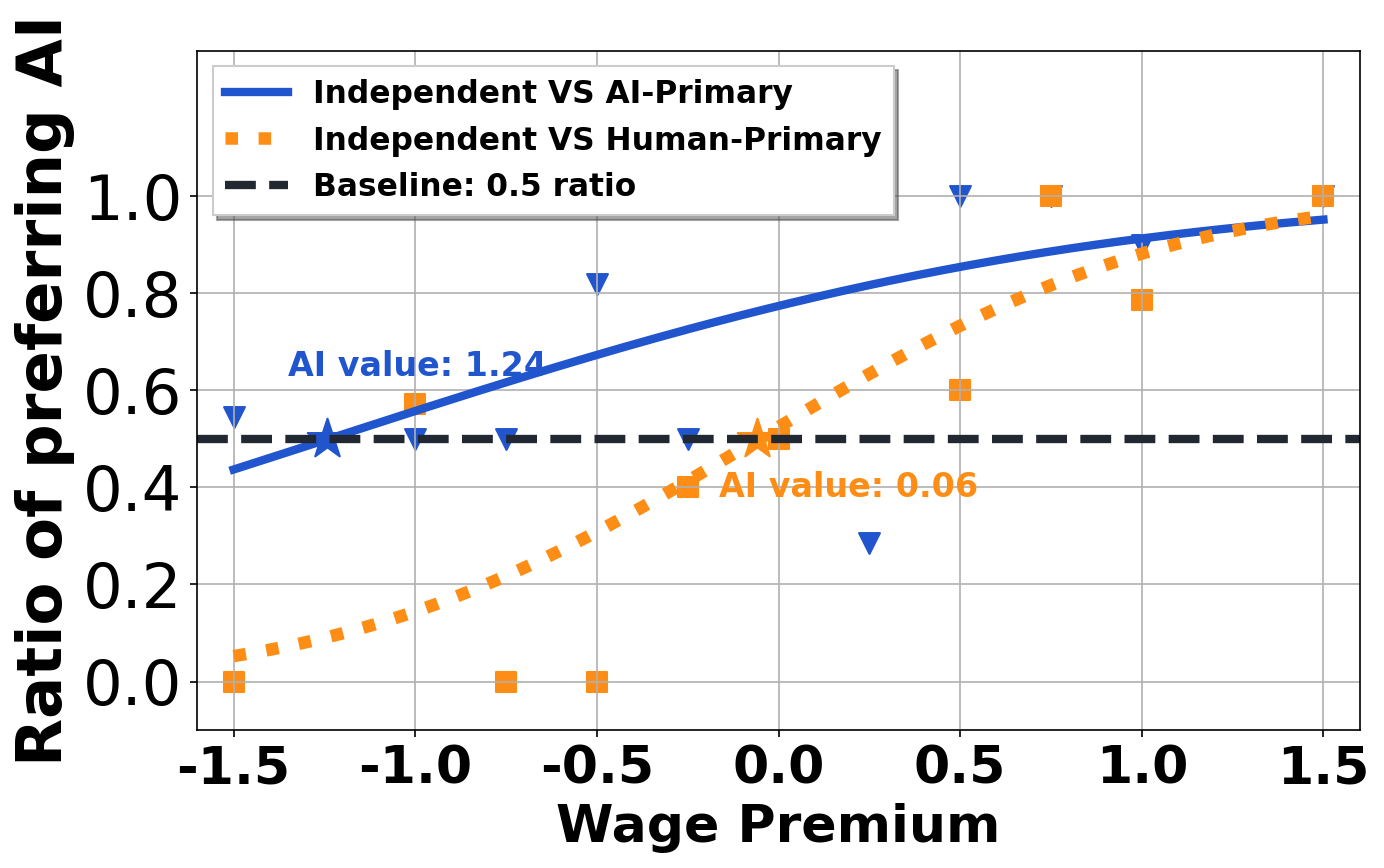}
\caption{Ratio of preferring AI: Creative story only}
\label{wtp_story}

\end{subfigure}

\caption{Comparing participants' probability of preferring an AI-assisted writing mode over the independent writing mode, (a) when writing either argumentative essays or creative stories, (b) when writing argumentative essays (arguments) only, or (c) when writing creative stories only.  
The wage premium is defined as the difference in the writing job payment participants would receive from selecting the AI-assisted writing mode, relative to the fixed payment of \$3 from the independent writing mode. Triangles (or squares) represent the actual fraction of participants who chose the AI-primary (or human-primary) writing mode over the independent writing mode in our dataset. The blue solid line (or the orange dotted line) depicts the predicted 
probability, derived from the fitted probit model, of participants opting for the AI-primary (or human-primary) writing mode over the independent writing mode. The AI assistance's value is the negative of the wage premium for the AI-assisted job that makes participants indifferent between the AI-assisted mode and the independent writing mode (i.e., choosing the AI-assisted writing mode with a probability of 0.5), which is marked by pentagrams.
}
\label{fig:wtp}
\end{figure*}

\subsection{Analysis Methods}
\label{sec:methods}
\paragraph{\textbf{Quantify the value of AI assistance.}}
We measure the value of AI assistance to participants following the econometric methods for estimating ``willingness to pay/accept'' ~\cite{rosen1986theory, mas2017valuing,liang2023hidden}, that is, the maximum amount participants would be willing to pay to acquire different kinds of AI assistance (e.g., editing assistance in the ``{\em human-primary}'' writing mode, or content generation assistance in the ``{\em AI-primary}'' writing mode).
Specifically, given the experimental data collected regarding participants' choices between the independent writing mode and a randomly assigned AI-assisted writing mode, we started by fitting a Probit regression model to predict how likely participants would prefer the AI-assisted writing mode over the independent writing mode, considering the wage difference between these two modes:  
\begin{equation}
    \text{Pr(select AI assistance=1)} =\phi(\Delta w)
\end{equation}
where $\Delta w$, defined by the equation $\Delta w = x - 3$, denotes the ``{\em wage premium}'' associated with the writing job offering AI assistance in comparison to the alternative independent writing job. 

Utilizing this fitted model, we then identified the threshold of wage premium $\Delta w^{*}$ such that $\phi(\Delta w^{*})=0.5$. In this scenario, participants would be indifferent between the independent writing job offer at a wage level of \$3 and the alternative job with AI assistance at a wage level of \$($3+\Delta w^{*}$). We can then estimate the value of AI assistance to participants as $-\Delta w^{*}$, representing the maximum amount of money they are willing to forgo in order to acquire the AI assistance in completing their writing jobs\footnote{In other words, compared to accepting the independent writing job offer at a wage level of \$3, participants are equally willing to adopt the AI assistance at a wage level of \$(3+$\Delta w^{*}$), and ``pay'' \$$(-\Delta w^{*})$ to the provision of the AI assistance. That is, participants' willingness to pay for the AI assistance is \$$(-\Delta w^{*})$.}. 




\paragraph{\textbf{Compare writing perceptions and performance across writing modes.}}
To understand how the AI assistance impacts participants' perceptions of the writing experience and their writing performance,  
we conducted regression analysis using our experimental data. In particular, the primary independent variable of the regressions was the writing mode selected by the participant. The dependent variables included a set of writing perception metrics that we obtained from participants' self-reports in the exit survey (e.g., participants' cognitive load, perceptions of the writing processes, perceptions of the final writing outcome, changes in writing confidence, perceptions of accountability), 
as well as a set of objective metrics of participants' writing performance (e.g., time spent, number of grammatical and spelling mistakes made, coherence of the article). 
To minimize the impact of potential confounding variables, we also accounted for a set of covariates in the regressions, including the participants' demographic background (age, gender, race, education), the writing job payment they received from their selected job offer, their confidence in writing the type of articles that they were asked to work on in the study, their familiarity with ChatGPT, and their frequency of using ChatGPT. Note that in our analysis, some key concepts were measured through multiple dependent variables (e.g., participants' perceptions of the final writing outcome was measured with respect to their perceived quality, ownership, pride, and uniqueness of the final article); in this case, we considered these dependent variables to belong to the same family and applied Bonferroni corrections to correct the $p$ values for multiple comparisons.  For clarity, we used {\em adjusted}-$p$ in the following to indicate the corrected $p$-values whenever Bonferroni corrections are applied. Finally, for visualization purposes, we plotted the predicted values of dependent variables for a participant with average characteristics at an average wage premium across three different writing modes.

\section{Results}
In total, 379 workers from Prolific took our study and passed the attention check. Among them, 183 were allocated to the ``{\em independent vs. human-primary}'' treatment, while the remaining 196 were placed in the ``{\em independent vs. AI-Primary}'' treatment\footnote{See Appendix~\ref{app-demo} for the demographic statistics of participants in our study.}. 
The average hourly payment workers received in our study was \$13.64.
In the following, we quantify the value participants place on AI assistance and analyze the impact of AI assistance on 
people's perception and performance in writing based on the experimental data we obtained from these workers.

\subsection{Estimating the value of AI assistance}

\noindent \textbf{\em People are willing to forgo financial payments to receive AI assistance in writing, and they value AI assistance more when AI can provide assistance in generating writing content. }
Figure~\ref{wtp_all} displays the fitted Probit regression models for predicting the likelihood of participants preferring a writing mode with AI assistance over the independent writing mode at the specified wage premium, when we do {\em not} differentiate the assigned type of writing jobs (i.e., argumentative essay writing and creative story writing). Clearly, we find that people attach a positive financial value to the writing assistance provided by ChatGPT, the
generative AI-powered assistant. Moreover, the value they attach to AI assistance is particularly salient when the AI model can directly generate content for them (as that in the ``{\em AI-primary}'' writing mode) beyond performing just text editing and polishing (as that in the ``{\em human-primary}'' writing mode)---Indeed, our data suggests that participants in our study were willing to forgo \$0.85 to receive ChatGPT's content generation assistance in the ``{\em AI-primary}'' writing mode. This was 28.3\% of the writing job payment that they would receive should they select the independent writing mode, and was much higher than the \$0.10 that they were willing to forgo to receive ChatGPT's editing assistance in the ``{\em human-primary}'' mode. 

To better contextualize the magnitude of the value that people attach to the writing assistance provided by ChatGPT, we translate the value into the amount of hourly payment participants were willing to give up in order to receive the AI assistance. To do so, we used our experimental data to learn a linear regression model to predict the amount of time participants would spend on completing a writing job, given the writing mode they chose as well as the amount of writing job payment they received from their chosen job offer. Our regression model also controlled for several covariates, including the participants' demographic background (e.g., age, gender, race, education), their confidence in the type of writing task that they were asked to complete, their familiarity with ChatGPT, and the frequency of their ChatGPT usage. Given this prediction model, we found that an average participant is predicted to  spend 13.7 minutes in completing a writing job in the ``{\em independent}'' mode to get the \$3 fixed writing job payment, resulting in an hourly wage of \$13.12. Meanwhile, when an average participant was willing to forgo \$0.85 to adopt ChatGPT's content generation assistance in the ``{\em AI-primary}'' mode, they are predicted to spend 11.3 minutes in completing a writing job to get the writing job payment of $\$2.15$ (i.e., $\$3-\$0.85=\$2.15$), leading to an hourly payment of \$11.41. In other words, the value of \$0.85 that participants in our study attached to ChatGPT's content generation assistance can be translated to an hourly wage of \$1.71 (i.e., \$13.12-\$11.41=\$1.71). As participants' average hourly wage in completing our entire study (including the writing job and surveys) was \$13.64, the \$1.71 hourly wage that they were willing to give up to receive ChatGPT's content generation assistance represents 12.5\% of their hourly payment, which is non-trivial.   Following a similar computation, we also found that the \$0.10 participants were willing to forgo to receive ChatGPT's editing assistance in the ``{\em human-primary}'' mode is effectively equivalent to giving up an hourly wage of \$0.42.


\vspace{2pt}
\noindent \textbf{\em The value of AI's editing assistance does not vary much with the nature of the writing job, but the value of AI's content generation assistance does. }
We then investigated into if people's value of AI assistance in writing differs across different types of writing tasks.
Figure~\ref{wtp_statement} and Figure~\ref{wtp_story} show the values that participants attributed to AI assistance when tasked with writing
an argumentative essay or a creative story, respectively. These values are derived from the Probit regression model 
as outlined previously. Here, we observed that in the {\em human-primary} writing mode, where ChatGPT was configured to exclusively polish the text generated by participants, the value participants attached to the AI assistance remained relatively stable as the type of writing jobs changed (i.e., the AI's editing assistance worth \$0.17 in the context of argumentative essay writing jobs and \$0.06 in the context of creative story writing jobs). In contrast, in the {\em AI-primary} writing mode, participants placed a much higher value on ChatGPT's content generation assistance when they were asked to write creative stories instead of argumentative essays (\$1.24 for creative stories and \$0.45 for argumentative essays). Regardless of the type of the writing tasks, we still found that people tend to value AI's content generation assistance more than its editing assistance. 



\begin{figure}[t]
\centering
\begin{subfigure}[b]{.235\textwidth}
\includegraphics[width=1\textwidth]{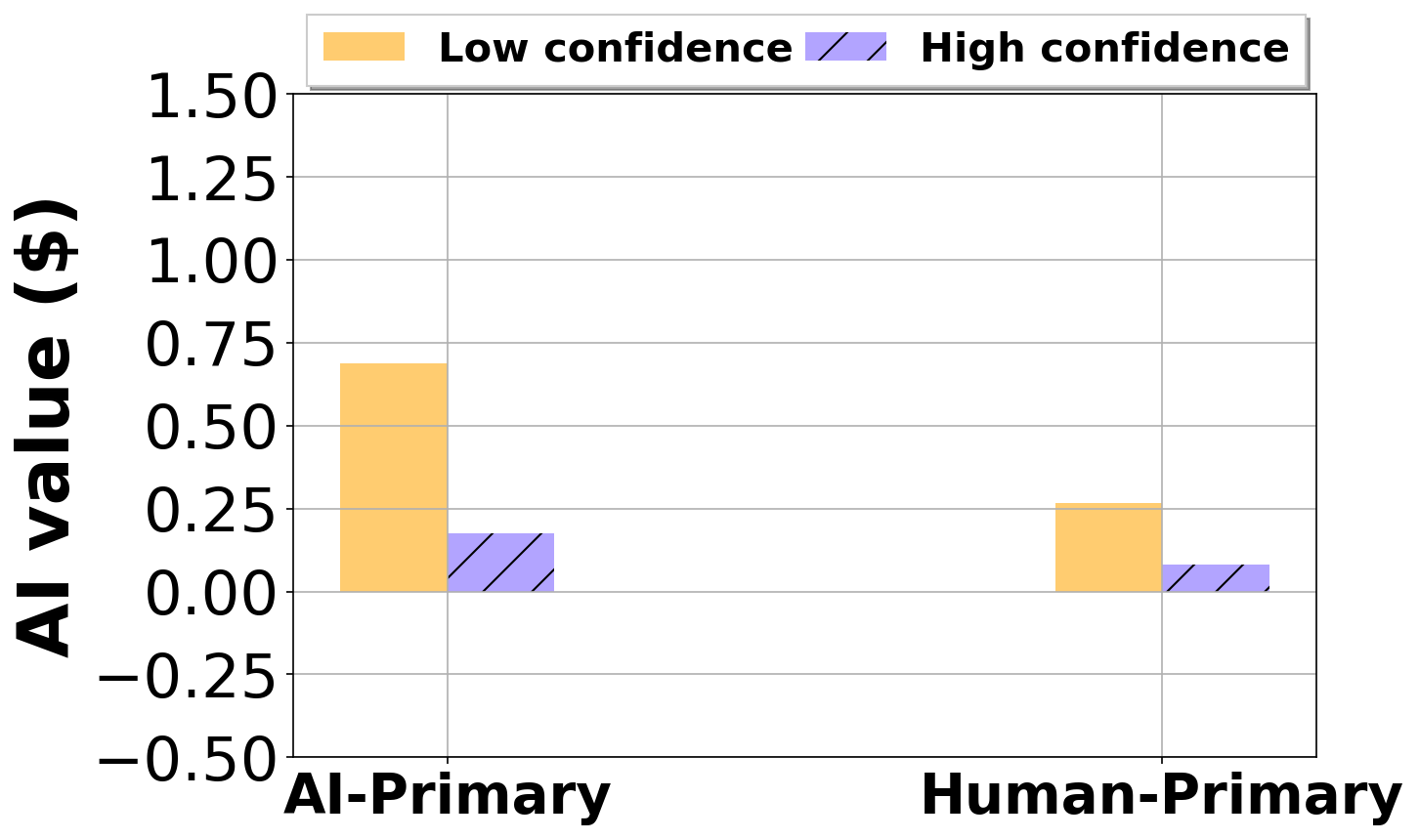}
\caption{Argumentative essay}
\label{fig:val_conf:argue}
\end{subfigure}
\centering
\begin{subfigure}[b]{.235\textwidth}
\includegraphics[width=1\textwidth]{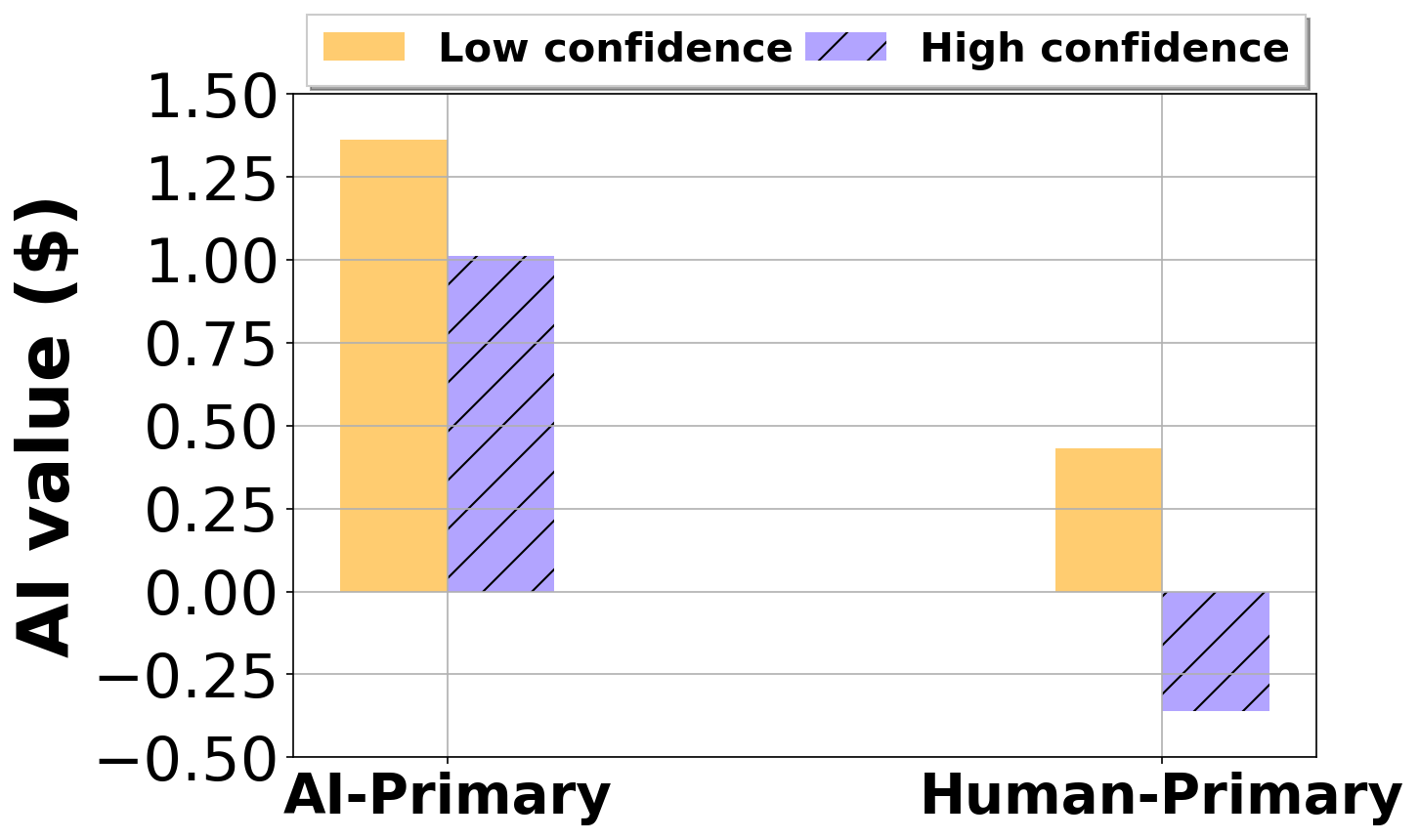}
\caption{Creative story}
\label{fig:val_conf:creative}
\end{subfigure}

\caption{Comparing the estimated value of AI assistance between participants with varying levels of writing confidence for different types of writing tasks. 
}
\vspace{-10pt}
\label{fig:value_confidence}
\end{figure}

\vspace{2pt}
\noindent \textbf{\em Writers with higher confidence generally attach lower value to AI assistance. } 
To understand whether people with different characteristics attach different values to AI assistance, we estimated the value of AI assistance for different subsets of people separately. First, we looked into the value of AI assistance for people with different levels of confidence in their writing skills. Specifically, for the set of participants who were asked to complete an argumentative essay writing job (or a creative story writing job), we divided them into two subsets based on a median split of participants' confidence in writing argumentative essay (or creative writing) that they self-reported at the beginning of our study. For each subset of participants, we used the Probit regression model to estimate their value for AI assistance. Figures~\ref{fig:val_conf:argue} and~\ref{fig:val_conf:creative} compare the average estimates of the AI assistance's value between participants with high and low confidence in completing the assigned type of writing tasks, for argumentative essay writing and creative story writing, respectively. We found that across both types of writing tasks,
people with high confidence in their own writing ability value the AI assistance less.  For example, participants who were more confident in their own argument-writing ability 
attached substantially less value to ChatGPT's content generation assistance  ($\$0.17$ for high-confident people vs. $\$0.68$ for low-confident people) and editing assistance ($\$0.08$ for high-confident people vs. $\$0.26$ for low-confident people)
for writing argumentative essays.
Similarly, participants who have high confidence in their own creative-writing ability also valued ChatGPT's content generation assistance  and editing assistance less in writing creative stories (content generation assistance: $\$1.01$ for high-confident people vs. $\$1.36$ for low-confident people; editing assistance: $-\$0.36$ for high-confident people vs. $\$0.43$ for low-confident people); in fact, they even attached a negative value to ChatGPT's editing assistance in writing creative stories. 

\begin{figure}[t]
\centering
\begin{subfigure}[b]{.235\textwidth}
\includegraphics[width=1\textwidth]{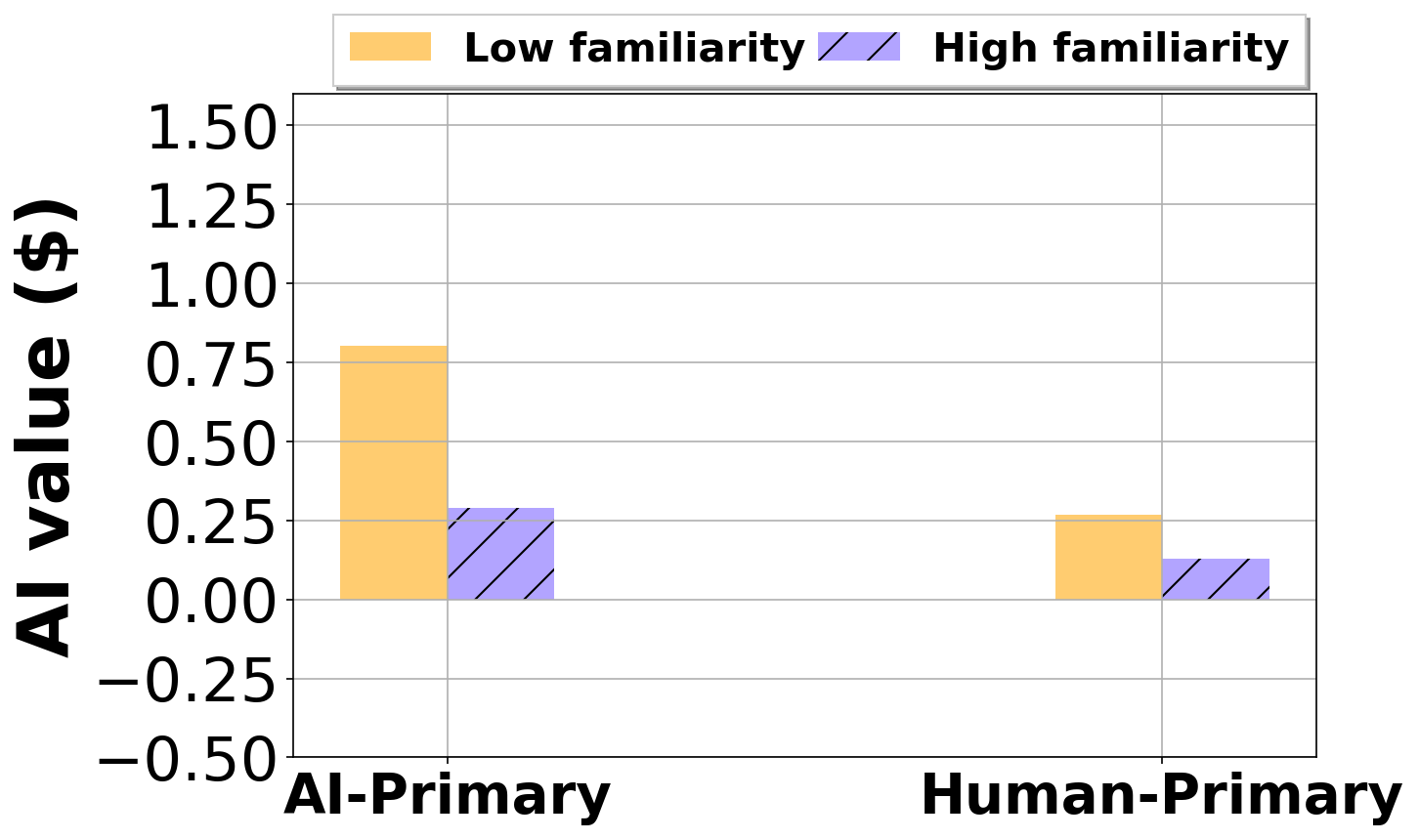}
\caption{Argumentative essay}
\end{subfigure}
\centering
\begin{subfigure}[b]{.235\textwidth}
\includegraphics[width=1\textwidth]{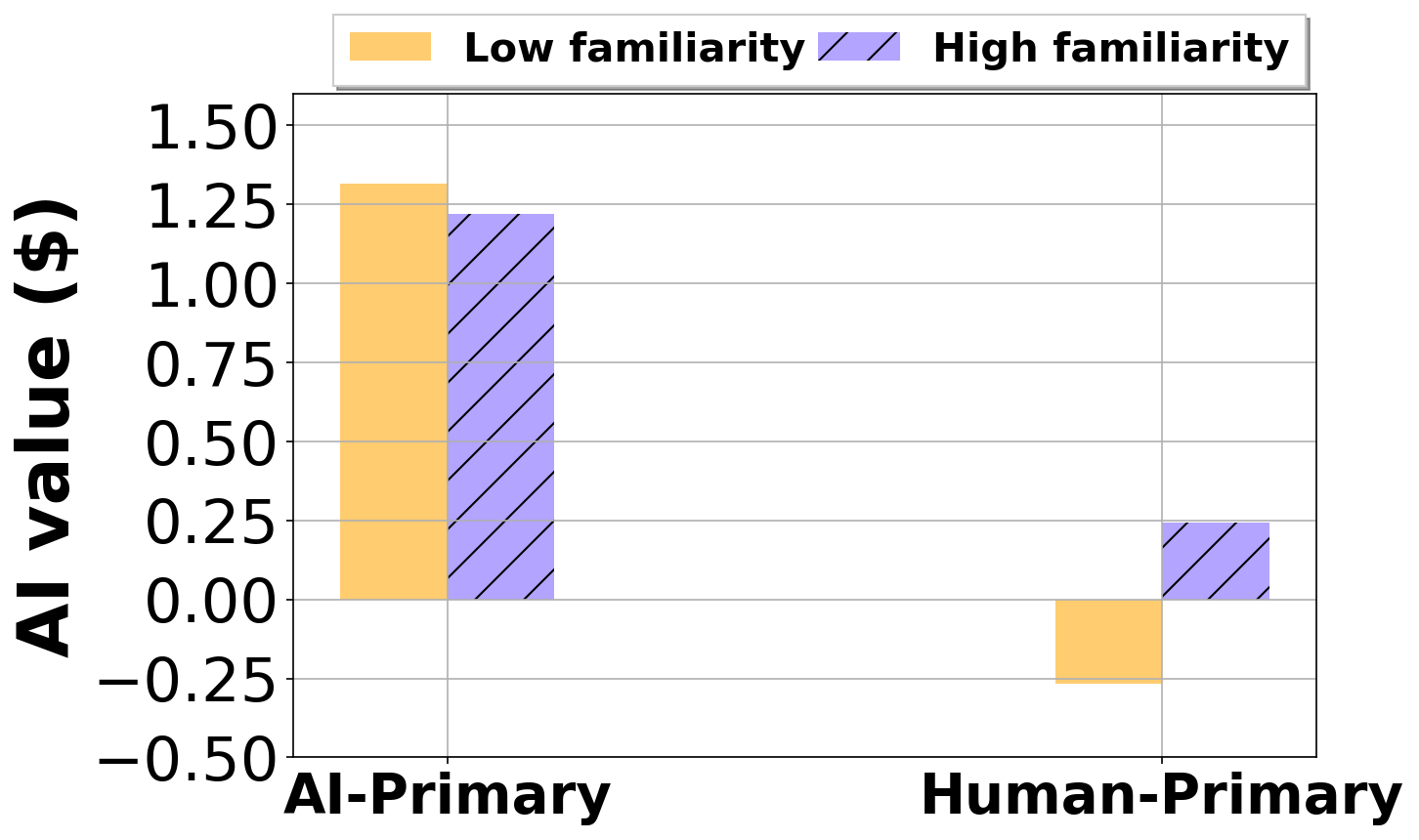}
\caption{Creative story}

\end{subfigure}

\caption{Comparing the estimated value of AI assistance between participants with varying levels of familiarity with ChatGPT for different types of writing tasks. 
}
\vspace{-10pt}
\label{fig:value_fam}
\end{figure}

\vspace{2pt}
\noindent \textbf{\em Writers who are more familiar with ChatGPT attach less value to its assistance in writing argumentative essays, but consider its editing assistance in creative writing to be more valuable.} Finally, we divided participants into two subsets based on a median split of their self-reported level of familiarity with ChatGPT at the end of our study. The average estimated values of different types of AI assistance for people who have high or low level of familiarity with ChatGPT are presented in Figure~\ref{fig:value_fam}. 
We noticed distinct patterns in how familiarity impacts people's perceived value of AI assistance across the two types of writing tasks in our study. For the argumentative essay writing task, regardless of whether ChatGPT provides content generation assistance or only provides editing assistance, those who were more familiar with ChatGPT always considered the value of ChatGPT's assistance to be lower 
than those who were less familiar with ChatGPT ($\$0.12$ for high-familiarity people vs. $\$0.27$ for low-familiarity people in AI editing assistance, and $\$0.29$ for high-familiarity people vs. $\$0.80$ for low-familiarity people in AI content generation assistance). On the other hand, for the creative story writing task, the value of ChatGPT's content generation assistance 
was similar across participants with different familiarity levels with ChatGPT ($\$1.22$ for high-familiarity people vs. $\$1.31$ for low-familiarity people in AI editing assistance).
However, participants with higher familiarity with ChatGPT were found to attach a greater value to ChatGPT's editing assistance in creative writing 
 ($\$0.24$ for high-familiarity people vs. $-\$0.27$ for low-familiarity people).

\subsection{Understanding the impacts of AI assistance in people's perceptions of the writing experience}

Next, we move on to examine how the incorporation of AI assistance in the writing processes changes people's perceptions of their writing experience, including their perceived cognitive load, their perceptions about the overall writing processes, their perceptions about the final writing outcome, changes in their writing confidence, and their accountability perceptions. As discussed in Section~\ref{sec:methods}, these analyses are based on regression models we fit from the data. The complete set of results of the fitted regression models can be found in Appendix~\ref{app-regression}.

\begin{figure}[t]
\centering
\begin{subfigure}[b]{.4\textwidth}
\includegraphics[width=1\textwidth]{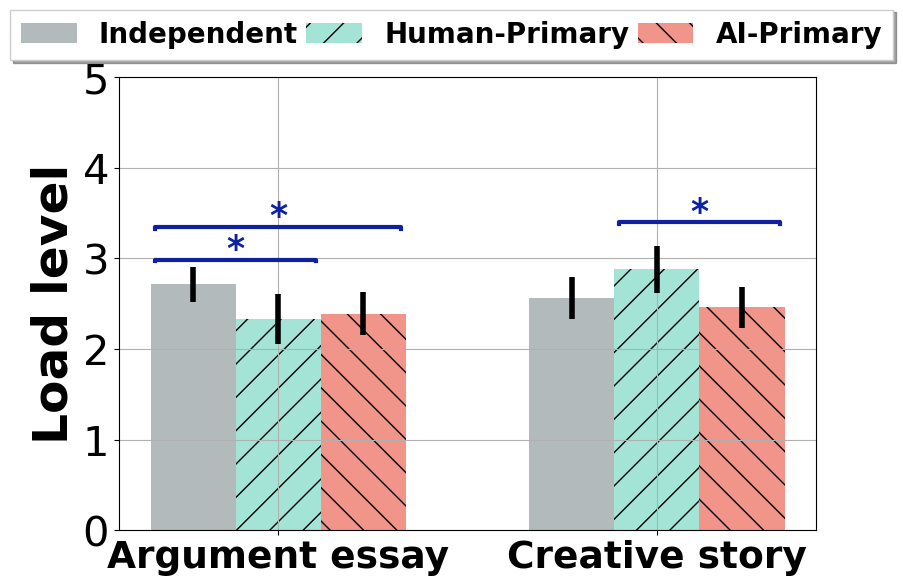}
\end{subfigure}

\caption{Comparing the predicted cognitive load that the average participant experienced for completing the two types of writing tasks (argumentative essay writing and creative story writing), across the {\em independent}, {\em human-primary}, and {\em AI-primary} writing modes. Error bars represent  the 95\% confidence intervals of the predicted cognitive load level.  
$\textsuperscript{*}$  denotes statistical significance levels of $0.05$.
}

\vspace{-10pt}
\label{fig:cognitive_load}
\end{figure}
\subsubsection{Cognitive load}

We computed the overall level of cognitive load that each participant experienced in completing the writing job in our study by averaging
their self-reported levels of mental demand, time pressure, amount of effort
taken, and frustration level, which were rated on a scale from 1 (very low) to 5 (very high). We then fitted a linear regression model to predict participants' cognitive load under the three writing modes.
Figure~\ref{fig:cognitive_load} illustrates the predicted cognitive load for a worker with average characteristics across various writing modes, when tasked with writing an argumentative essay or a creative story.

\vspace{2pt}
\noindent \textbf{\em The provision of AI assistance in writing leads to significant cognitive load decreases in argumentative essay writing but not creative story writing.}
As illustrated in Figure~\ref{fig:cognitive_load}, for the argumentative essay writing task, participants who worked independently generally exhibited higher cognitive loads compared to  those in the ``{\em human-primary}'' writing mode ($p=0.032$) and those in the ``{\em AI-primary}'' writing mode ($p=0.042$). This implies that one of the benefits brought up by AI assistance is that it decreases people's perceived cognitive load in writing argumentative essays. However, we did not observe similar benefits from AI assistance for the creative story writing tasks---among those tasked with writing creative stories, there was no significant difference between the level of cognitive load perceived by participants who completed the task independently and those who received assistance from ChatGPT. 
In fact, those in the ``{\em human-primary}'' mode who received editing assistance from ChatGPT actually reported the highest level of cognitive load, which was significantly higher than that reported by those in the ``{AI-primary}'' writing mode ($p=0.022$).

\begin{figure*}[t]
\centering
\begin{subfigure}[b]{.235\textwidth}
\includegraphics[width=1\textwidth]{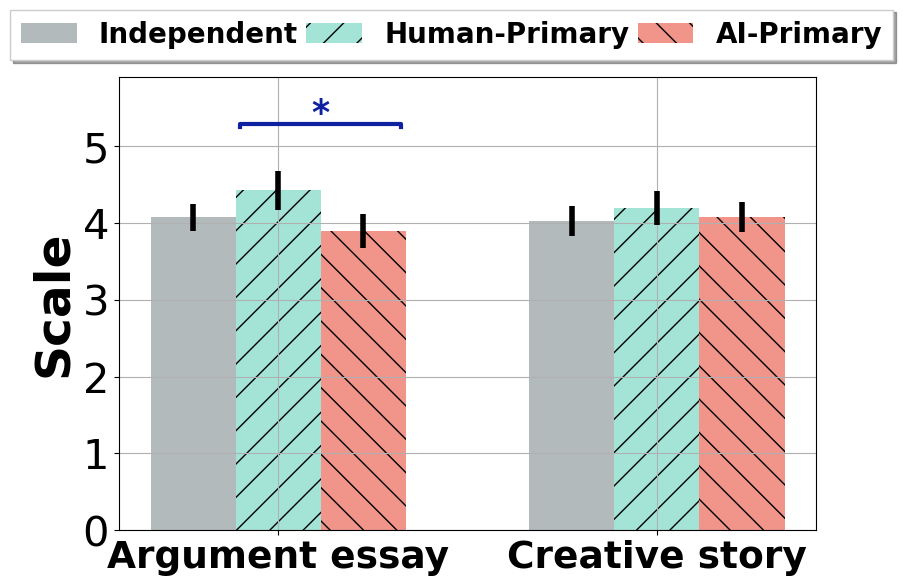}
\caption{Satisfaction}
\label{satisfaction}
\end{subfigure}
\centering
\begin{subfigure}[b]{.235\textwidth}
\includegraphics[width=1\textwidth]{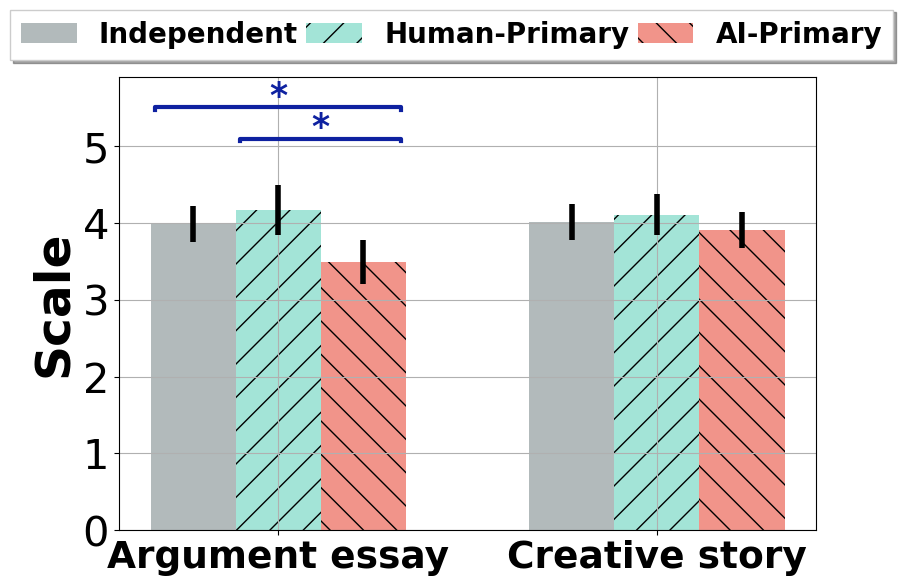}
\caption{Enjoyment}
\label{enjoyment}
\end{subfigure}
\centering
\begin{subfigure}[b]{.235\textwidth}
\includegraphics[width=1\textwidth]{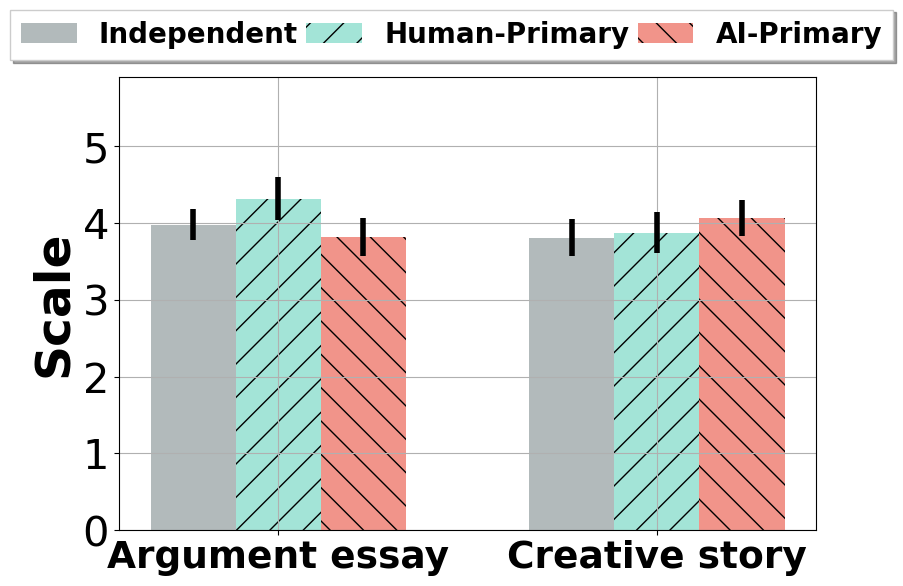}
\caption{Ease}
\label{ease}
\end{subfigure}
\centering
\begin{subfigure}[b]{.235\textwidth}
\includegraphics[width=1\textwidth]{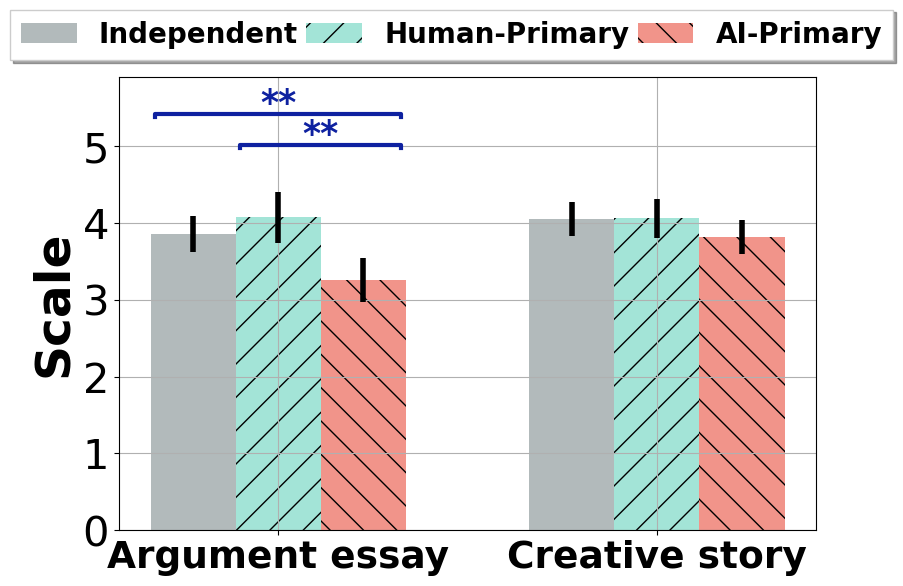}
\caption{Ability of self-expression}
\label{express}
\end{subfigure}
\vspace{-5pt}
\caption{Comparing the predicted perceptions of the average participant regarding the overall writing process for completing the two types of writing tasks (argumentative essay writing and creative story writing), across the {\em independent}, {\em human-primary}, and {\em AI-primary} writing modes. Error bars represent the 95\% confidence intervals.  
$\textsuperscript{*}$ and $\textsuperscript{**}$  denote statistical significance levels of $0.05$ and $0.01$, respectively.
}
\label{fig:feeling_exp}
\end{figure*}

\begin{figure*}[t]
\centering
\begin{subfigure}[b]{.235\textwidth}
\includegraphics[width=1\textwidth]{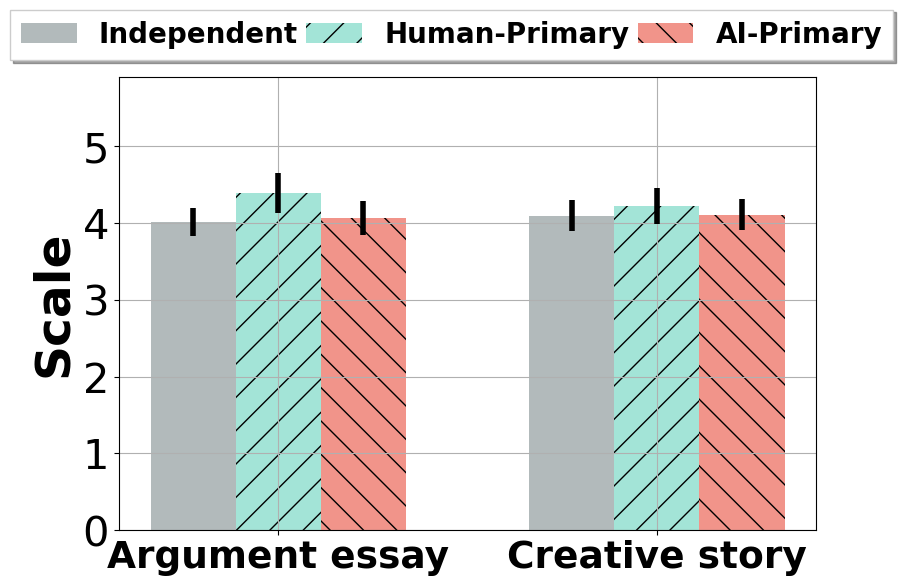}
\caption{Quality}
\label{quality}
\end{subfigure}
\centering
\begin{subfigure}[b]{.235\textwidth}
\includegraphics[width=1\textwidth]{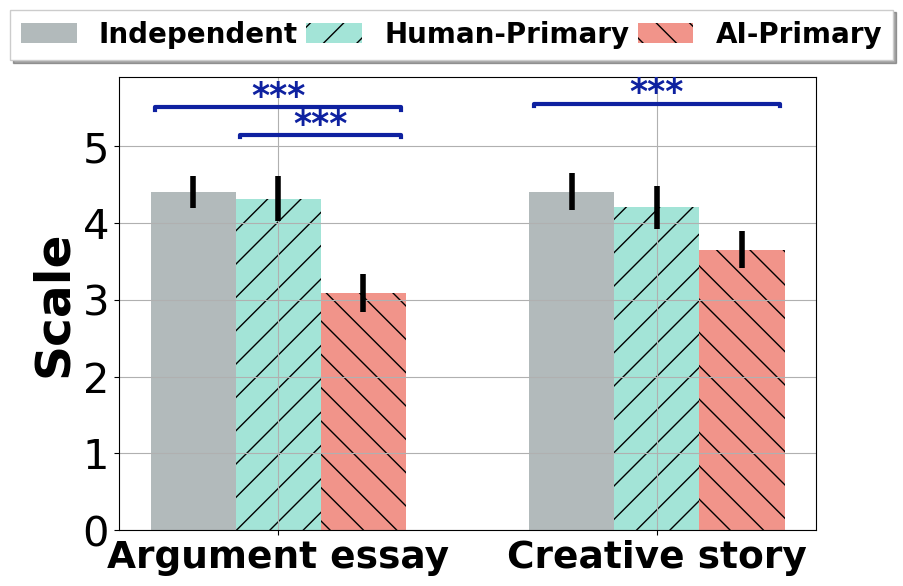}
\caption{Ownership}
\label{ownership}
\end{subfigure}
\centering
\begin{subfigure}[b]{.235\textwidth}
\includegraphics[width=1\textwidth]{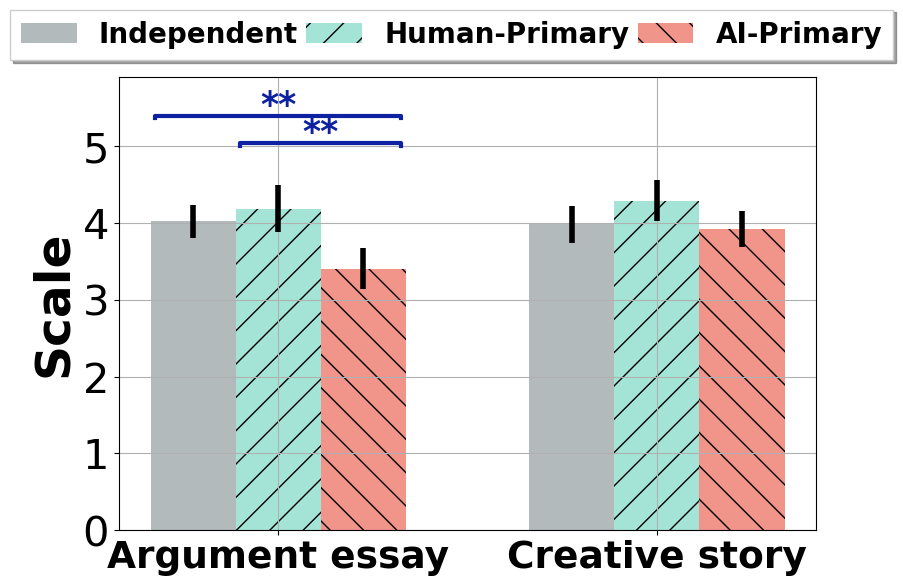}
\caption{Pride}
\label{pride}
\end{subfigure}
\centering
\begin{subfigure}[b]{.235\textwidth}
\includegraphics[width=1\textwidth]{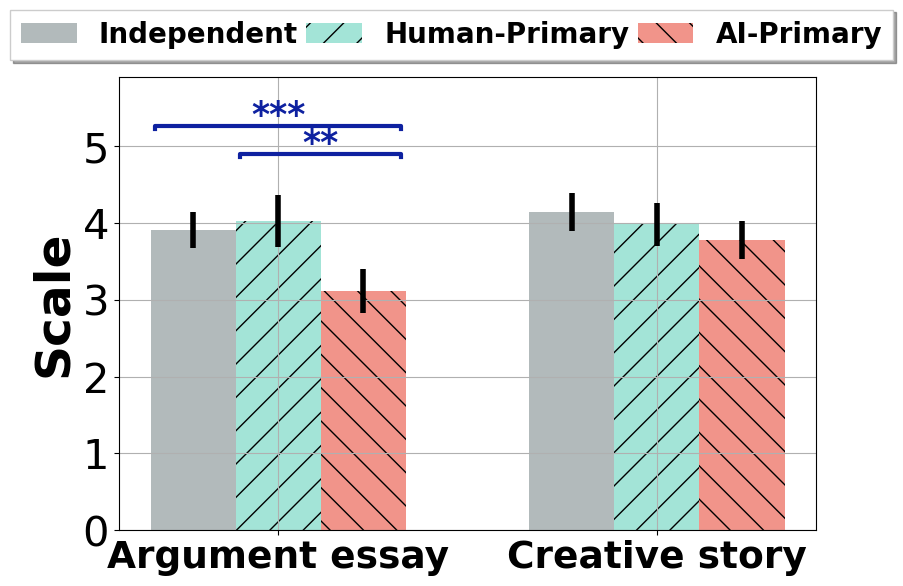}
\caption{Uniqueness}
\label{unique}
\end{subfigure}
\vspace{-5pt}
\caption{Comparing participants' predicted perceptions of the final writing outcome for two types of writing tasks (argumentative essay writing and creative story writing), across the {\em independent}, {\em human-primary}, and {\em AI-primary} writing modes. Error bars represent the 95\% confidence intervals. $\textsuperscript{**}$ and $\textsuperscript{***}$ denote statistical significance levels of  $0.01$ and $0.001$ respectively.
}
\label{fig:feeling_article}
\end{figure*}


\subsubsection{Perceptions about the writing process}
We then examined how participants' perceptions of the overall writing process vary across different writing modes. 
In Figure~\ref{fig:feeling_exp}, we compared participants' perceptions on four aspects of the writing processes---their satisfaction with the writing processes, the enjoyment they experienced during writing, the ease of completing the writing processes, and their perceived ability to express themselves during writing. As discussed in Section~\ref{sec:methods}, these four dependent variables were considered to belong to the same family, thus Bonferroni correction was applied when examining the statistical significance of the treatment effects.

\vspace{2pt}
\noindent \textbf{\em In argumentative essay writing, the provision of content generation assistance by AI significantly decreases people's enjoyment of the writing processes, and it also limits people's ability to express their creative goals.}
A visual inspection of Figure~\ref{fig:feeling_exp} suggests a highly consistent pattern---ChatGPT's editing assistance in the {\em human-primary} mode almost always resulted in participants' highest levels of ratings on various aspects of the writing processes, while ChatGPT's content generation assistance in the {\em AI-primary} mode often led to the lowest levels of ratings.  
These differences were particularly salient when we examine participants' perceptions of the degree to which they enjoyed the writing processes and their ability to express their creative goals while composing argumentative essays.  
For example, we found that when writing argumentative essays, 
participants who received ChatGPT's content generation assistance in the {\em AI-primary} mode reported a significantly lower level of enjoyment during the writing processes compared to those who wrote independently ({\em adjusted}-$p=0.037$) or those who only received editing assistance from ChatGPT in the {\em human-primary} mode ({\em adjusted}-$p=0.044$). 
Similarly, participants who wrote their argumentative essays in the {\em AI-primary} mode also reported a significantly lower level of 
ability to express their creative goals compared to those in the {\em independent} ({\em adjusted}-$p=0.007$) or {\em human-primary} writing modes ({\em adjusted}-$p=0.006$). An additional significant difference that we detected was that participants in the {\em AI-primary} writing mode expressed significantly lower satisfaction with the writing processes than participants in the {\em human-primary} writing mode ({\em adjusted}-$p=0.042$).


\begin{figure*}[t]
\centering
\begin{subfigure}[b]{.33\textwidth}
\includegraphics[width=.95\textwidth]{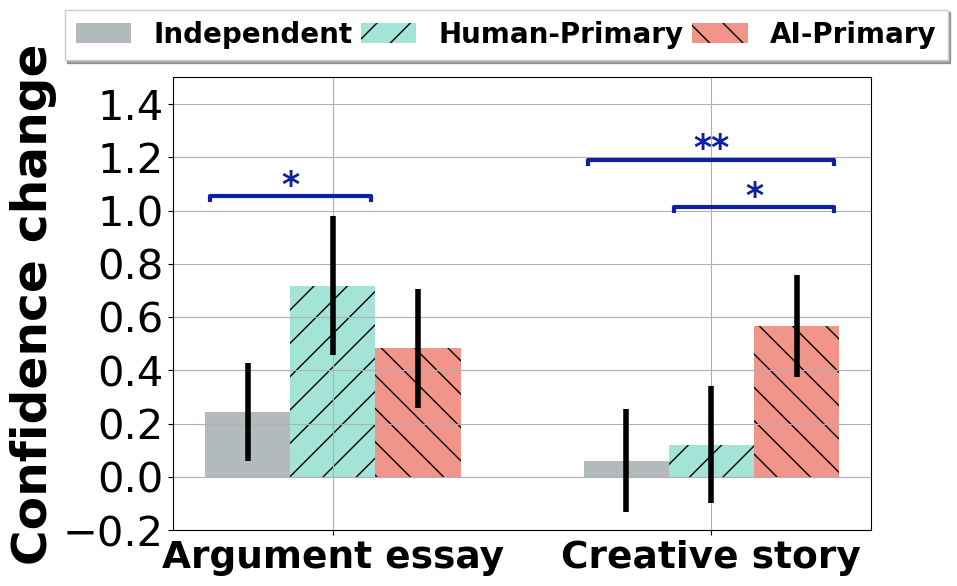}
\caption{Confidence boost in argument writing}
\label{fig:confidence_boost:argue}
\end{subfigure}
\centering
\begin{subfigure}[b]{.33\textwidth}
\includegraphics[width=.95\textwidth]{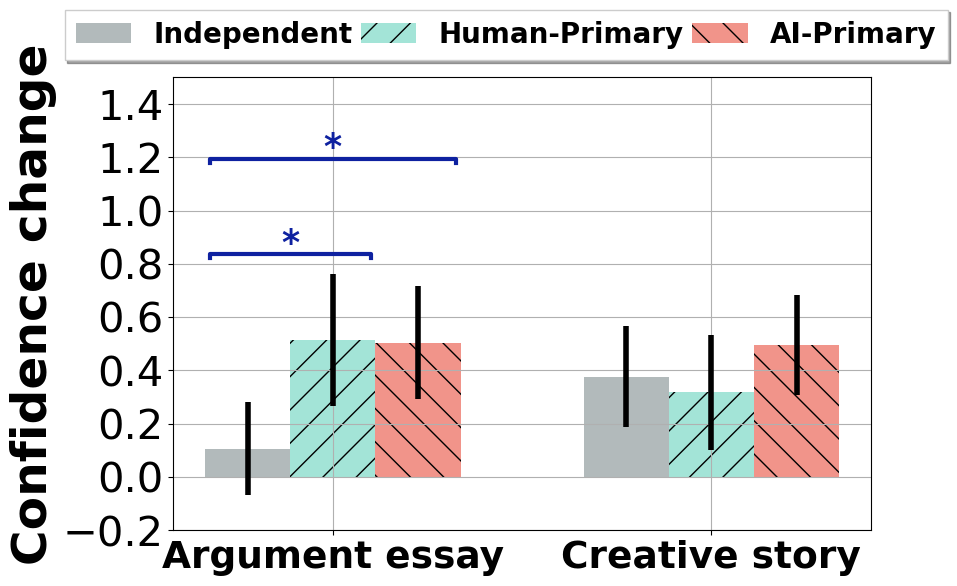}
\caption{Confidence boost in creative writing}
\label{fig:confidence_boost:creative}
\end{subfigure}
\centering
\begin{subfigure}[b]{.33\textwidth}
\includegraphics[width=.95\textwidth]{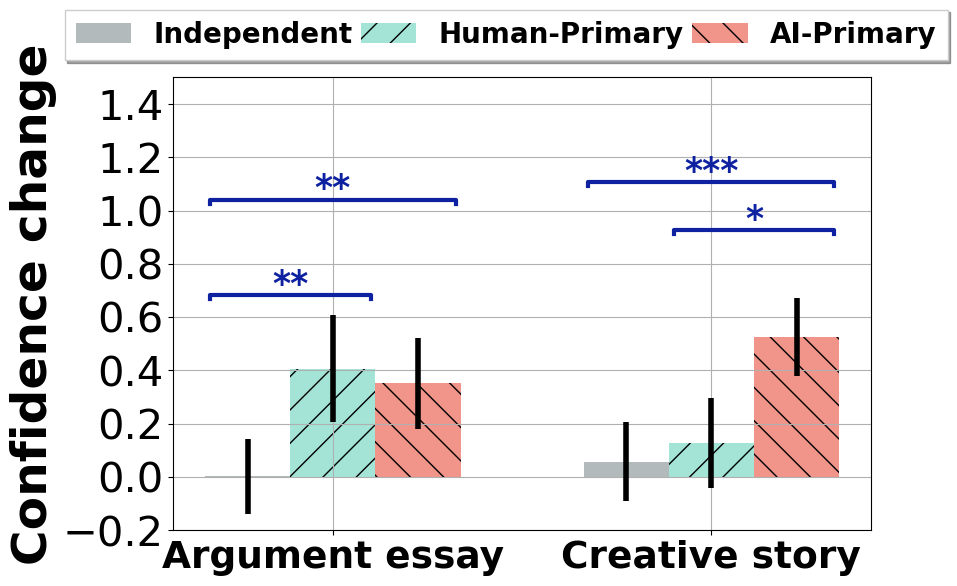}
\caption{Confidence boost in other writing}
\label{fig:confidence_boost:other}
\end{subfigure}
\vspace{-10pt}
\caption{Comparing the predicted change in the confidence of an average participant in writing (a) argumentative essays, (b) creative novel or stories, and (c) other types of articles (e.g., emails/letters, product/book reviews, business reports/proposals, blogs), should they have future opportunities to complete these writing tasks in the same writing mode as they experienced in our study, after they completed argumentative essay or creative story writing in our study. 
Error bars represent the 95\% confidence intervals. $\textsuperscript{*}$, $\textsuperscript{**}$, and $\textsuperscript{***}$ denote statistical significance levels of $0.05$, $0.01$, and $0.001$ respectively.
}
\label{fig:confidence_boost}
\end{figure*}

\subsubsection{Perceptions of the final writing outcome}
Figure~\ref{fig:feeling_article} compares participants' perceptions of four aspects of the final writing outcome---their perceived quality of the article, their perceived ownership of the article, their perceived pride in the article, and their perceived uniqueness of the article, upon completing the writing tasks in different writing modes. Bonferroni corrections were implemented to account for multiple statistical tests in our analysis of
this family of dependent variables. Visually, it appears that participants in the {\em AI-primary} writing mode often reported the lowest level of ratings across various aspects of the final writing outcome. In particular, consistent with previous findings~\cite{biermann2022tool}, we find that \textbf{\em the provision of content generation assistance by AI significantly decreases people's perceived ownership of the final writing outcome} (see Figure~\ref{ownership}; {\em adjusted}-$p<0.001$ for the comparisons between {\em independent} and {\em AI-primary}). Moreover, when participants worked on argumentative essay writing tasks, we also find that \textbf{\em the provision of content generation assistance by AI significantly decreases how much they take pride in the final writing outcome and how much they consider it as unique} (see Figure~\ref{pride} and~\ref{unique}; {\em adjusted}-$p<0.01$ for the comparisons between {\em independent} and {\em AI-primary}). For more detailed results, see Appendix~\ref{app:more_results}.

\begin{figure*}[t]
\centering
\centering
\begin{subfigure}[b]{.235\textwidth}
\includegraphics[width=1\textwidth]{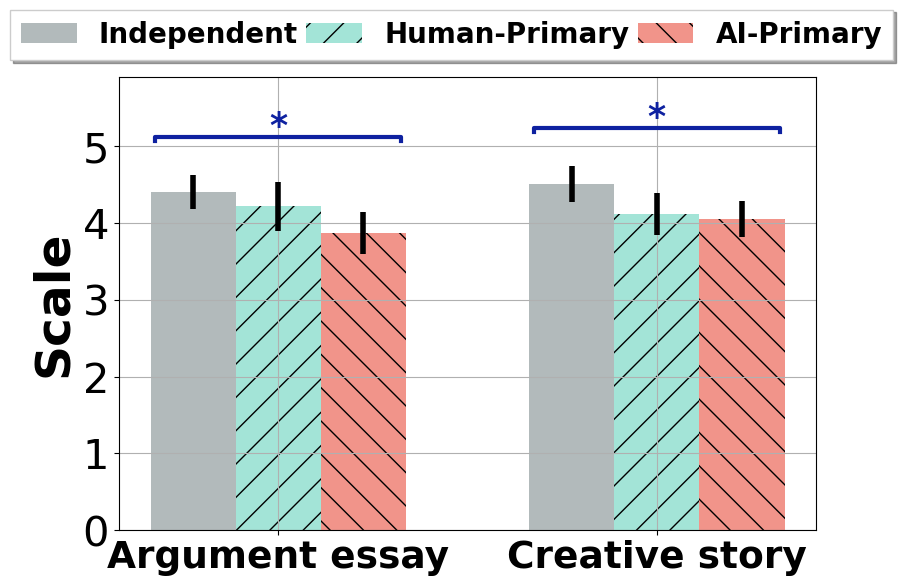}
\caption{Deceptive content}
\end{subfigure}
\centering
\begin{subfigure}[b]{.235\textwidth}
\includegraphics[width=1\textwidth]{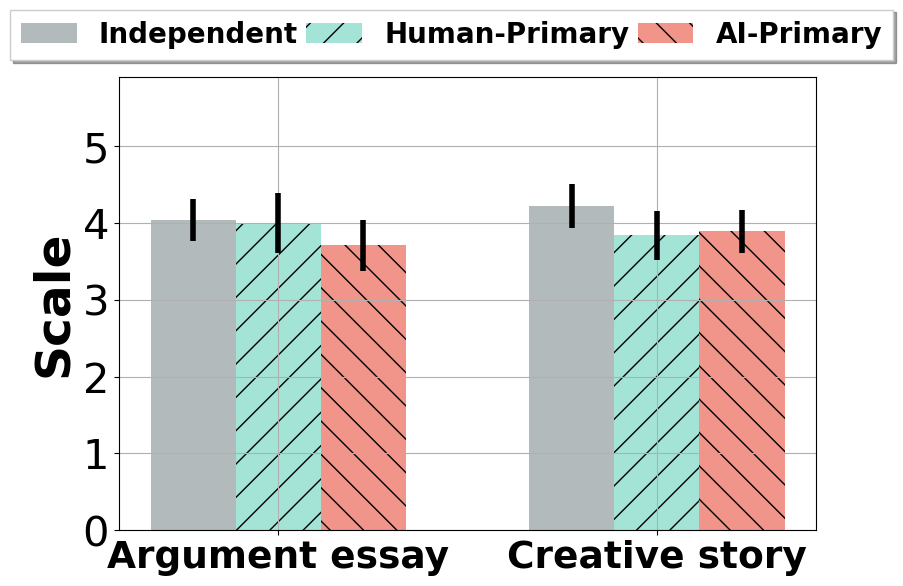}
\caption{Plagiarism}
\end{subfigure}
\centering
\begin{subfigure}[b]{.235\textwidth}
\includegraphics[width=1\textwidth]{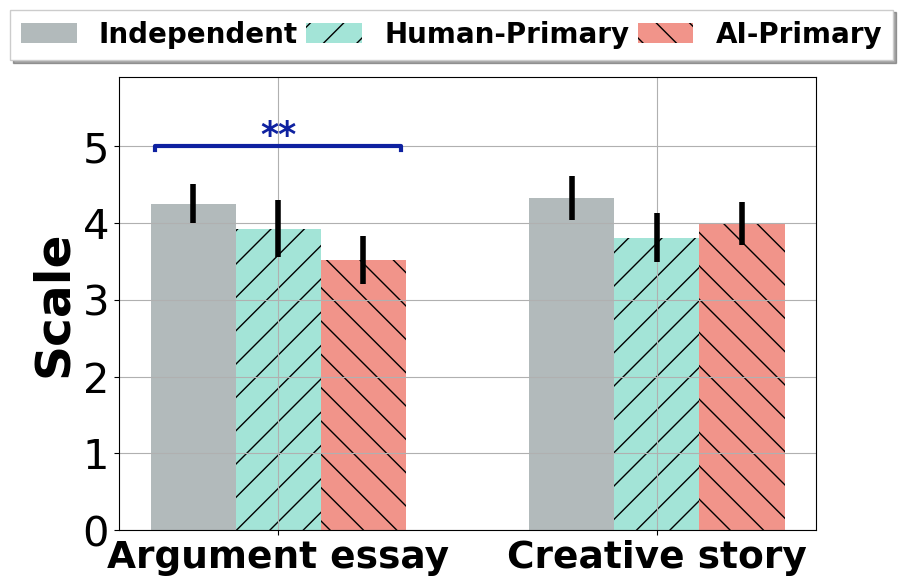}
\caption{Privacy invasion}
\end{subfigure}
\centering
\begin{subfigure}[b]{.235\textwidth}
\includegraphics[width=1\textwidth]{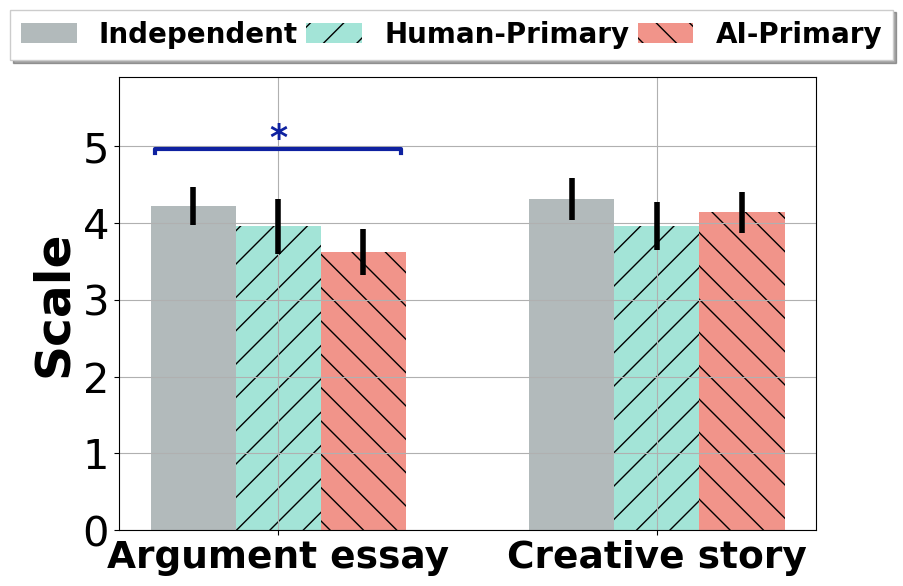}
\caption{Discrimination}
\end{subfigure}
\vspace{-10pt}
\caption{Comparing the predicted willingness of an average participant to take responsibility for potential issues in their articles across the {\em independent}, {\em human-primary}, and {\em AI-primary} modes after completing argumentative essay or creative story writing tasks. Error bars represent the 95\% confidence intervals. $\textsuperscript{*}$ and $\textsuperscript{**}$  denote statistical significance levels of $0.05$ and $0.01$ respectively.
}
\label{fig:issue}
\end{figure*}

\subsubsection{Change of confidence in writing}
Next, we look into whether AI assistance can increase participant's writing confidence for future writing tasks, after they experienced generative AI-powered assistance in writing. 
Recall that we asked participants to report their confidence in six types of writing tasks (e.g., argumentative essay, creative novel or story, email or letter, etc.) at the beginning of our study. In addition, after they had finished their writing job in their selected writing mode, participants again reported their confidence in these six types of writing tasks assuming that they would work on them in the future in the same writing mode as they just experienced. For each type of writing tasks, we were interested in the {\em change} in writing confidence (or ``{\em confidence boost}'') after participants tried out different writing modes (i.e., {\em independent}, {\em human-primary}, {\em AI-primary}) on different writing tasks (i.e., argumentative essay writing or creative story writing). 
Figures~\ref{fig:confidence_boost:argue}--\ref{fig:confidence_boost:other} compare participants' confidence boost in completing future argumentative essay writing, creative story writing, and the other four types of writing tasks\footnote{We calculated the pooled average of the confidence increase observed across the four types of writing tasks other than argumentative essay writing and creative story writing (i.e., emails/letters, product/book reviews, business reports/proposals, blogs).}, respectively, after they used different writing modes in our study to write argumentative essays or creative stories.

\vspace{2pt}
\noindent \textbf{\em After writing argumentative essays with AI assistance, people generally have higher confidence in completing all types of writing tasks in the future.} When focusing on participants who completed argumentative essay writing in our study, we found that the provision of ChatGPT's writing assistance to them---both its editing assistance and content generation assistance---generally made them become more confident in completing future writing tasks. Interestingly, this confidence increase often occurs not only for completing future argumentative essay writing tasks, but also for 
completing all types of other writing tasks. 
For example, compared to those who wrote the argumentative essay independently, participants in the {\em human-primary} mode reported a significantly larger increase in their confidence in completing future argumentative essay writing ({\em adjusted}-$p=0.014$),  
creative writing ({\em adjusted}-$p=0.030$), and other writing tasks ({\em adjusted}-$p=0.005$). Similarly, participants in the {\em AI-primary} mode also had a significantly larger confidence boost in completing future creative writing ({\em adjusted}-$p=0.016$) and other writing tasks ({\em adjusted}-$p=0.007$). 

\vspace{2pt}
\noindent \textbf{\em After writing creative stories with AI assistance, only if AI provides content generation assistance will people increase their confidence in writing, but not for creative writing.} In contrast, when focusing on participants who completed creative story writing in our study, we found less evidence of boost of writing confidence brought up by AI assistance. Participants in the {\em human-primary} mode who only received editing assistance from ChatGPT in writing creative stories did not show any significant difference than those who wrote the stories on their own in terms of their confidence in completing any type of future writing tasks. On the other hand, the content generation assistance of ChatGPT did make participants who wrote creative stories in the {\em AI-primary} mode report significantly larger increases in their writing confidence than independent writers, but only for tasks other than creative writing (for future argumentative essay writing: {\em adjusted}-$p=0.001$; for other writing tasks: {\em adjusted}-$p<0.001$).

\subsubsection{Perceptions of accountability}
Finally, we examined how writing with AI assistance influence people's willingness to take responsibility if their final articles face criticism for issues such as deceptive content, plagiarism, privacy invasion, and discrimination. Figure~\ref{fig:issue} compares the willingness to assume responsibility for these issues 
across participants in the three writing modes.  

\vspace{2pt}
\noindent \textbf{\em People who receive content generation assistance from AI are less willing to take responsibility for criticisms of the final writing outcome.}
Visually, we note that when people receive AI assistance in their writing, they are generally less willing to take responsibility on all four potential criticisms of the final writing outcome. This trend is particularly salient when AI provides content generation assistance during the writing processes. For example, we found that after participants completed argumentative essay writing in the {\em AI-primary} mode, they were less likely to take responsibility for criticisms on potential deceptive content ({\em adjusted}-$p=0.015$), privacy invasion ({\em adjusted}-$p=0.002$), or bias/discrimination issues ({\em adjusted}-$p=0.014$) in their final articles, compared to participants who wrote independently without AI assistance. In addition, after writing creative stories in the {\em AI-primary} mode, participants also became less willing to take the responsibility for any criticism of deceptive content in their final articles than the participants in the {\em independent} writing mode ({\em adjusted}-$p=0.028$).


\subsection{Understanding the impacts of AI assistance on people's writing performance}

\begin{figure}[t]
\centering
\begin{subfigure}[b]{.4\textwidth}
\includegraphics[width=.95\textwidth]{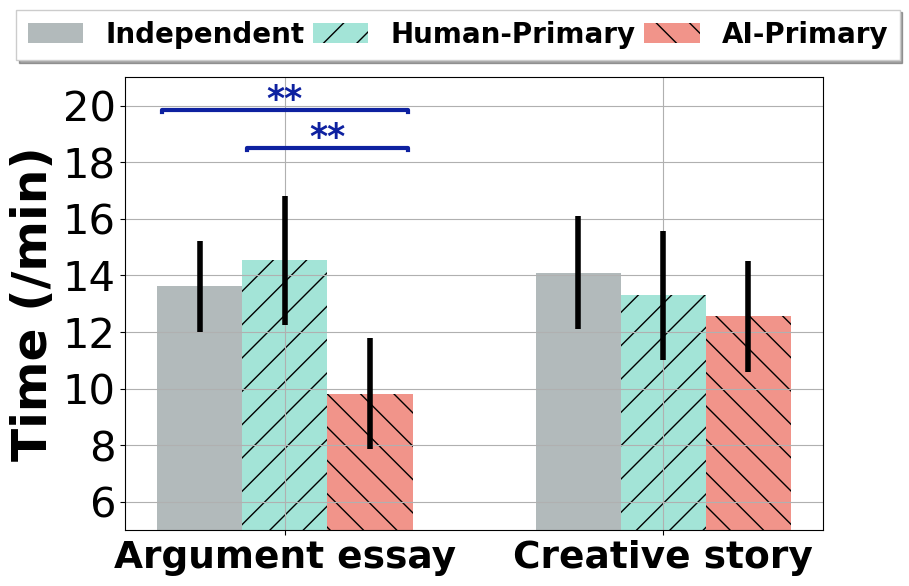}

\end{subfigure}

\caption{Comparing the predicted time taken by participants to complete the two types of writing tasks (argumentative essay and creative story writing), across the {\em independent}, {\em human-primary}, and {\em AI-primary} writing modes. Error bars represent the 95\% confidence intervals. $\textsuperscript{**}$ denote statistical significance levels of $0.01$, respectively.
}
\label{fig:completion_time}
\end{figure}

Lastly, we investigate into how the incorporation of AI assistance in the writing processes affect people's writing performance. In this analysis, we focus on a set of objective metrics to quantify people's writing performance from different perspectives. Unless specified otherwise, this analysis is conducted based on regression models that we fit from the data, and the results of the full set of regression models can be found in Appendix~\ref{app-regression}.

\subsubsection{Completion time}

We start by examining people's time efficiency in completing the writing job. Therefore, we compare how much time it took for participants to complete the main writing task in our study across participants in the three writing modes, and results are presented in
Figure~\ref{fig:completion_time}. 

\vspace{2pt}
\noindent \textbf{\em Content generation assistance provided by AI decreases the amount of time it takes for people to complete argumentative essay writing.}
As shown in Figure~\ref{fig:completion_time}, while editing assistance provided by ChatGPT does not appear to significantly influence participants' efficiency in completing their writing job, the provision of content generation assistance by ChatGPT appears to increase participants' writing efficiency. Specifically,  
we found that when writing argumentative essays, participants in the {\em AI-primary} writing mode took significantly less time to complete the writing compared to those working independently
($p=0.004$), and those in the {\em human-primary} writing mode ($p=0.004$). 




\begin{figure}[t]
\centering
\begin{subfigure}[b]{.235\textwidth}
\includegraphics[width=1\textwidth]{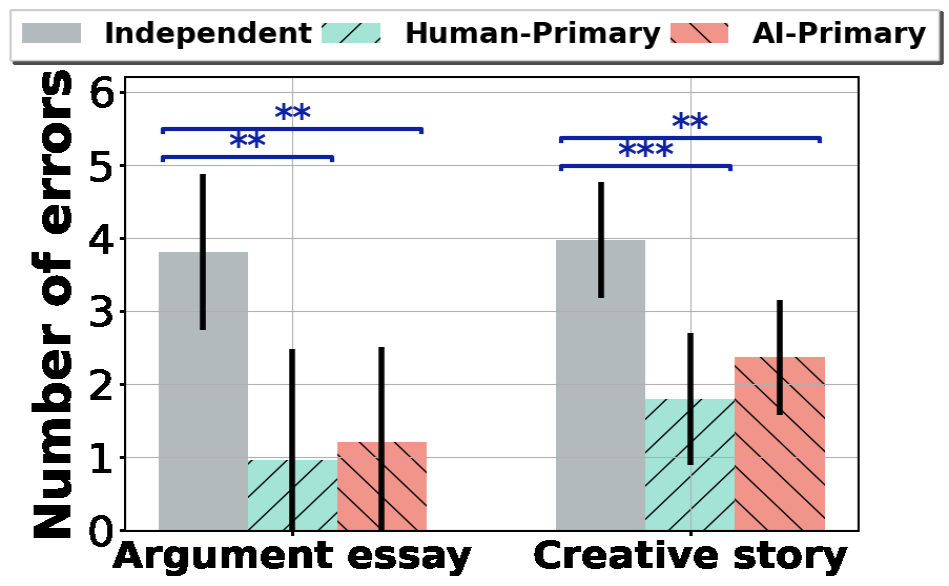}
\caption{Grammar and spelling}
\label{fig:grammar}
\end{subfigure}
\hfill
\begin{subfigure}[b]{.235\textwidth}
\includegraphics[width=1\textwidth]{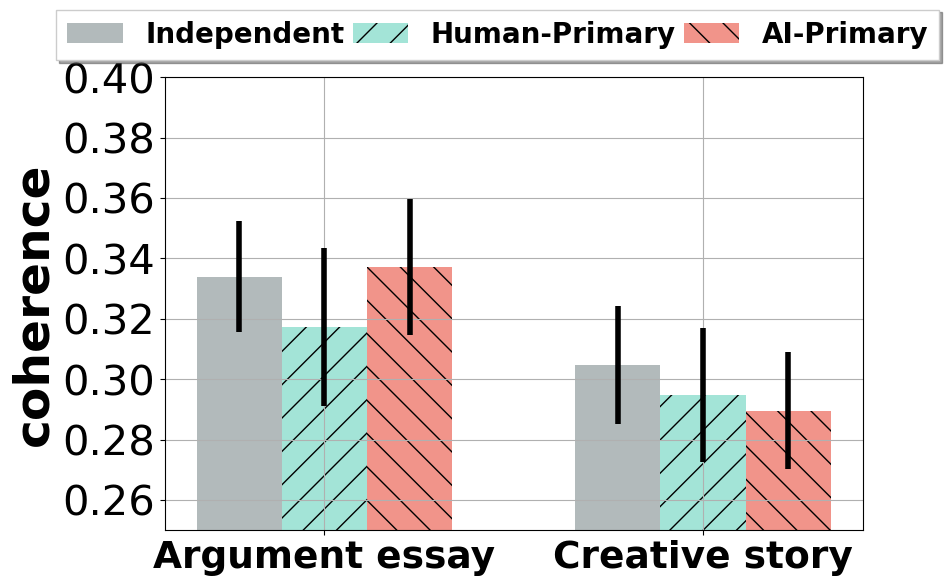}
\caption{Coherence}
\label{fig:coherence}
\end{subfigure}

\caption{Comparing the predicted number of grammar and spelling mistakes in the final article and the coherence of the article in two types of writing tasks (argumentative essay and creative story writing) for an average participant, across the {\em independent}, {\em human-primary}, and {\em AI-primary} writing modes. Error bars represent the 95\% confidence intervals. $\textsuperscript{**}$ and $\textsuperscript{***}$ denote statistical significance levels of $0.01$ and $0.001$, respectively.
}
\vspace{-10pt}
\label{fig:grammar_coherence}
\end{figure}

\subsubsection{Grammar and spelling}
To see if the AI assistance affects people's writing performance with respect to the proper usage of language in their final article, we utilized LanguageTool\footnote{https://languagetool.org/.}, an open-sourced grammar, style, and spell checker, to conduct a grammar and spelling error analysis on the final articles submitted by participants in different writing modes. 
Figure~\ref{fig:grammar} compares the predicted average number of grammar and spelling mistakes per article when participants completed argumentative essay or creative story writing tasks  across the three writing modes. Not surprisingly, we find that \textbf{\em both AI's editing assistance and content generation assistance leads to significantly decreased number of grammar and spelling mistakes in people's writing} ($p<0.01$ for all comparisons between independent and an AI-assisted writing mode). For more detailed results, see Appendix~\ref{app:grammar}.


\subsubsection{Content coherence} Another aspect of writing performance we measured was the coherence of the content in participants' final articles. Arguably, an article is of higher quality if it contains multiple sets of more coherent sentences, such that sentences in each set address the same topic and could support one another. To compute the ``coherence score'' of an article, 
following previous work~\cite{roder2015exploring}, we first used Latent Dirichlet Allocation (LDA)~\cite{blei2003latent}---a common topic modelling algorithm---to identify latent topics in the article, and then computed the semantic similarity between highly-scored words within each topic based on word co-occurrence probabilities. Intuitively, the higher the coherence score, the better.  
Figure~\ref{fig:coherence} presents the predicted average coherence score of articles that participants produced across the three 
writing modes. 
Here, we do not find any evidence suggesting that the incorporation of AI assistance during the writing processes significantly changes the coherence of people's final writing outcomes.  

\begin{figure}[t]
\centering
\begin{subfigure}[b]{.235\textwidth}
\includegraphics[width=1\textwidth]{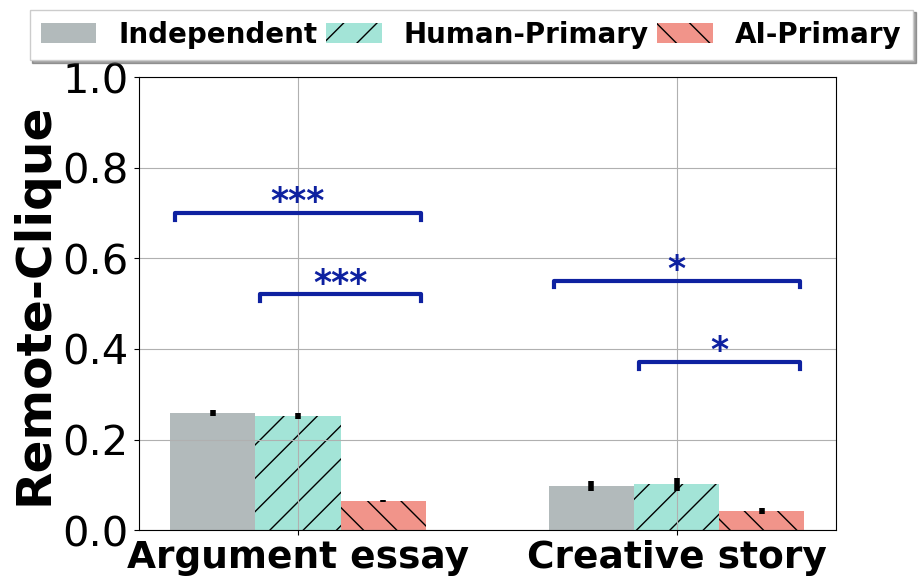}
\caption{Remote clique score}
\label{fig:remote}
\end{subfigure}
\hfill
\begin{subfigure}[b]{.235\textwidth}
\includegraphics[width=1\textwidth]{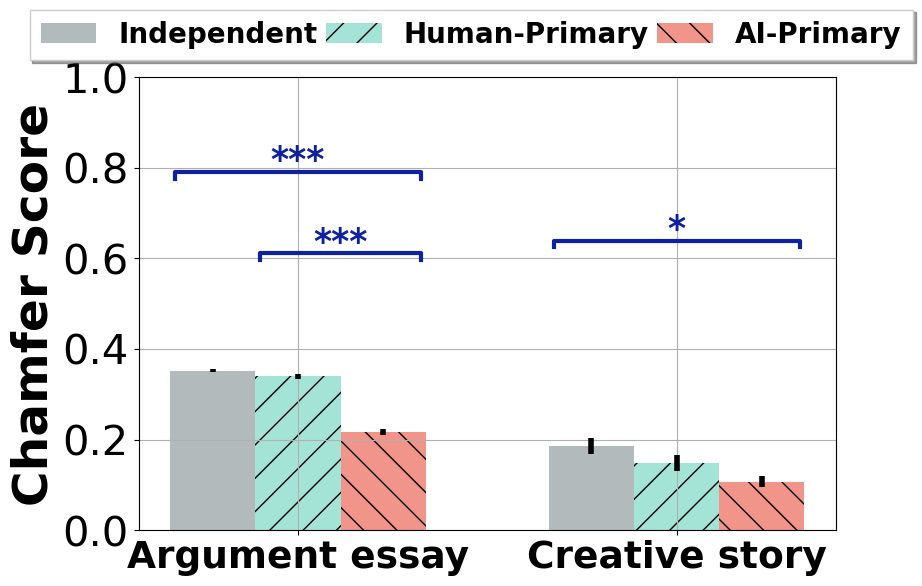}
\caption{Chamfer distance score}
\label{fig:chamfer}
\end{subfigure}
\caption{Comparing the diversity of final articles on the same topic for two types of writing tasks (argumentative essay and creative story writing), across the {\em independent}, {\em human-primary}, and {\em AI-primary} writing modes. Error bars represent standard errors of the mean. $\textsuperscript{*}$ and $\textsuperscript{***}$ denote statistical significance levels of $0.05$ and $0.001$, respectively.
}
\vspace{-10pt}
\label{fig:diversity}
\end{figure}

\subsubsection{Content diversity}
Finally, we examine how the incorporation of AI assistance changes the diversity of writings generated by different individuals on the same topic. One can also consider this metric as a proxy of the ``uniqueness'' of the writing---a lower level of diversity across multiple writings on the same topic usually implies that each writing in the group is less unique.  
For our study, given a set of articles submitted by participants on the same topic (i.e., argumentative essays for the same statement or creative stories for the same prompt) using the same writing mode, we used \emph{remote clique score} and \emph{Chamfer distance score} to quantify the diversity of these articles following
the prior literature ~\cite{rhys2021directed}. Specifically, the remote clique score is computed as the average mean 
distance of an article to other articles in the set, while 
the Chamfer distance score is defined as the average minimum distance of an article to other articles in the set. 
For both 
metrics, higher values indicate greater diversity. To determine the distance between different articles, we utilized a pre-trained Sentence Transformer model~\cite{sentence_embed} to convert each article into a vector embedding and then computed the cosine distance between each pair of articles. 
Figure~\ref{fig:diversity} compares the diversity of articles written on the same topics across the three writing modes. Since our diversity metrics are computed for each writing topic instead of for each participant, in the following, we did not fit regression models to analyze the data. Instead, one-way ANOVA was used to determine if the mean value of diversity metrics was statistically the same across the three writing modes, and post-hoc Tukey HSD tests were used to detect significant pairwise comparisons.

\vspace{2pt}
\noindent \textbf{\em AI's content generation assistance significantly reduces the diversity of people's writing.}
Visually, it is clear that when participants received content generation assistance from ChatGPT, the diversity of the articles they produced on a given topic is consistently lower than other participants, irrespective of the specific writing tasks they undertook. 
Specifically, when writing argumentative essays, the diversity of articles generated from the {\em AI-primary} writing mode was significantly lower compared to those generated from the {\em independent} or {\em human-primary} modes, which was evident in both the {\em remote clique score} and {\em Chamfer distance score} metrics 
($p<0.001$ for all comparisons). Similarly, when writing creative stories, the diversity of stories produced from the {\em AI-primary} writing mode was also significantly lower compared to those produced from the {\em independent} writing mode (remote clique score: $p=0.021$; 
Chamfer distance score: $p=0.043$) and the {\em human-primary} mode (remote clique score: $p=0.042$).

{
\renewcommand{\arraystretch}{1.55}
\begin{table*}
\centering
\small
\resizebox{\textwidth}{!}{
\begin{tabular}{cc|c|c||c|c} 
\hline
\multicolumn{2}{c|}{\multirow{2}{*}{\textbf{Dependent Variable}}}                           & \multicolumn{2}{c||}{\textbf{Argumentative essay }} & \multicolumn{2}{c}{\textbf{Creative story}}   \\ 
\cline{3-6}
\multicolumn{2}{c|}{}                                                                       & \textbf{Human-Primary} & \textbf{AI-Primary}      & \textbf{Human-Primary} & \textbf{AI-Primary}  \\ 
\cline{1-6}
\multirow{5}{*}{Value of AI assistance}                & All population                    & \$0.17                 & \$0.45                   & \$0.06                 & \$1.24               \\
                                                       & Low writing confidence            & $\$0.26 $              & $ \$0.68 $               & $\$0.43 $              & $\$1.36 $            \\
                                                       & High writing confidence           & $ \$0.08$              & $ \$0.17 $               & $-\$0.36$              & $\$1.01$             \\
                                                       & Low familiarity with ChatGPT      & $\$0.27 $              & $ \$0.80 $               & $ -\$0.27 $            & $\$1.31 $            \\
                                                       & High familiarity with ChatGPT     & $\$0.12$               & $\$0.29 $                & $\$0.24 $              & $ \$1.22$            \\ 
\cline{1-6}
Cognitive load                                         & NASA Task Load Index              & \textcolor{green}{\ensuremath{\downarrow}}    & \textcolor{green}{\ensuremath{\downarrow}}      & -                      & -                    \\ 
\hline
\multirow{4}{*}{\tabincell{c}{Perceptions about \\the writing process}} & Satisfaction                      & -                      & -                        & -                      & -                    \\
                                                       & Enjoyment                         & -                      & \textcolor{red}{\ensuremath{\downarrow}}        & -                      & -                    \\
                                                       & Ease                              & -                      & -                        & -                      & -                    \\
                                                       & Ability of self-expression        & -                      & \textcolor{red}{\ensuremath{\downarrow}}        & -                      & -                    \\ 
\cline{1-6}
\multirow{4}{*}{\tabincell{c}{Perceptions about \\the final outcome}}   & Quality                           & -                      & -                        & -                      & -                    \\
                                                       & Ownership                         & -                      & \textcolor{red}{\ensuremath{\downarrow}}        & -                      & \textcolor{red}{\ensuremath{\downarrow}}    \\
                                                       & Pride                             & -                      & \textcolor{red}{\ensuremath{\downarrow}}        & -                      & -                    \\
                                                       & Uniqueness                        & -                      & \textcolor{red}{\ensuremath{\downarrow}}        & -                      & -                    \\ 
\cline{1-6}
\multirow{3}{*}{Change of confidence}                  & Argument writing                  & \textcolor{green}{\ensuremath{\uparrow}}    & -                        & -                      & \textcolor{green}{\ensuremath{\uparrow}}  \\
                                                       & Creative writing                  & \textcolor{green}{\ensuremath{\uparrow}}    & \textcolor{green}{\ensuremath{\uparrow}}      & -                      & -                    \\
                                                       & Other writing                     & \textcolor{green}{\ensuremath{\uparrow}}    & \textcolor{green}{\ensuremath{\uparrow}}      & -                      & \textcolor{green}{\ensuremath{\uparrow}}  \\ 
\cline{1-6}
\multirow{4}{*}{Perceptions of accountability}         & Deceptive content                 & -                      & \textcolor{red}{\ensuremath{\downarrow}}        & -                      & \textcolor{red}{\ensuremath{\downarrow}}    \\
                                                       & Plagiarism                        & -                      & -                        & -                      & -                    \\
                                                       & Privacy invasion                  & -                      & \textcolor{red}{\ensuremath{\downarrow}}        & -                      & -                    \\
                                                       & Discrimination                    & -                      & \textcolor{red}{\ensuremath{\downarrow}}        & -                      & -                    \\ 
\cline{1-6}
\multirow{5}{*}{Writing performance}                   & Completion time                   & -                      & \textcolor{green}{\ensuremath{\downarrow}}      & -                      & -                    \\
                                                       & Grammar and spelling              & \textcolor{green}{\ensuremath{\downarrow}}    & \textcolor{green}{\ensuremath{\downarrow}}      & \textcolor{green}{\ensuremath{\downarrow}}    & \textcolor{green}{\ensuremath{\downarrow}}  \\
                                                       & Content coherence                 & -                      & -                        & -                      & -                    \\
                                                       & Diversity: Remote clique score    & -                      & \textcolor{red}{\ensuremath{\downarrow}}        & -                      & \textcolor{red}{\ensuremath{\downarrow}}    \\
                                                       & Diversity: Chamfer distance score & -                      & \textcolor{red}{\ensuremath{\downarrow}}        & -                      & \textcolor{red}{\ensuremath{\downarrow}}    \\
\cline{1-6}
\end{tabular}
}
\caption{A summary of the key findings of this study. In the ``Value of AI assistance'' row, the average value different subsets of people attach to AI assistance is reported. In other rows, we compare the dependent variables measured in both the {\em human-primary} and {\em AI-primary} modes with those obtained in the {\em independent} mode. The symbols \ensuremath{\downarrow} and \ensuremath{\uparrow} indicate significantly lower and higher values, respectively, in comparison to the {\em independent} mode. $-$ denotes no statistical significance was detected. \textcolor{green}{\ensuremath{\downarrow}} (\textcolor{green}{\ensuremath{\uparrow}}) denotes desirable changes while \textcolor{red}{\ensuremath{\downarrow}} (\textcolor{red}{\ensuremath{\uparrow}}) denotes undesirable changes. }~\label{tab:summary}
\end{table*}

}

\section{Discussions}
Table~\ref{tab:summary} summarizes the key findings of our study. Below, we provide a few explanations to some of our findings, discuss design implications of our results, and outline limitations and future work.

\subsection{The moderating role of writing task nature}

As shown in Table~\ref{tab:summary}, both people's value of AI assistance and the AI assistance's impacts on people's writing experience and performance are moderated by the nature of the writing tasks to some extent. For example, participants generally attached a higher value to the content generation assistance provided by LLMs for creative story writing than for argumentative essay writing. In addition, the AI assistance appears to mostly bring about productivity benefits on argumentative essay writing tasks---it reduces participants' perceived cognitive load and decreases their completion time in writing only when they wrote argumentative essays but not creative stories.    
We conjecture that these differences can be 
attributed to the distinct nature and objectives of these two writing tasks. 

For example, 
for argumentative essay writing, the main cognitive load and the most time-consuming part for writers is to construct the logical structure of the essay and identify the appropriate evidence. Generative AI models can easily provide such information and generate reasonably logical sentences for writers, thus can greatly boost writers' productivity. 
However, for creative story writing, the primary cognitive burden for writers lies in the process of utilizing their imagination to create compelling characters and engaging plots, and they often need to intentionally weave conflicts and surprises into the storyline. Compared to argumentative essay writing, creative story writing places higher demand on creativity and originality, and it may lack the kind of ``templates'' that often turn out to be effective for writing argumentative essays. In this sense, it is not surprising that people tend to attach higher value to generative AI-powered content generation assistance for creative writing than for argumentative essay writing, because such assistance is especially useful to inspire the writers and help them think ``out of the box'' in creative writing. 
Nevertheless, the fact that AI assistance does not significantly affect people's cognitive load and writing time in creative story writing suggests that today's LLMs may still fall short in providing the kind of assistance that people need and expect in creative writing tasks. For instance, LLMs may still be struggling on creating truly novel plots or appropriately adjust the content they generate to align with writers' feedback.

\subsection{Potential mismatch between human attached value to AI assistance and the ``true value'' of AI assistance}

Another somewhat puzzling finding of our study is that considering the different types of writing assistance provided by generative AI, while people are consistently willing to pay a higher price to receive the content generation assistance from LLMs than the editing assistance, the former type of assistance actually brings about a much larger set of undesirable impacts on people's writing experience and performance than the latter (e.g., all red arrows, representing undesirable impacts, are in ``{\em AI-primary}'' columns in Table~\ref{tab:summary}). This seems to indicate a degree of mismatch between the price people pay for AI assistance, and the ``true value'' of AI assistance. We provide a few plausible explanations for this finding.  

One possible explanation is the media portrayal of generative AI models like ChatGPT as revolutionary and next-generation tools has potentially inflated people's expectations. The general public might not have a deep understanding of how these AI models operate, leading to assumptions that the generative AI-powered writing assistant tools are more capable than they truly are, especially on their content generation and adaptation capabilities that is new compared to traditional writing assistant tools. This limited understanding, coupled with intensive media coverage, might create an overly optimistic perception of AI's capabilities. Furthermore, most people may not recognize that LLMs are trained on extensive datasets and might, at times, generate content that does not perfectly meet their specific needs, intentions, or style. These deviations between expectation and reality might contribute to the possibly overly high financial value that people attach to content generation assistance of LLMs. In fact, this conjecture can be supported by one interesting observation in our study---After participants experienced content generation assistance of LLMs in argumentative essay writing (or creative story writing) tasks, their confidence boost in completing future argumentative essay writing (or creative story writing) tasks are {\em not} significantly different from those participants who completed the writing tasks independently, although their confidence boost in other types of writing tasks was significantly larger. This implies that upon directly experiencing the content generation capabilities of the generative AI-powered assistant tools and encountering their limitations, people may recalibrate their assessment of the AI assistance's true value, at least on the type of writing tasks they have worked on. In fact, the contrast between people's initial high expectation of AI's content generation assistance and the practical output they observe may even inadvertently 
amplify their hope that AI's content generation capabilities might truly shine in other unexplored tasks, which explains their writing confidence boost in other types of writing tasks. 

Another possible explanation is that the financial value people attach to different types of AI assistance may be primarily driven by their productivity-related considerations. Indeed, people may expect the incorporation of AI-powered content generation assistance in their writing workflow to significantly increase their time efficiency than the editing assistance. In fact, we did find that participants were more likely to complete their writing tasks in a shorter amount of time when they were in the ``{\em AI-primary}'' writing mode than in the ``{\em human-primary}'' mode. So, if people determine the price that they are willing to pay for AI assistance primarily, or even solely based on the productivity benefits the assistance is expected to bring, then it is not surprising that people attach a higher value to generative AI's content generation assistance, despite they might be aware of its undesirable impacts on the writing experience and outcome. 

Finally, this finding may also relate to the fact that our experiment is of relatively low stakes. The low-stake nature of our experimental task may contribute to the potential price-value mismatch of AI assistance via two ways: First, people may care less about the undesirable impacts on writing experience and performance caused by AI's content generation assistance due to the low stakes of the tasks, effectively making productivity-related considerations to be the primary driver in people's valuation of AI assistance. Second, due to the relatively low stakes of the task, people may also exhibit a degree of ``over-reliance'' on the content generation assistance provided by LLMs and show less motivation to fully explore how to best utilize the AI assistance and unlock their true value.

\subsection{Implications for designing AI writing assistance }
Our findings in this study suggest several implications for designing future AI-powered writing
assistance.  First, as people are found to generally value LLMs' content generation assistance more than their editing assistance, a dynamic pricing model could potentially be adopted for future generative AI-powered writing assistants to align their price with users' perceived value of the AI assistant in different scenarios. For example, instead of pricing a generative AI-powered writing assistant at a fixed rate based on the length of the output that it produces, one may consider to adjust its price based on the type of output it produces---a higher price can be used when the assistant generates new content from scratch (as that in the ``{\em AI-primary}'' writing mode), while editing user-generated content (as that in the ``{\em human-primary}'' writing mode) can be associated with a lower price.

Secondly, we observed in our study that individuals with varying characteristics, such as different levels of confidence in writing,
may have distinct preferences for AI assistance, and the value they place on AI assistance also varies with the type of writing tasks. These findings suggest that AI writing assistant systems could  
consider providing different kinds of customized assistance that tailors to the characteristics of the users and the needs of the tasks.  For example, future generative AI-powered writing assistants could allow for personalized settings 
and offer granular controls to users to choose 
the type of AI assistance 
they prefer. In addition, the generative AI-powered writing assistant could also actively adapt its assistance to the profiles and interaction history of users, as well as the inferred user intent of writing. In other words, given a specific user and a writing context, the AI writing assistant should strive to provide the kind of assistance that the user values the most and avoid the kind of assistance that the user dislikes in the current context. For example, as high-confident writers appear to value editing assistance from AI negatively when writing creative stories, the AI writing assistant could learn to avoid providing text editing and polishing suggestions to them in this writing context by default.
As the AI writing assistant identifies the users' writing intent (e.g., creative storytelling vs. fact-based reporting) on the fly, it could also  dynamically adjust the range of assistance that it provides to users and prioritize them based on the user's preferences. 

Furthermore, our study results showed that when generative AI models provide content generation assistance, it can often diminish the writer's satisfaction with both the writing process and the final product, e.g., resulting in a decrease in writers' perceived ability to express themselves. This suggests that when designing AI writing assistants, it's crucial to allow writers to maintain control over the AI assistant's content generation features. To ensure this, future AI writing assistants could incorporate options like ``on-demand suggestions'' (i.e., the AI assistant only provides suggestions when specifically requested by the user) or ``AI pause'' (i.e., users are allowed to temporarily pause the AI's active content suggestions). 
Another useful feature could be ``content generation intensity sliders'', in which users can drag the sliders to specify their requirements for the content generated by AI on different dimensions (e.g., length, formality, emotional degree, writing styles) to better fit their needs. Since people do not seem to decrease their satisfaction with their writing process and performance when AI assistants simply polish their own writing, one may also envision a new kind of content generation assistance that may have limited negative impacts on user perceptions---AI assistants can provide ``imitating writing'' assistance based on writing examples that users provide, which may or may not be written by the user themselves. To counteract the decreased level of enjoyment in writing and ownership of the writing outcome caused by AI's content generation assistance, visual annotations or indicators can be incorporated to distinguish between AI-generated content and human edits, hence highlighting the human contributions to the final writing and emphasizing the collaborative nature of the final output and writing process. Finally, as the current content generation assistance offered by AI assistants appears to significantly decrease the diversity in people's writing, future designs should explore ways to decrease users' over-reliance on the AI-generated content and to encourage users to explore new and creative ways to collaborate with the AI assistant.

\subsection{The potential influence of study setup}
We acknowledge that our findings may be specific to the experimental setup used in this study. First, as previously discussed, our experiment was based on a relatively low-stake and domain-knowledge-free writing scenario. It does not represent all writing contexts that people may encounter, such as those involving high stakes and require substantial domain expertise (e.g., writing critical business proposals, where domain knowledge is needed and poor quality can lead to proposal rejections). 
We speculate that in those high-stake, domain-knowledge-required writing scenarios, people may prioritize quality-related considerations over productivity-related considerations. Thus, without being fully convinced that the content produced by generative AI-powered assistant can significantly boost the quality of their writing, people may place less value in AI's content generation assistance. In contrast, given the common belief that AI's editing assistance can effectively reduce grammar errors and enhance the overall writing quality, the perceived value of AI’s editing assistance may be elevated, aligning with people’s priority on quality considerations.
Meanwhile, we may also expect a degree of heterogeneity in how changes in factors like  stakes affect people's value of AI assistance. For example, for those individuals with very low confidence in their own writing, they may even associate a higher value with AI's content generation assistance as the stake of the writing task becomes higher. Future studies should be conducted to gain deeper understandings of the extent to which our findings can generalize to other writing scenarios with different characteristics.

Another feature of our study setup that may have influenced our finding is that our experiment created a writing scenario that emphasizes on people's {\em extrinsic motivations} for writing---participants in our study completed writing jobs to get paid. However, in the real world, there are people who write to satisfy their {\em intrinsic motivations}, such as expressing themselves in a creative way and seeing their identity through the writing process. The extent to which people find intrinsic motivations in writing may change how much they value AI assistance and how the provision of AI assistance affects their writing process and outcome. For example, it is possible that when people attach intrinsic value to writing themselves, they may find the generative AI-powered writing assistants, especially those that provide content generation assistance, to erode their intrinsic motivation. If this is the case, we may expect people with high intrinsic motivation for writing to place a lower value to AI assistance, especially on those creative writing tasks that allow them to fully express themselves. Moreover, when people with high intrinsic motivation for writing are provided with AI assistance, they may also show greater motivation in exploring creative ways to steer AI to generate the type of assistance that they would find truly inspiring and useful---if they are successful in doing so, the incorporation of AI assistance in their writing flows may result in an improved perception of the writing experience and outcome for them.

Related to this, it's important to note that our experiment was conducted on an online platform, Prolific. As such, participants of our study might be primarily driven by factors like task completion efficiency, and they may have the tendency to satisfice. That is, the population of participants in our study may heavily focus on their extrinsic motivations rather than their intrinsic ones. This means that while the paid writing job design of our study already emphasizes on people's extrinsic motivation for writing, the fact that our study was conducted on an online platform may have amplified this emphasis. Future work should be conducted to explore to what extent our findings can be generalized to a study setup where people place high intrinsic motivation in writing.    

\subsection{Limitations and future work}
We acknowledge a few limitations of our study. Firstly, the two types of writing tasks we selected may not encompass the full range of real-world writing activities. Our findings suggest that human's value for AI assistance, as well as the influence of AI assistance on people's writing perceptions and performance, is significantly influenced by the nature of the writing task.  Therefore, conclusions drawn from our study concerning argumentative essay and creative story writing tasks may not generalize to other writing tasks. 
For example, when the writing task requires a high level of personal touch, 
such as writing letters to friends,  
it is unclear whether people would still be willing to place substantial values on the AI assistance. 
Future research should delve into various writing tasks to re-examine the questions we have asked in this study, in order to provide a more comprehensive understanding of the nuanced differences in how people perceive the AI assistance and how the AI assistance impacts people in different writing domains.

In addition, we explored two types of AI assistance in this paper: (1) text editing and polishing assistance, 
and (2) free-form content generation assistance.  While they capture two representative formats of assistance offered by AI-powered writing assistants, 
our study might not encompass more complicated forms of AI assistance, such as those kinds of assistance enabled by advanced interactive interfaces that facilitate the communication between humans and LLMs.
Consequently, our conclusions are primarily reflective of the two types of AI assistance we considered, manifested through the interfaces we used and the generative AI model we deployed in this study (i.e., ChatGPT). 
However,  we believe that some design implications revealed by 
our study could still be relevant for other AI writing assistants that are powered by a different model, or present different kinds of content generation capabilities. 
For instance, the needs for customizing the type of AI assistance based on user characteristics and task contexts, 
and the critical importance of maintaining human control over the writing process are likely still true 
across a variety of generative AI-powered writing assistants. In the future, it's essential to investigate AI writing assistants that operate on different AI models, use different interfaces, or provide different functionalities, to comprehensively cover the landscape of AI writing assistants.  These investigations will be instrumental in revealing both commonalities and differences in how various forms of AI assistance are perceived by users and affect users in diverse writing contexts, and may further enhance our understanding of how to choose from different forms of AI writing assistants in different scenarios. 

Furthermore, we note that the aspects of writing experience perceptions and performance covered in our study are not exhaustive. So, generative AI-powered writing assistants may bring about additional benefits or risks to people that are not measured in this study. For example, it is possible that writers' viewpoints get biased or even mislead by the discriminatory content or misinformation generated by the AI assistant. Future research could further expand these aspects to obtain a more comprehensive understanding of the effects of generative AI-powered writing assistants.

Lastly, our study was limited in its exploration of the {\em long-term effects} of AI assistant usage on humans' writing skills and habits, as well as how this prolonged interaction might influence individuals' perceptions and valuation of AI writing assistance. It remains unclear whether a sustained engagement with AI writing assistants would lead to increased human reliance on AI and a corresponding rise in the perceived financial value of AI assistance, or whether it would enable users to recalibrate their beliefs about the capabilities of AI assistants, potentially decreasing their perceived value. Longitudinal studies are needed in the future to understand how the long-term usage of AI writing assistants influences people's writing process and performance over time, and how it alters people's perceived value of such AI assistance in the long run. 
\section{Conclusion}
In this paper, we present a study to obtain a systematic understanding of whether and how much value people attach to AI assistance, and how the incorporation of AI assistance in writing workflows changes people's writing perceptions and performance. Our results show that people are willing to forgo financial payments to receive writing assistance,  especially if AI can provide direct content generation assistance and the writing task is highly creative. In addition, our results highlight that although the generative-AI powered assistance is found to offer benefits in increasing people's productivity and confidence in writing, direct content generation assistance offered by AI also comes with risks,
including decreasing people’s sense of accountability and diversity in writing. 

Our work provides important implications for the design of human-AI collaborative writing systems. For example, future generative AI-powered writing assistants could dynamically adjust their assistance based on the characteristics of the user and the writing task to better align with what users value the most in the specific context. In addition, to minimize the negative impacts of AI's content generation assistance on people's writing perceptions and performance, 
actions need to be taken to ensure users' control over the content generated by AI and the final writing product. To gain a more comprehensive understanding of the commonality and differences in people's attitude towards and experience with generative AI-powered writing assistants in various scenarios, future work may look into the generalizability of our findings across a wider range of writing tasks, a broader spectrum of AI-powered writing assistants, and among different user populations.

\bibliographystyle{ACM-Reference-Format}
\bibliography{reference}
\clearpage
\newpage

\appendix
\setcounter{table}{0}
\renewcommand{\thetable}{A\arabic{table}}
\setcounter{figure}{0}
\renewcommand{\thefigure}{A\arabic{figure}}
\section{Task interfaces}
\begin{figure*}[t]
\centering
\begin{subfigure}[b]{.99\textwidth}
\includegraphics[width=.99\textwidth]{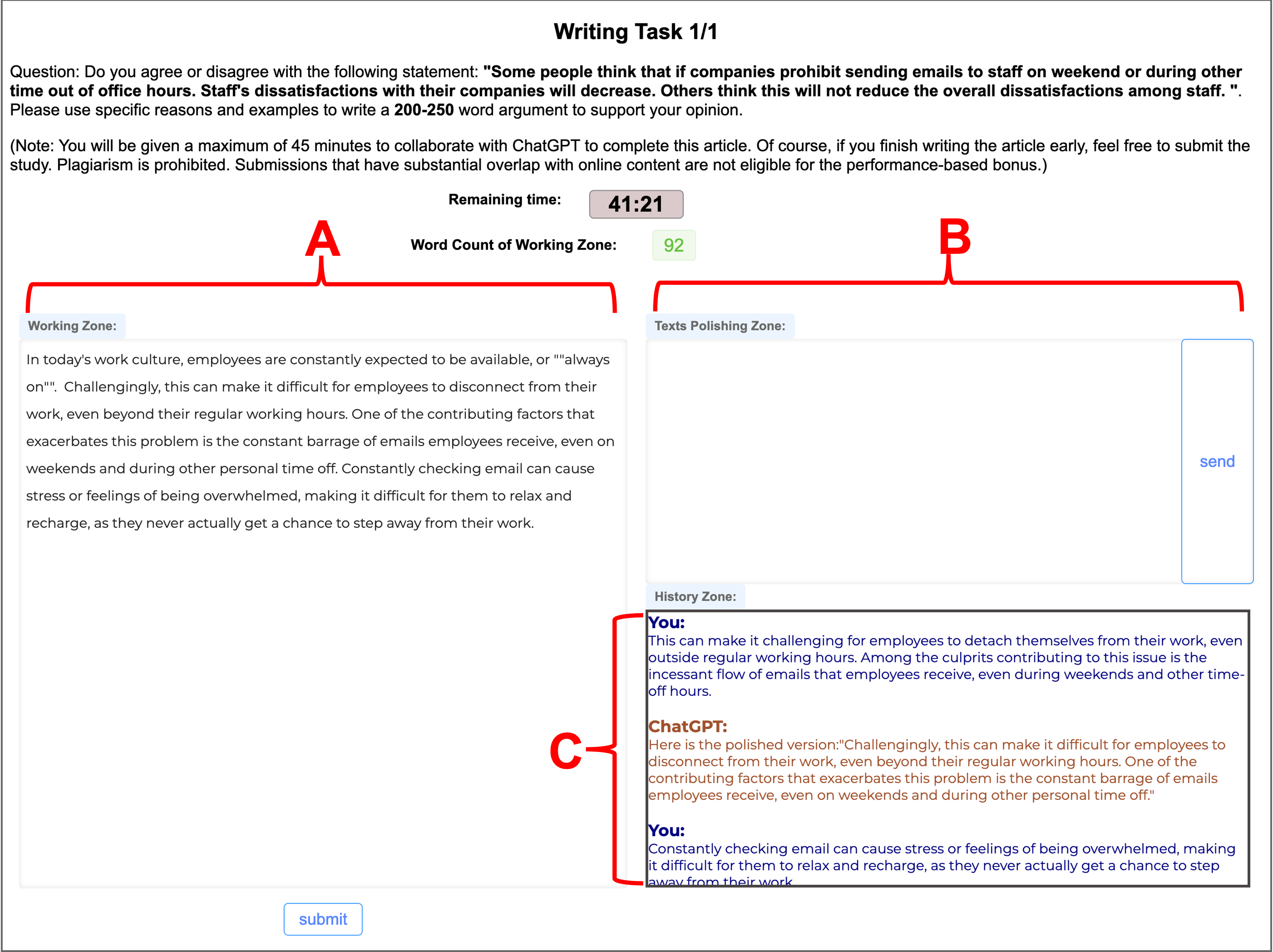}
\end{subfigure}
\caption{An example of the writing interface for participants in the \textbf{Human-Primary} writing mode. \textbf{Working Zone (Part A)}: This is where participants can compose the article and integrate any ChatGPT-polished text into their own writing. Once they are satisfied with the version of the article in this zone, they can proceed to submit the current version by clicking the ``submit'' button.
\textbf{Text Polishing Zone (Part B)}: This is where participants can enter the texts that they want ChatGPT to polish, with the option to directly copy from any part of the texts in the working zone. After entering the desired text for editing assistance, participants can click the ``send'' button to forward it to ChatGPT.
\textbf{History Zone (Part C)}: This zone shows the entire interaction history between the participant and ChatGPT, which includes the original text the participant send to ChatGPT for polish and the polished version of the text produced by ChatGPT. Participants can compare both versions of the text and decide whether and how to integrate the ChatGPT-polished text into their articles in the working zone. 
}
\vspace{-10pt}
\label{fig:interface_human_primary}
\end{figure*}

\begin{figure*}[t]
\centering
\begin{subfigure}[b]{.99\textwidth}
\includegraphics[width=.99\textwidth]{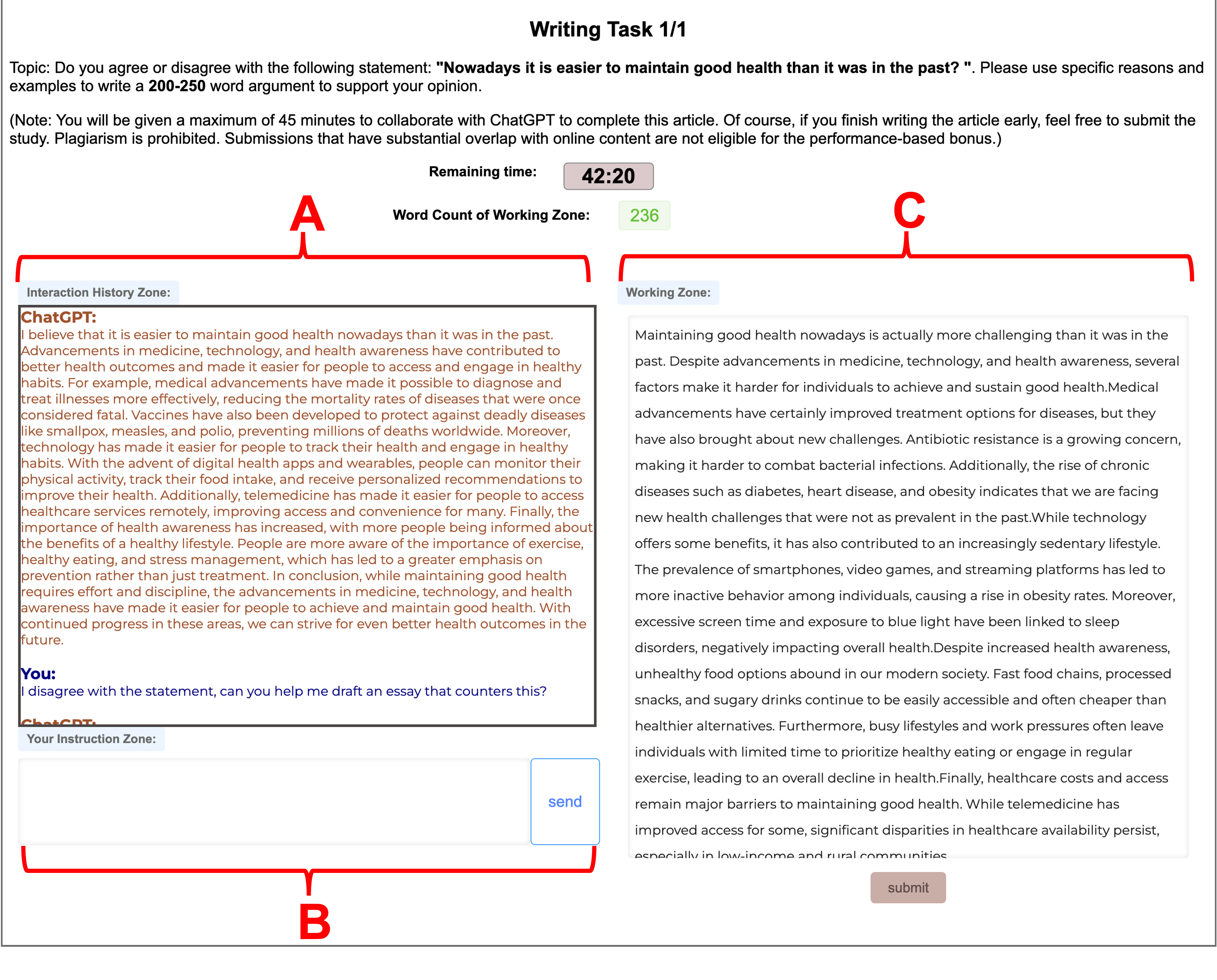}
\end{subfigure}
\caption{An example of the writing interface for participants in the \textbf{AI-Primary} mode. \textbf{Interaction History Zone (Part A)}: This is where participants will first see the initial draft of the article that ChatGPT generates for them. Given the initial draft, participants can use the Instruction Zone below to offer feedback. ChatGPT will then adjust and re-generate the article accordingly. All interactions between the participant and ChatGPT will be logged within this zone, allowing the participant to review their interaction history whenever needed.
\textbf{Instruction Zone (Part B)}: This is where the participant can send instructions or feedback to ChatGPT for it to improve the draft. There are no restrictions on what kinds of  instructions participants can provide to ChatGPT.
\textbf{Working Zone (Part C)}: This is where participants can compose and submit their final article. If participants want to use any section of the text generated by ChatGPT, they can directly copy it from the Interaction History Zone and paste it into here. Participants can also directly write a part of the article on their own or manually edit the ChatGPT-generated text. Once they are satisfied with the version of the article in this zone, they can proceed to submit the current version by clicking the ``submit'' button.
}
\vspace{-10pt}
\label{fig:interface_ai_primary}
\end{figure*}

Figures~\ref{fig:interface_human_primary} and~\ref{fig:interface_ai_primary} show the task interfaces that participants in the ``{\em human-primary}'' and ``{\em AI-primary}'' writing modes used, respectively.

\section{Demographic Information of Participants}
\label{app-demo}
In total, 379 workers from Prolific took our study and passed the attention check. Among them, 183 were allocated to the ``{\em independent vs. human-primary}'' treatment, while the remaining 196 were placed in the ``{\em independent vs. AI-Primary}'' treatment.
The full demographic information of participants is shown in Table~\ref{demo-info}.

\begin{table*}[]
\begin{tabular}{cc|c|c}
\toprule
\multicolumn{2}{c|}{\textbf{Demographics}}                           & \begin{tabular}[c]{@{}c@{}} \textbf{{\em independent vs. AI-Primary}}\\  ($N=196$)\end{tabular} &\begin{tabular}[c]{@{}c@{}} \textbf{{\em independent vs. human-primary}}\\ ($N=183$)\end{tabular}\\ \midrule
\multirow{3}{*}{Gender}& Male                        &                $52.0\%$                                                                     &  $56.3\%$                                                                        \\
& Female                        &        $45.9\%$                                                                       &     $41.5\%$                                                                          \\
& Others or prefer not to say                          &     $2.1\%$                                                                            &   $2.2\%$                                                                           \\
\hline
\multirow{3}{*}{Age}& Below 35                           &    $38.3\%$                                                                            &      $31.7\%$                                                                         \\
& 35--44                              &   $33.7\%$                                                                               &   $32.2\%$                                                                          \\
& 45 or above                           &  $28.0\%$                                                                               &   $36.1\%$                                                                         \\
\hline
\multirow{4}{*}{Race}& White            &  $74.5\%$                                                                             &    $78.6\%$                                                                           \\
& Black                             &   $12.2\%$                                                                              &    $6.6\%$                                                                         \\
& Hispanic                             &   $4.1\%$                                                                              &    $4.4\%$                                                                          \\
& Others                             &   $9.2\%$                                                                              &    $10.4\%$                                                                          \\
\hline
\multirow{4}{*}{Education}& High school or lower          & $16.3\%$                                                                                &  $20.2\%$                                                                             \\
& Some college                  & $28.0\%$                                                                                &  $23.5\%$                                                                         \\
& Bachelor degree               &   $38.3\%$                                                                              &   $40.4\%$                                                                            \\
& Graduate school or higher     &  $17.4\%$                                                                               &  $15.9\%$                                                                            \\
\hline

\multirow{3}{*}{\tabincell{c}{Average\\writing confidence\\(Pre-task)}} & Argumentative essay & 3.14                                                                                &     3.28                                                                          \\
& Creative story      & 3.32                                                                             &     3.46                                                                          \\
& Other writing tasks              & 3.21                                                                              &    3.32 \\
\hline
\multirow{5}{*}{\tabincell{c}{ChatGPT usage\\ frequency} }& Never        & $29.1\%$                                                                                &  $27.9\%$                                                                             \\
& Rarely  & $35.7\%$                                                                                &  $38.3\%$                                                                         \\
&  Occasionally             &   $20.4\%$                                                                              &   $22.4\%$                                                                            \\
& Frequently    &  $11.7\%$                                                                               &  $7.6\%$                                                                            \\ & Very frequently      &  $3.1\%$                                                                               &  $3.8\%$ \\ 
\hline
\multicolumn{2}{c|}{Average familiarity with ChatGPT}                  &  3.76                                                                              &        3.78                                                                       \\ 
 \bottomrule
\end{tabular}
\caption{Details of the demographic backgrounds of our study participants.}
\label{demo-info}
\end{table*}


\section{Full Regression Results}
\label{app-regression}
In this study, we investigate 19 dependent variables which are listed as:
\begin{enumerate}
    \item Dependent Variable 1 (DV1): Average cognitive load participants experience in terms of mental load, time pressure, effort required, and the  frustration level when completing the writing task.
    \item Dependent Variable 2 (DV2): Satisfaction with the writing process.
    \item Dependent Variable 3 (DV3): Enjoyment during the writing process.
    \item Dependent Variable 4 (DV4): Sense of ease  during the writing process.
    \item Dependent Variable 5 (DV5): Ability to express creative goals during the writing process.
    \item Dependent Variable 6 (DV6): Sense of satisfaction with the quality of the final article.
    \item Dependent Variable 7 (DV7): Sense of ownership over the final article.
    \item Dependent Variable 8 (DV8): Sense of pride in the final article.
    \item Dependent Variable 9 (DV9): Sense of uniqueness of the final article.
    \item Dependent Variable 10 (DV10): Confidence boost in writing argumentative essays.
    \item Dependent Variable 11 (DV11): Confidence boost in writing creative stories.
    \item Dependent Variable 12 (DV12): The pooled average confidence boost in other writing tasks.
    \item Dependent Variable 13 (DV13): The willingness to take responsibility if the article is criticized for containing deceptive content (e.g., misinformation).
    \item Dependent Variable 14 (DV14): The willingness to take responsibility if the article is criticized for containing content that is highly similar to someone else's writing.
    \item Dependent Variable 15 (DV15): The willingness to take responsibility if the article is criticized for invading other's privacy.
    \item Dependent Variable 16 (DV16): The willingness to take responsibility if the article is criticized as exhibiting \\bias/discrimination.
    \item Dependent Variable 17 (DV17): Completion time of the writing task.
    \item Dependent Variable 18 (DV18): Number of grammar and spelling errors in the final article.
    \item Dependent Variable 19 (DV19): Coherence score of the final article.
\end{enumerate}
In the regression models, the primary independent variable is the writing mode the participants are engaged in, and the ``{\em independent}'' writing mode is used as the reference. Additionally, several covariates are incoporated into the regression models, including
\begin{enumerate}
    \item Demographics: \begin{itemize}
        \item Gender: ``Male'', ``Female'', and ``Others''. The reference category is ``Male''.
        \item Age: ``below 35'', ``35-44'', and ``above 45''. The reference category is ``below 35''.
        \item Race: ``White'', ``Black or African American'', ``Hispanic or Latino'', and ``Others''. The reference category is ``White''.
        \item Education: ``High School or lower'', ``Some College'', ``Bachelor's Degree'', and ``Graduate School or higher''. "High School or lower" is the reference category.
    \end{itemize}
    \item Payment: The writing job payment associated with their selected job offer.
    \item Writing Confidence: Participants' confidence in their own writing abilities.
    \item Familiarity with ChatGPT.
    \item Usage frequency of ChatGPT: The raw five levels are ``Never'', ``Rarely (once a week or less)'', ``Occasionally (a few times a week)'', ``Frequently (once a day)'', and ``Very Frequently (more than once a day)''. According to our data, the median category selected by our participants is ``Rarely''. Based on this, the categories 'Occasionally', 'Frequently', and 'Very Frequently' are grouped as the ``high'' level whereas the ``Never'' and ``Rarely'' categories are grouped as the ``Low'' level.
    The ``Low'' level is set as the reference.
\end{enumerate}
The complete regression results for participants who completed the argumentative writing task are shown in Table~\ref{regression_essay1} and Table~\ref{regression_essay2}.
The complete regression results for participants who completed the creative story task are shown in Table~\ref{regression_story1} and Table~\ref{regression_story2}.

\begin{table*}[]
\resizebox{\textwidth}{!}{
\centering
\begin{tabular}{|c|c|c|c|c|c|c|c|c|c|} 
\hline
                                          & DV1                                           & DV2                 & DV3                     &DV4                 & DV5        & DV6            &DV7            & DV8              & DV9               \\ 
\hline
Human-Primary                             & \begin{tabular}[c]{@{}c@{}}-0.38\\ (0.19)\end{tabular}  & \begin{tabular}[c]{@{}c@{}}0.35*\\ (0.16)\end{tabular}    & \begin{tabular}[c]{@{}c@{}}0.17\\ (0.20)\end{tabular}    & \begin{tabular}[c]{@{}c@{}}0.33\\ (0.17)\end{tabular}    & \begin{tabular}[c]{@{}c@{}}0.21\\ (0.20)\end{tabular}     & \begin{tabular}[c]{@{}c@{}}0.37*\\ (0.16)\end{tabular}    & \begin{tabular}[c]{@{}c@{}}-0.08\\ (0.18)\end{tabular}    & \begin{tabular}[c]{@{}c@{}}0.16\\ (0.19)\end{tabular}     & \begin{tabular}[c]{@{}c@{}}0.12\\ (0.21)\end{tabular}      \\ 
\hline
AI-Primary                                & \begin{tabular}[c]{@{}c@{}}-0.31*\\ (0.15)\end{tabular}  & \begin{tabular}[c]{@{}c@{}}-0.17\\ (0.14)\end{tabular}    & \begin{tabular}[c]{@{}c@{}}-0.48**\\ (0.18)\end{tabular} & \begin{tabular}[c]{@{}c@{}}-0.15\\ (0.15)\end{tabular}   & \begin{tabular}[c]{@{}c@{}}-0.59**\\ (0.18)\end{tabular}  & \begin{tabular}[c]{@{}c@{}}0.05\\ (0.14)\end{tabular}     & \begin{tabular}[c]{@{}c@{}}-1.32***\\ (0.16)\end{tabular} & \begin{tabular}[c]{@{}c@{}}-0.60***\\ (0.17)\end{tabular} & \begin{tabular}[c]{@{}c@{}}-0.79***\\ (0.18)\end{tabular}  \\ 
\hline
ChatGPT usage frequency (high)            & \begin{tabular}[c]{@{}c@{}}-0.37*\\ (0.15)\end{tabular}  & \begin{tabular}[c]{@{}c@{}}0.04\\ (0.14)\end{tabular}     & \begin{tabular}[c]{@{}c@{}}0.10\\ (0.18)\end{tabular}    & \begin{tabular}[c]{@{}c@{}}0.37*\\ (0.15)\end{tabular}   & \begin{tabular}[c]{@{}c@{}}-0.21\\ (0.18)\end{tabular}    & \begin{tabular}[c]{@{}c@{}}-0.07\\ (0.14)\end{tabular}    & \begin{tabular}[c]{@{}c@{}}0.01\\ (0.16)\end{tabular}     & \begin{tabular}[c]{@{}c@{}}-0.15\\ (0.17)\end{tabular}    & \begin{tabular}[c]{@{}c@{}}-0.02\\ (0.18)\end{tabular}     \\ 
\hline
Gender (female)                          & \begin{tabular}[c]{@{}c@{}}-0.07\\ (0.13)\end{tabular}   & \begin{tabular}[c]{@{}c@{}}0.01\\ (0.12)\end{tabular}     & \begin{tabular}[c]{@{}c@{}}0.05\\ (0.16)\end{tabular}    & \begin{tabular}[c]{@{}c@{}}0.21\\ (0.14)\end{tabular}    & \begin{tabular}[c]{@{}c@{}}-0.27\\ (0.16)\end{tabular}    & \begin{tabular}[c]{@{}c@{}}0.19\\ (0.12)\end{tabular}     & \begin{tabular}[c]{@{}c@{}}0\\ (0.14)\end{tabular}        & \begin{tabular}[c]{@{}c@{}}-0.02\\ (0.15)\end{tabular}    & \begin{tabular}[c]{@{}c@{}}0.06\\ (0.16)\end{tabular}      \\ 
\hline
Gender (others)                          & \begin{tabular}[c]{@{}c@{}}0.27 \\ (0.39)\end{tabular}   & \begin{tabular}[c]{@{}c@{}}0.38\\ (0.36)\end{tabular}     & \begin{tabular}[c]{@{}c@{}}-0.25\\ (0.46)\end{tabular}   & \begin{tabular}[c]{@{}c@{}}0.17\\ (0.40)\end{tabular}    & \begin{tabular}[c]{@{}c@{}}-0.31\\ (0.47)\end{tabular}    & \begin{tabular}[c]{@{}c@{}}0.44\\ (0.37)\end{tabular}     & \begin{tabular}[c]{@{}c@{}}-0.63\\ (0.41)\end{tabular}    & \begin{tabular}[c]{@{}c@{}}0.10\\ (0.44)\end{tabular}     & \begin{tabular}[c]{@{}c@{}}0\\ (0.47)\end{tabular}         \\ 
\hline
Age (35-44)                              & \begin{tabular}[c]{@{}c@{}}0.34*\\ (0.16)\end{tabular}   & \begin{tabular}[c]{@{}c@{}}-0.01\\ (0.14)\end{tabular}    & \begin{tabular}[c]{@{}c@{}}-0.06\\ (0.19)\end{tabular}   & \begin{tabular}[c]{@{}c@{}}-0.31\\ (0.16)\end{tabular}   & \begin{tabular}[c]{@{}c@{}}-0.37\\ (0.19)\end{tabular}    & \begin{tabular}[c]{@{}c@{}}-0.07\\ (0.15)\end{tabular}    & \begin{tabular}[c]{@{}c@{}}0.14\\ (0.16)\end{tabular}     & \begin{tabular}[c]{@{}c@{}}-0.09\\ (0.18)\end{tabular}    & \begin{tabular}[c]{@{}c@{}}0.04\\ (0.19)\end{tabular}      \\ 
\hline
Age (above 45)                         & \begin{tabular}[c]{@{}c@{}}0.27\\ (0.18)\end{tabular}    & \begin{tabular}[c]{@{}c@{}}0.28\\ (0.16)\end{tabular}     & \begin{tabular}[c]{@{}c@{}}0.08\\ (0.21)\end{tabular}    & \begin{tabular}[c]{@{}c@{}}-0.35\\ (0.18)\end{tabular}   & \begin{tabular}[c]{@{}c@{}}-0.19\\ (0.21)\end{tabular}    & \begin{tabular}[c]{@{}c@{}}0.15\\ (0.17)\end{tabular}     & \begin{tabular}[c]{@{}c@{}}0.14\\ (0.19)\end{tabular}     & \begin{tabular}[c]{@{}c@{}}0.09\\ (0.20)\end{tabular}     & \begin{tabular}[c]{@{}c@{}}-0.11\\ (0.21)\end{tabular}     \\ 
\hline
Race (Black or African American)         & \begin{tabular}[c]{@{}c@{}}0.33\\ (0.22)\end{tabular}    & \begin{tabular}[c]{@{}c@{}}-0.04\\ (0.20)\end{tabular}    & \begin{tabular}[c]{@{}c@{}}-0.14\\ (0.27)\end{tabular}   & \begin{tabular}[c]{@{}c@{}}-0.07\\ (0.22)\end{tabular}   & \begin{tabular}[c]{@{}c@{}}-0.36\\ (0.26)\end{tabular}    & \begin{tabular}[c]{@{}c@{}}-0.04\\ (0.20)\end{tabular}    & \begin{tabular}[c]{@{}c@{}}-0.32\\ (0.23)\end{tabular}    & \begin{tabular}[c]{@{}c@{}}0.08\\ (0.24)\end{tabular}     & \begin{tabular}[c]{@{}c@{}}-0.51\\ (0.26)\end{tabular}     \\ 
\hline
Race (Hispanic or Latino)                 & \begin{tabular}[c]{@{}c@{}}0.10\\ (0.26)\end{tabular}    & \begin{tabular}[c]{@{}c@{}}0.03\\ (0.24)\end{tabular}     & \begin{tabular}[c]{@{}c@{}}-0.43\\ (0.31)\end{tabular}   & \begin{tabular}[c]{@{}c@{}}-0.46\\ (0.27)\end{tabular}   & \begin{tabular}[c]{@{}c@{}}-0.62\\ (0.31)\end{tabular}    & \begin{tabular}[c]{@{}c@{}}-0.17\\ (0.24)\end{tabular}    & \begin{tabular}[c]{@{}c@{}}0.27\\ (0.27)\end{tabular}     & \begin{tabular}[c]{@{}c@{}}-0.04\\ (0.29)\end{tabular}    & \begin{tabular}[c]{@{}c@{}}-0.17\\ (0.31)\end{tabular}     \\ 
\hline
Race (Others)                             & \begin{tabular}[c]{@{}c@{}}0.23\\ (0.26)\end{tabular}    & \begin{tabular}[c]{@{}c@{}}-0.85***\\ (0.24)\end{tabular} & \begin{tabular}[c]{@{}c@{}}-0.78*\\ (0.31)\end{tabular}  & \begin{tabular}[c]{@{}c@{}}-0.87**\\ (0.27)\end{tabular} & \begin{tabular}[c]{@{}c@{}}-1.13***\\ (0.31)\end{tabular} & \begin{tabular}[c]{@{}c@{}}-0.85**\\ (0.24)\end{tabular} & \begin{tabular}[c]{@{}c@{}}-0.82**\\ (0.27)\end{tabular}  & \begin{tabular}[c]{@{}c@{}}-1.01***\\ (0.29)\end{tabular} & \begin{tabular}[c]{@{}c@{}}-1.17***\\ (0.31)\end{tabular}  \\ 
\hline
Education (Some college)               & \begin{tabular}[c]{@{}c@{}}0.08\\ (0.20)\end{tabular}    & \begin{tabular}[c]{@{}c@{}}0.28\\ (0.18)\end{tabular}     & \begin{tabular}[c]{@{}c@{}}0.04\\ (0.24)\end{tabular}    & \begin{tabular}[c]{@{}c@{}}0.38\\ (0.20)\end{tabular}    & \begin{tabular}[c]{@{}c@{}}-0.11\\ (0.24)\end{tabular}    & \begin{tabular}[c]{@{}c@{}}0.28\\ (0.19)\end{tabular}     & \begin{tabular}[c]{@{}c@{}}0.35\\ (0.21)\end{tabular}     & \begin{tabular}[c]{@{}c@{}}0.15\\ (0.22)\end{tabular}     & \begin{tabular}[c]{@{}c@{}}0.16\\ (0.24)\end{tabular}      \\ 
\hline
Education (Bachelor degree)               & \begin{tabular}[c]{@{}c@{}}0.17\\ (0.17)\end{tabular}    & \begin{tabular}[c]{@{}c@{}}0.07\\ (0.16)\end{tabular}     & \begin{tabular}[c]{@{}c@{}}-0.17\\ (0.21)\end{tabular}   & \begin{tabular}[c]{@{}c@{}}0.19\\ (0.18)\end{tabular}    & \begin{tabular}[c]{@{}c@{}}-0.30\\ (0.21)\end{tabular}    & \begin{tabular}[c]{@{}c@{}}0.23\\ (0.16)\end{tabular}     & \begin{tabular}[c]{@{}c@{}}0.27\\ (0.18)\end{tabular}     & \begin{tabular}[c]{@{}c@{}}-0.02\\ (0.19)\end{tabular}    & \begin{tabular}[c]{@{}c@{}}0.28\\ (0.21)\end{tabular}      \\ 
\hline
Education (Graduate school or higher)     & \begin{tabular}[c]{@{}c@{}}0.18\\ (0.23)\end{tabular}    & \begin{tabular}[c]{@{}c@{}}-0.02\\ (0.22)\end{tabular}    & \begin{tabular}[c]{@{}c@{}}-0.02\\ (0.28)\end{tabular}   & \begin{tabular}[c]{@{}c@{}}0.05\\ (0.24)\end{tabular}    & \begin{tabular}[c]{@{}c@{}}-0.13\\ (0.28)\end{tabular}    & \begin{tabular}[c]{@{}c@{}}0.26\\ (0.22)\end{tabular}     & \begin{tabular}[c]{@{}c@{}}0.50*\\ (0.25)\end{tabular}    & \begin{tabular}[c]{@{}c@{}}0.11\\ (0.26)\end{tabular}     & \begin{tabular}[c]{@{}c@{}}0.50\\ (0.28)\end{tabular}      \\ 
\hline
Payment                                   & \begin{tabular}[c]{@{}c@{}}0.04\\ (0.10)\end{tabular}    & \begin{tabular}[c]{@{}c@{}}0.01\\ (0.09)\end{tabular}     & \begin{tabular}[c]{@{}c@{}}-0.04\\ (0.12)\end{tabular}   & \begin{tabular}[c]{@{}c@{}}0.02\\ (0.06)\end{tabular}    & \begin{tabular}[c]{@{}c@{}}-0.05\\ (0.12)\end{tabular}    & \begin{tabular}[c]{@{}c@{}}-0.03\\ (0.09)\end{tabular}    & \begin{tabular}[c]{@{}c@{}}0.11\\ (0.10)\end{tabular}     & \begin{tabular}[c]{@{}c@{}}-0.08\\ (0.11)\end{tabular}    & \begin{tabular}[c]{@{}c@{}}0.03\\ (0.12)\end{tabular}      \\ 
\hline
Writing confidence in argumentative essay & \begin{tabular}[c]{@{}c@{}}-0.03\\ (0.05\end{tabular}    & \begin{tabular}[c]{@{}c@{}}0.10\\ (0.05)\end{tabular}     & \begin{tabular}[c]{@{}c@{}}0.16*\\ (0.07)\end{tabular}   & \begin{tabular}[c]{@{}c@{}}0.21***\\ (0.06)\end{tabular} & \begin{tabular}[c]{@{}c@{}}0.09\\ (0.07)\end{tabular}     & \begin{tabular}[c]{@{}c@{}}0.08\\ (0.05)\end{tabular}     & \begin{tabular}[c]{@{}c@{}}0.10\\ (0.06)\end{tabular}     & \begin{tabular}[c]{@{}c@{}}0.06\\ (0.06)\end{tabular}     & \begin{tabular}[c]{@{}c@{}}0.10\\ (0.07)\end{tabular}      \\ 
\hline
Familiarity with ChatGPT                   & \begin{tabular}[c]{@{}c@{}}-0.14*\\ (0.06)\end{tabular}  & \begin{tabular}[c]{@{}c@{}}0.17**\\ (0.06)\end{tabular}   & \begin{tabular}[c]{@{}c@{}}0.15\\ (0.08)\end{tabular}    & \begin{tabular}[c]{@{}c@{}}0.27***\\ (0.06)\end{tabular} & \begin{tabular}[c]{@{}c@{}}0.16*\\ (0.08)\end{tabular}    & \begin{tabular}[c]{@{}c@{}}0.14*\\ (0.06)\end{tabular}    & \begin{tabular}[c]{@{}c@{}}-0.01\\ (0.07)\end{tabular}    & \begin{tabular}[c]{@{}c@{}}0.14*\\ (0.07)\end{tabular}    & \begin{tabular}[c]{@{}c@{}}0.12\\ (0.08)\end{tabular}      \\ 
\hline
Constant                                  & \begin{tabular}[c]{@{}c@{}}3.09***\\ (0.51)\end{tabular} & \begin{tabular}[c]{@{}c@{}}2.88***\\ (0.47)\end{tabular}  & \begin{tabular}[c]{@{}c@{}}3.11***\\ (0.61)\end{tabular} & \begin{tabular}[c]{@{}c@{}}1.98***\\ (0.52)\end{tabular} & \begin{tabular}[c]{@{}c@{}}3.84***\\ (0.61)\end{tabular}  & \begin{tabular}[c]{@{}c@{}}3.09***\\ (0.48)\end{tabular}  & \begin{tabular}[c]{@{}c@{}}3.48***\\ (0.53)\end{tabular}  & \begin{tabular}[c]{@{}c@{}}3.62***\\ (0.57)\end{tabular}  & \begin{tabular}[c]{@{}c@{}}2.90***\\ (0.61)\end{tabular}   \\
\hline
\end{tabular}
}
\caption{Regression models predicting experiences and perceptions of writers who completed argumentative essay writing tasks, considering the writing mode they are in, their demographic information, the payment they could earn in their writing modes, their own writing confidence, and their familiarity with and usage frequency of ChatGPT. Coefficients and standard errors (in parentheses) are reported. $*$,$**$, and $***$ represent significance levels of 0.05, 0.01, and 0.001, respectively ($p$-values are unadjusted).}
\label{regression_essay1}
\end{table*}

\begin{table*}[]
\resizebox{\textwidth}{!}{
\centering
\begin{tabular}{|c|c|c|c|c|c|c|c|c|c|c|} 
\hline
                                          & DV10                              & DV11                             & DV12                                & DV13                                                   & DV14                                                   & DV15                                                    & DV16                            & DV17                 & DV18                     & DV19                         \\ 
\hline
Human-Primary                             & \begin{tabular}[c]{@{}c@{}}0.47**\\ (0.16)\end{tabular}   & \begin{tabular}[c]{@{}c@{}}0.40**\\ (0.15)\end{tabular} & \begin{tabular}[c]{@{}c@{}}0.40**\\ (0.12)\end{tabular}  & \begin{tabular}[c]{@{}c@{}}-0.18\\ (0.20)\end{tabular}   & \begin{tabular}[c]{@{}c@{}}-0.04\\ (0.24)\end{tabular}   & \begin{tabular}[c]{@{}c@{}}-0.32\\ (0.23)\end{tabular}    & \begin{tabular}[c]{@{}c@{}}-0.26\\ (0.22)\end{tabular}   & \begin{tabular}[c]{@{}c@{}}0.92\\ (1.44)\end{tabular}    & \begin{tabular}[c]{@{}c@{}}-2.86**\\ (0.95)\end{tabular} & \begin{tabular}[c]{@{}c@{}}-0.018\\ (0.016)\end{tabular}  \\ 
\hline
AI-Primary                                & \begin{tabular}[c]{@{}c@{}}0.24\\ (0.14)\end{tabular}     & \begin{tabular}[c]{@{}c@{}}0.39**\\ (0.13)\end{tabular} & \begin{tabular}[c]{@{}c@{}}0.34**\\ (0.10)\end{tabular}   & \begin{tabular}[c]{@{}c@{}}-0.53**\\ (0.17)\end{tabular} & \begin{tabular}[c]{@{}c@{}}-0.33\\ (0.22)\end{tabular}   & \begin{tabular}[c]{@{}c@{}}-0.73***\\ (0.20)\end{tabular} & \begin{tabular}[c]{@{}c@{}}-0.60**\\ (0.19)\end{tabular} & \begin{tabular}[c]{@{}c@{}}-3.79**\\ (1.29)\end{tabular} & \begin{tabular}[c]{@{}c@{}}-2.56**\\ (0.85)\end{tabular} & \begin{tabular}[c]{@{}c@{}}0.01\\ (0.01)\end{tabular}     \\ 
\hline
ChatGPT usage frequency (high)            & \begin{tabular}[c]{@{}c@{}}0.35*\\ (0.14)\end{tabular}    & \begin{tabular}[c]{@{}c@{}}-0.08\\ (0.13)\end{tabular}  & \begin{tabular}[c]{@{}c@{}}0.07\\ (0.10)\end{tabular}     & \begin{tabular}[c]{@{}c@{}}0.18\\ (0.17)\end{tabular}    & \begin{tabular}[c]{@{}c@{}}0.16\\ (0.21)\end{tabular}    & \begin{tabular}[c]{@{}c@{}}0.30\\ (0.20)\end{tabular}     & \begin{tabular}[c]{@{}c@{}}0.33\\ (0.19)\end{tabular}    & \begin{tabular}[c]{@{}c@{}}0.52\\ (1.27)\end{tabular}    & \begin{tabular}[c]{@{}c@{}}-1.71*\\ (0.84)\end{tabular}  & \begin{tabular}[c]{@{}c@{}}-0.01\\ (0.01)\end{tabular}    \\ 
\hline
Gender (Female)                         & \begin{tabular}[c]{@{}c@{}}0.20\\ (0.12)\end{tabular}     & \begin{tabular}[c]{@{}c@{}}-0.19\\ (0.12)\end{tabular}  & \begin{tabular}[c]{@{}c@{}}-0.11\\ (0.09)\end{tabular}    & \begin{tabular}[c]{@{}c@{}}-0.10\\ (0.16)\end{tabular}   & \begin{tabular}[c]{@{}c@{}}-0.24\\ (0.19)\end{tabular}   & \begin{tabular}[c]{@{}c@{}}-0.33\\ (0.18)\end{tabular}    & \begin{tabular}[c]{@{}c@{}}-0.14\\ (0.17)\end{tabular}   & \begin{tabular}[c]{@{}c@{}}-0.04\\ (1.13)\end{tabular}   & \begin{tabular}[c]{@{}c@{}}-0.50\\ (0.75)\end{tabular}   & \begin{tabular}[c]{@{}c@{}}0.02\\ (0.02)\end{tabular}     \\ 
\hline
Gender (Other)                             & \begin{tabular}[c]{@{}c@{}}-0.01\\ (0.37)\end{tabular}    & \begin{tabular}[c]{@{}c@{}}-0.32\\ (0.35)\end{tabular}  & \begin{tabular}[c]{@{}c@{}}-0.04\\ (0.28)\end{tabular}    & \begin{tabular}[c]{@{}c@{}}-0.45\\ (0.45)\end{tabular}   & \begin{tabular}[c]{@{}c@{}}-0.10\\ (0.55)\end{tabular}   & \begin{tabular}[c]{@{}c@{}}-0.05\\ (0.52)\end{tabular}    & \begin{tabular}[c]{@{}c@{}}-0.16\\ (0.50)\end{tabular}   & \begin{tabular}[c]{@{}c@{}}2.11\\ (3.26)\end{tabular}    & \begin{tabular}[c]{@{}c@{}}2.62\\ (2.16)\end{tabular}    & \begin{tabular}[c]{@{}c@{}}0.03*\\ (0.02)\end{tabular}    \\ 
\hline
Age (35-44)                              & \begin{tabular}[c]{@{}c@{}}-0.07\\ (0.15)\end{tabular}    & \begin{tabular}[c]{@{}c@{}}0.08\\ (0.14)\end{tabular}   & \begin{tabular}[c]{@{}c@{}}-0.05\\ (0.11)\end{tabular}    & \begin{tabular}[c]{@{}c@{}}0.08\\ (0.18)\end{tabular}    & \begin{tabular}[c]{@{}c@{}}0.11\\ (0.22)\end{tabular}    & \begin{tabular}[c]{@{}c@{}}0.41\\ (0.21)\end{tabular}     & \begin{tabular}[c]{@{}c@{}}0.28\\ (0.20)\end{tabular}    & \begin{tabular}[c]{@{}c@{}}2.21\\ (1.33)\end{tabular}    & \begin{tabular}[c]{@{}c@{}}1.61\\ (0.88)\end{tabular}    & \begin{tabular}[c]{@{}c@{}}0.02\\ (0.01)\end{tabular}     \\ 
\hline
Age (above 45)                            & \begin{tabular}[c]{@{}c@{}}0.12\\ (0.17)\end{tabular}     & \begin{tabular}[c]{@{}c@{}}0.31\\ (0.16)\end{tabular}   & \begin{tabular}[c]{@{}c@{}}0.06\\ (0.12)\end{tabular}     & \begin{tabular}[c]{@{}c@{}}0.12\\ (0.20)\end{tabular}    & \begin{tabular}[c]{@{}c@{}}0.15\\ (0.25)\end{tabular}    & \begin{tabular}[c]{@{}c@{}}0.48*\\ (0.24)\end{tabular}    & \begin{tabular}[c]{@{}c@{}}0.45\\ (0.23)\end{tabular}    & \begin{tabular}[c]{@{}c@{}}3.70\\ (1.50)\end{tabular}    & \begin{tabular}[c]{@{}c@{}}0.22\\ (1.00)\end{tabular}    & \begin{tabular}[c]{@{}c@{}}0.01\\ (0.02)\end{tabular}     \\ 
\hline
 Race (Black or African American)        & \begin{tabular}[c]{@{}c@{}}0.37\\ (0.20)\end{tabular}     & \begin{tabular}[c]{@{}c@{}}0.11\\ (0.19)\end{tabular}   & \begin{tabular}[c]{@{}c@{}}0.11\\ (0.15)\end{tabular}     & \begin{tabular}[c]{@{}c@{}}0.03\\ (0.25)\end{tabular}    & \begin{tabular}[c]{@{}c@{}}-0.04\\ (0.31)\end{tabular}   & \begin{tabular}[c]{@{}c@{}}-0.02\\ (0.29)\end{tabular}    & \begin{tabular}[c]{@{}c@{}}0.13\\ (0.28)\end{tabular}    & \begin{tabular}[c]{@{}c@{}}-2.53\\ (1.83)\end{tabular}   & \begin{tabular}[c]{@{}c@{}}-0.36\\ (1.22)\end{tabular}   & \begin{tabular}[c]{@{}c@{}}-0.01\\ (0.03)\end{tabular}    \\ 
\hline
Race (Hispanic or Latino)                 & \begin{tabular}[c]{@{}c@{}}-0.08\\ (0.24)\end{tabular}    & \begin{tabular}[c]{@{}c@{}}-0.01\\ (0.23)\end{tabular}  & \begin{tabular}[c]{@{}c@{}}0.06\\ (0.16)\end{tabular}     & \begin{tabular}[c]{@{}c@{}}0.22\\ (0.30)\end{tabular}    & \begin{tabular}[c]{@{}c@{}}-0.02\\ (0.37)\end{tabular}   & \begin{tabular}[c]{@{}c@{}}-0.06\\ (0.35)\end{tabular}    & \begin{tabular}[c]{@{}c@{}}0.20\\ (0.33)\end{tabular}    & \begin{tabular}[c]{@{}c@{}}-0.24\\ (2.19)\end{tabular}   & \begin{tabular}[c]{@{}c@{}}-1.12\\ (1.45)\end{tabular}   & \begin{tabular}[c]{@{}c@{}}0.02\\ (0.03)\end{tabular}     \\ 
\hline
Race (Others)                              & \begin{tabular}[c]{@{}c@{}}-0.31\\ (0.24)\end{tabular}    & \begin{tabular}[c]{@{}c@{}}-0.24\\ (0.23)\end{tabular}  & \begin{tabular}[c]{@{}c@{}}-0.38*\\ (0.18)\end{tabular}   & \begin{tabular}[c]{@{}c@{}}0.34\\ (0.30)\end{tabular}    & \begin{tabular}[c]{@{}c@{}}0.22\\ (0.37)\end{tabular}    & \begin{tabular}[c]{@{}c@{}}0.55\\ (0.35)\end{tabular}     & \begin{tabular}[c]{@{}c@{}}0.63\\ (0.35)\end{tabular}    & \begin{tabular}[c]{@{}c@{}}2.22\\ (2.27)\end{tabular}    & \begin{tabular}[c]{@{}c@{}}-0.95\\ (1.45)\end{tabular}   & \begin{tabular}[c]{@{}c@{}}0.01\\ (0.02)\end{tabular}     \\ 
\hline
Education (Some college)                  & \begin{tabular}[c]{@{}c@{}}-0.08\\ (0.16)\end{tabular}    & \begin{tabular}[c]{@{}c@{}}0.11\\ (0.18)\end{tabular}   & \begin{tabular}[c]{@{}c@{}}0.11\\ (0.14)\end{tabular}     & \begin{tabular}[c]{@{}c@{}}0.21\\ (0.23)\end{tabular}    & \begin{tabular}[c]{@{}c@{}}0.20\\ (0.28)\end{tabular}    & \begin{tabular}[c]{@{}c@{}}-0.25\\ (0.26)\end{tabular}    & \begin{tabular}[c]{@{}c@{}}0.21\\ (0.25)\end{tabular}    & \begin{tabular}[c]{@{}c@{}}-0.04\\ (1.67)\end{tabular}   & \begin{tabular}[c]{@{}c@{}}-1.86\\ (1.11)\end{tabular}   & \begin{tabular}[c]{@{}c@{}}-0.01\\ (0.02)\end{tabular}    \\ 
\hline
Education (Bachelor degree)               & \begin{tabular}[c]{@{}c@{}}-0.14\\ (0.22)\end{tabular}    & \begin{tabular}[c]{@{}c@{}}-0.31\\ (0.15)\end{tabular}  & \begin{tabular}[c]{@{}c@{}}-0.04\\ (0.12)\end{tabular}    & \begin{tabular}[c]{@{}c@{}}0.04\\ (0.20)\end{tabular}    & \begin{tabular}[c]{@{}c@{}}0.03\\ (0.25)\end{tabular}    & \begin{tabular}[c]{@{}c@{}}0.13\\ (0.23)\end{tabular}     & \begin{tabular}[c]{@{}c@{}}0.19\\ (0.22)\end{tabular}    & \begin{tabular}[c]{@{}c@{}}-0.20\\ (1.46)\end{tabular}   & \begin{tabular}[c]{@{}c@{}}-1.34\\ (0.97)\end{tabular}   & \begin{tabular}[c]{@{}c@{}}-0.03\\ (0.02)\end{tabular}    \\ 
\hline
Education (Graduate school or higher)      & \begin{tabular}[c]{@{}c@{}}0.17\\ (0.19)\end{tabular}     & \begin{tabular}[c]{@{}c@{}}-0.26\\ (0.21)\end{tabular}  & \begin{tabular}[c]{@{}c@{}}-0.06\\ (0.17)\end{tabular}    & \begin{tabular}[c]{@{}c@{}}0.30\\ (0.27)\end{tabular}    & \begin{tabular}[c]{@{}c@{}}0.16\\ (0.33)\end{tabular}    & \begin{tabular}[c]{@{}c@{}}-0.23\\ (0.31)\end{tabular}    & \begin{tabular}[c]{@{}c@{}}0.35\\ (0.30)\end{tabular}    & \begin{tabular}[c]{@{}c@{}}-0.70\\ (1.97)\end{tabular}   & \begin{tabular}[c]{@{}c@{}}-2.69*\\ (1.31)\end{tabular}  & \begin{tabular}[c]{@{}c@{}}0.01\\ (0.01)\end{tabular}     \\ 
\hline
Payment                                   & \begin{tabular}[c]{@{}c@{}}0.02\\ (0.09)\end{tabular}     & \begin{tabular}[c]{@{}c@{}}-0.16\\ (0.09)\end{tabular}  & \begin{tabular}[c]{@{}c@{}}0.01\\ (0.07)\end{tabular}     & \begin{tabular}[c]{@{}c@{}}0.20\\ (0.12)\end{tabular}    & \begin{tabular}[c]{@{}c@{}}0.16\\ (0.14)\end{tabular}    & \begin{tabular}[c]{@{}c@{}}0.20\\ (0.13)\end{tabular}     & \begin{tabular}[c]{@{}c@{}}0.34*\\ (0.13)\end{tabular}   & \begin{tabular}[c]{@{}c@{}}0.94\\ (0.86)\end{tabular}    & \begin{tabular}[c]{@{}c@{}}-0.01\\ (0.57)\end{tabular}   & \begin{tabular}[c]{@{}c@{}}0.01\\ (0.01)\end{tabular}     \\ 
\hline
Writing confidence in argumentative essay & \begin{tabular}[c]{@{}c@{}}-0.40***\\ (0.05)\end{tabular} & \begin{tabular}[c]{@{}c@{}}-0.08\\ (0.05)\end{tabular}  &                                                           & \begin{tabular}[c]{@{}c@{}}0.06\\ (0.06)\end{tabular}    & \begin{tabular}[c]{@{}c@{}}0.05\\ (0.08)\end{tabular}    & \begin{tabular}[c]{@{}c@{}}0.03\\ (0.07)\end{tabular}     & \begin{tabular}[c]{@{}c@{}}-0.02\\ (0.07)\end{tabular}   & \begin{tabular}[c]{@{}c@{}}0.21\\ (0.48)\end{tabular}    & \begin{tabular}[c]{@{}c@{}}0.11\\ (0.32)\end{tabular}    & \begin{tabular}[c]{@{}c@{}}0.01\\ (0.02)\end{tabular}     \\ 
\hline
Average confidence in others              &                                                           &                                                         & \begin{tabular}[c]{@{}c@{}}-0.12**\\ (0.04)\end{tabular} &                                                          &                                                          &                                                           &                                                          &                                                          &                                                          &                                                           \\ 
\hline
familiarity with ChatGPT                   & \begin{tabular}[c]{@{}c@{}}0.12*\\ (0.06)\end{tabular}    & \begin{tabular}[c]{@{}c@{}}0.07\\ (0.05)\end{tabular}   & \begin{tabular}[c]{@{}c@{}}0.07\\ (0.04)\end{tabular}     & \begin{tabular}[c]{@{}c@{}}0.09\\ (0.07)\end{tabular}    & \begin{tabular}[c]{@{}c@{}}0.04\\ (0.09)\end{tabular}    & \begin{tabular}[c]{@{}c@{}}0.10\\ (0.08)\end{tabular}     & \begin{tabular}[c]{@{}c@{}}0.08\\ (0.08)\end{tabular}    & \begin{tabular}[c]{@{}c@{}}-0.53\\ (0.48)\end{tabular}   & \begin{tabular}[c]{@{}c@{}}-0.32\\ (0.36)\end{tabular}   & \begin{tabular}[c]{@{}c@{}}0.02\\ (0.05)\end{tabular}     \\ 
\hline
Constant                                  & \begin{tabular}[c]{@{}c@{}}0.70\\ (0.48)\end{tabular}     & \begin{tabular}[c]{@{}c@{}}0.83\\ (0.45)\end{tabular}   & \begin{tabular}[c]{@{}c@{}}0.32\\ (0.37)\end{tabular}     & \begin{tabular}[c]{@{}c@{}}3.23***\\ (0.59)\end{tabular} & \begin{tabular}[c]{@{}c@{}}3.02***\\ (0.72)\end{tabular} & \begin{tabular}[c]{@{}c@{}}2.81***\\ (0.67)\end{tabular}  & \begin{tabular}[c]{@{}c@{}}2.23***\\ (0.65)\end{tabular} & \begin{tabular}[c]{@{}c@{}}10.14*\\ (4.22)\end{tabular}   & \begin{tabular}[c]{@{}c@{}}6.84*\\ (2.80)\end{tabular}   & \begin{tabular}[c]{@{}c@{}}0.26***\\ (0.05)\end{tabular}  \\
\hline
\end{tabular}
}
\caption{Regression models predicting experiences and perceptions of writers who completed argumentative essay writing tasks, considering the writing mode they are in, their demographic information, the payment they could earn in their writing modes, their own writing confidence, and their familiarity with and usage frequency of ChatGPT. Coefficients and standard errors (in parentheses) are reported. $*$,$**$, and $***$ represent significance levels of 0.05, 0.01, and 0.001, respectively ($p$-values are unadjusted).}
\label{regression_essay2}
\end{table*}

\begin{table*}[]
\resizebox{\textwidth}{!}{
\centering
\begin{tabular}{|c|c|c|c|c|c|c|c|c|c|}
\hline

                                      & DV1                                            & DV2                          & DV3                                                & DV4                           & DV5                         & DV6                                 & DV7                                                 & DV8                                                    & DV9                                    
                \\
                \hline
Human-Primary                         & \begin{tabular}[c]{@{}c@{}}0.31\\ (0.18)\end{tabular}     & \begin{tabular}[c]{@{}c@{}}0.16\\ (0.15)\end{tabular}    & \begin{tabular}[c]{@{}c@{}}0.09\\ (0.18)\end{tabular}    & \begin{tabular}[c]{@{}c@{}}0.06\\ (0.19)\end{tabular}     & \begin{tabular}[c]{@{}c@{}}0.01\\ (0.17)\end{tabular}    & \begin{tabular}[c]{@{}c@{}}0.12\\ (0.16)\end{tabular}    & \begin{tabular}[c]{@{}c@{}}-0.20\\ (0.19)\end{tabular}    & \begin{tabular}[c]{@{}c@{}}0.30\\ (0.18)\end{tabular}    & \begin{tabular}[c]{@{}c@{}}-0.15\\ (0.19)\end{tabular}    \\
\hline

AI-Primary                            & \begin{tabular}[c]{@{}c@{}}-0.10\\ (0.16)\end{tabular}    & \begin{tabular}[c]{@{}c@{}}0.04\\ (0.13)\end{tabular}    & \begin{tabular}[c]{@{}c@{}}-0.10\\ (0.16)\end{tabular}   & \begin{tabular}[c]{@{}c@{}}0.25\\ (0.16)\end{tabular}     & \begin{tabular}[c]{@{}c@{}}-0.24\\ (0.15)\end{tabular}   & \begin{tabular}[c]{@{}c@{}}0.01\\ (0.14)\end{tabular}    & \begin{tabular}[c]{@{}c@{}}-0.75***\\ (0.17)\end{tabular} & \begin{tabular}[c]{@{}c@{}}-0.06\\ (0.16)\end{tabular}   & \begin{tabular}[c]{@{}c@{}}-0.35*\\ (0.17)\end{tabular}   \\
\hline

ChatGPT usage frequency (High)        & \begin{tabular}[c]{@{}c@{}}-0.10\\ (0.15)\end{tabular}    & \begin{tabular}[c]{@{}c@{}}0.02\\ (0.13)\end{tabular}    & \begin{tabular}[c]{@{}c@{}}-0.03\\ (0.15)\end{tabular}   & \begin{tabular}[c]{@{}c@{}}-0.01\\ (0.16)\end{tabular}    & \begin{tabular}[c]{@{}c@{}}0.05\\ (0.15)\end{tabular}    & \begin{tabular}[c]{@{}c@{}}0.01\\ (0.13)\end{tabular}    & \begin{tabular}[c]{@{}c@{}}-0.17\\ (0.16)\end{tabular}    & \begin{tabular}[c]{@{}c@{}}0.04\\ (0.15)\end{tabular}    & \begin{tabular}[c]{@{}c@{}}0.16\\ (0.16)\end{tabular}     \\
\hline

Gender (Female)                     & \begin{tabular}[c]{@{}c@{}}-0.23\\ (0.14)\end{tabular}    & \begin{tabular}[c]{@{}c@{}}0.21\\ (0.12)\end{tabular}    & \begin{tabular}[c]{@{}c@{}}0.13\\ (0.14)\end{tabular}    & \begin{tabular}[c]{@{}c@{}}0.03\\ (0.14)\end{tabular}     & \begin{tabular}[c]{@{}c@{}}0.03\\ (0.13)\end{tabular}    & \begin{tabular}[c]{@{}c@{}}0.07\\ (0.12)\end{tabular}    & \begin{tabular}[c]{@{}c@{}}-0.01\\ (0.15)\end{tabular}    & \begin{tabular}[c]{@{}c@{}}0.10\\ (0.14)\end{tabular}    & \begin{tabular}[c]{@{}c@{}}0.13\\ (0.15)\end{tabular}     \\
\hline

Gender (Others)                      & \begin{tabular}[c]{@{}c@{}}-0.41\\ (0.67)\end{tabular}    & \begin{tabular}[c]{@{}c@{}}-0.73\\ (0.56)\end{tabular}   & \begin{tabular}[c]{@{}c@{}}-0.22\\ (0.68)\end{tabular}   & \begin{tabular}[c]{@{}c@{}}0.37\\ (0.68)\end{tabular}     & \begin{tabular}[c]{@{}c@{}}-0.29\\ (0.64)\end{tabular}   & \begin{tabular}[c]{@{}c@{}}-0.92\\ (0.59)\end{tabular}   & \begin{tabular}[c]{@{}c@{}}-1.24\\ (0.71)\end{tabular}    & \begin{tabular}[c]{@{}c@{}}-0.91\\ (0.68)\end{tabular}   & \begin{tabular}[c]{@{}c@{}}-0.87\\ (0.72)\end{tabular}    \\
\hline

Age (35-44)                           & \begin{tabular}[c]{@{}c@{}}-0.04\\ (0.17)\end{tabular}    & \begin{tabular}[c]{@{}c@{}}0.04\\ (0.14)\end{tabular}    & \begin{tabular}[c]{@{}c@{}}-0.06\\ (0.17)\end{tabular}   & \begin{tabular}[c]{@{}c@{}}0\\ (0.17)\end{tabular}        & \begin{tabular}[c]{@{}c@{}}-0.20\\ (0.17)\end{tabular}   & \begin{tabular}[c]{@{}c@{}}-0.03\\ (0.15)\end{tabular}   & \begin{tabular}[c]{@{}c@{}}-0.30\\ (0.18)\end{tabular}    & \begin{tabular}[c]{@{}c@{}}-0.22\\ (0.17)\end{tabular}   & \begin{tabular}[c]{@{}c@{}}-0.07\\ (0.18)\end{tabular}    \\
\hline

Age (above 45)                        & \begin{tabular}[c]{@{}c@{}}0.14\\ (0.17)\end{tabular}     & \begin{tabular}[c]{@{}c@{}}0.01\\ (0.14)\end{tabular}    & \begin{tabular}[c]{@{}c@{}}0.17\\ (0.17)\end{tabular}    & \begin{tabular}[c]{@{}c@{}}0.01\\ (0.17)\end{tabular}     & \begin{tabular}[c]{@{}c@{}}0.08\\ (0.16)\end{tabular}    & \begin{tabular}[c]{@{}c@{}}-0.01\\ (0.15)\end{tabular}   & \begin{tabular}[c]{@{}c@{}}0.14\\ (0.18)\end{tabular}     & \begin{tabular}[c]{@{}c@{}}0.04\\ (0.17)\end{tabular}    & \begin{tabular}[c]{@{}c@{}}0.14\\ (0.18)\end{tabular}     \\
\hline

Race (Black or African American)      & \begin{tabular}[c]{@{}c@{}}0.71**\\ (0.24)\end{tabular}   & \begin{tabular}[c]{@{}c@{}}-0.24\\ (0.20)\end{tabular}   & \begin{tabular}[c]{@{}c@{}}-0.18\\ (0.24)\end{tabular}   & \begin{tabular}[c]{@{}c@{}}-0.18\\ (0.25)\end{tabular}    & \begin{tabular}[c]{@{}c@{}}-0.42\\ (0.23)\end{tabular}   & \begin{tabular}[c]{@{}c@{}}-0.16\\ (0.21)\end{tabular}   & \begin{tabular}[c]{@{}c@{}}-0.39\\ (0.25)\end{tabular}    & \begin{tabular}[c]{@{}c@{}}-0.27\\ (0.24)\end{tabular}   & \begin{tabular}[c]{@{}c@{}}-0.26\\ (0.26)\end{tabular}    \\
\hline

Race (Hispanic or Latino)            & \begin{tabular}[c]{@{}c@{}}-0.35\\ (0.28)\end{tabular}    & \begin{tabular}[c]{@{}c@{}}-0.36\\ (0.23)\end{tabular}   & \begin{tabular}[c]{@{}c@{}}-0.21\\ (0.28)\end{tabular}   & \begin{tabular}[c]{@{}c@{}}-0.19\\ (0.28)\end{tabular}    & \begin{tabular}[c]{@{}c@{}}-0.48\\ (0.27)\end{tabular}   & \begin{tabular}[c]{@{}c@{}}-0.31\\ (0.24)\end{tabular}   & \begin{tabular}[c]{@{}c@{}}-0.08\\ (0.29)\end{tabular}    & \begin{tabular}[c]{@{}c@{}}-0.42\\ (0.28)\end{tabular}   & \begin{tabular}[c]{@{}c@{}}-0.13\\ (0.30)\end{tabular}    \\
\hline

Race (Others)                         & \begin{tabular}[c]{@{}c@{}}-0.10\\ (0.26)\end{tabular}    & \begin{tabular}[c]{@{}c@{}}-0.05\\ (0.22)\end{tabular}   & \begin{tabular}[c]{@{}c@{}}0.07\\ (0.27)\end{tabular}    & \begin{tabular}[c]{@{}c@{}}0.08\\ (0.27)\end{tabular}     & \begin{tabular}[c]{@{}c@{}}-0.01\\ (0.25)\end{tabular}   & \begin{tabular}[c]{@{}c@{}}-0.19\\ (0.23)\end{tabular}   & \begin{tabular}[c]{@{}c@{}}0.17\\ (0.28)\end{tabular}     & \begin{tabular}[c]{@{}c@{}}-0.13\\ (0.27)\end{tabular}   & \begin{tabular}[c]{@{}c@{}}0.29\\ (0.28)\end{tabular}     \\

\hline

Education (Some college)              & \begin{tabular}[c]{@{}c@{}}0.30\\ (0.21)\end{tabular}     & \begin{tabular}[c]{@{}c@{}}-0.12\\ (0.18)\end{tabular}   & \begin{tabular}[c]{@{}c@{}}-0.15\\ (0.21)\end{tabular}   & \begin{tabular}[c]{@{}c@{}}-0.02\\ (0.22)\end{tabular}    & \begin{tabular}[c]{@{}c@{}}0.01\\ (0.20)\end{tabular}    & \begin{tabular}[c]{@{}c@{}}-0.06\\ (0.18)\end{tabular}   & \begin{tabular}[c]{@{}c@{}}0.04\\ (0.22)\end{tabular}     & \begin{tabular}[c]{@{}c@{}}-0.13\\ (0.21)\end{tabular}   & \begin{tabular}[c]{@{}c@{}}-0.15\\ (0.23)\end{tabular}    \\

\hline

Education (Bachelor degree)            & \begin{tabular}[c]{@{}c@{}}0.36\\ (0.21)\end{tabular}     & \begin{tabular}[c]{@{}c@{}}-0.32\\ (0.17)\end{tabular}   & \begin{tabular}[c]{@{}c@{}}-0.33\\ (0.21)\end{tabular}   & \begin{tabular}[c]{@{}c@{}}-0.31\\ (0.21)\end{tabular}    & \begin{tabular}[c]{@{}c@{}}-0.16\\ (0.20)\end{tabular}   & \begin{tabular}[c]{@{}c@{}}-0.06\\ (0.18)\end{tabular}   & \begin{tabular}[c]{@{}c@{}}0.09\\ (0.22)\end{tabular}     & \begin{tabular}[c]{@{}c@{}}-0.09\\ (0.21)\end{tabular}   & \begin{tabular}[c]{@{}c@{}}-0.22\\ (0.22)\end{tabular}    \\
\hline

Education (Graduate school or higher)  & \begin{tabular}[c]{@{}c@{}}0.36\\ (0.23)\end{tabular}     & \begin{tabular}[c]{@{}c@{}}-0.35\\ (0.19)\end{tabular}   & \begin{tabular}[c]{@{}c@{}}-0.26\\ (0.23)\end{tabular}   & \begin{tabular}[c]{@{}c@{}}-0.02\\ (0.23)\end{tabular}    & \begin{tabular}[c]{@{}c@{}}-0.25\\ (0.22)\end{tabular}   & \begin{tabular}[c]{@{}c@{}}-0.14\\ (0.20)\end{tabular}   & \begin{tabular}[c]{@{}c@{}}0.04\\ (0.24)\end{tabular}     & \begin{tabular}[c]{@{}c@{}}-0.39\\ (0.23)\end{tabular}   & \begin{tabular}[c]{@{}c@{}}-0.34\\ (0.24)\end{tabular}    \\
\hline

Payment                               & \begin{tabular}[c]{@{}c@{}}-0.09\\ (0.10)\end{tabular}    & \begin{tabular}[c]{@{}c@{}}0.01\\ (0.08)\end{tabular}    & \begin{tabular}[c]{@{}c@{}}0.04\\ (0.10)\end{tabular}    & \begin{tabular}[c]{@{}c@{}}0.06\\ (0.10)\end{tabular}     & \begin{tabular}[c]{@{}c@{}}0.08\\ (0.10)\end{tabular}    & \begin{tabular}[c]{@{}c@{}}0.08\\ (0.09)\end{tabular}    & \begin{tabular}[c]{@{}c@{}}-0.04\\ (0.24)\end{tabular}    & \begin{tabular}[c]{@{}c@{}}-0.01\\ (0.10)\end{tabular}   & \begin{tabular}[c]{@{}c@{}}0.25*\\ (0.11)\end{tabular}    \\
\hline

Writing confidence in creative story  & \begin{tabular}[c]{@{}c@{}}0.03\\ (0.06)\end{tabular}     & \begin{tabular}[c]{@{}c@{}}0.25***\\ (0.05)\end{tabular} & \begin{tabular}[c]{@{}c@{}}0.26***\\ (0.06)\end{tabular} & \begin{tabular}[c]{@{}c@{}}0.35***\\ (0.06)\end{tabular}  & \begin{tabular}[c]{@{}c@{}}0.31***\\ (0.06)\end{tabular} & \begin{tabular}[c]{@{}c@{}}0.16**\\ (0.05)\end{tabular}  & \begin{tabular}[c]{@{}c@{}}0.19**\\ (0.06)\end{tabular}   & \begin{tabular}[c]{@{}c@{}}0.20**\\ (0.06)\end{tabular}  & \begin{tabular}[c]{@{}c@{}}0.25***\\ (0.07)\end{tabular}  \\
\hline

Average confidence in others          &                                                           &                                                          &                                                          &                                                           &                                                          &                                                          &                                                           &                                                          &   
\\
\hline

familiarity with ChatGPT               & \begin{tabular}[c]{@{}c@{}}-0.03\\ (0.07)\end{tabular}    & \begin{tabular}[c]{@{}c@{}}0.13*\\ (0.06)\end{tabular}   & \begin{tabular}[c]{@{}c@{}}0.16*\\ (0.07)\end{tabular}   & \begin{tabular}[c]{@{}c@{}}0.10\\ (0.07)\end{tabular}     & \begin{tabular}[c]{@{}c@{}}0.26***\\ (0.07)\end{tabular} & \begin{tabular}[c]{@{}c@{}}0.16*\\ (0.06)\end{tabular}   & \begin{tabular}[c]{@{}c@{}}0.01\\ (0.08)\end{tabular}     & \begin{tabular}[c]{@{}c@{}}0.15\\ (0.07)\end{tabular}    & \begin{tabular}[c]{@{}c@{}}0.23**\\ (0.08)\end{tabular}   \\
\hline

Constant                              & \begin{tabular}[c]{@{}c@{}}2.71***\\ (0.558)\end{tabular} & \begin{tabular}[c]{@{}c@{}}2.81***\\ (0.46)\end{tabular} & \begin{tabular}[c]{@{}c@{}}2.50***\\ (0.56)\end{tabular} & \begin{tabular}[c]{@{}c@{}}2.13***\\ (0.572)\end{tabular} & \begin{tabular}[c]{@{}c@{}}1.87***\\ (0.53)\end{tabular} & \begin{tabular}[c]{@{}c@{}}2.73***\\ (0.49)\end{tabular} & \begin{tabular}[c]{@{}c@{}}4.00***\\ (0.59)\end{tabular}  & \begin{tabular}[c]{@{}c@{}}2.90***\\ (0.56)\end{tabular} & \begin{tabular}[c]{@{}c@{}}1.60**\\ (0.60) \end{tabular}  
\\
\hline
\end{tabular}
}
\caption{Regression models predicting experiences and perceptions of writers who completed creative story writing tasks, considering the writing mode they are in, their demographic information, the payment they could earn in their writing modes, their own writing confidence, and their familiarity with and usage frequency of ChatGPT. Coefficients and standard errors (in parentheses) are reported. $*$,$**$, and $***$ represent significance levels of 0.05, 0.01, and 0.001, respectively ($p$-values are unadjusted).}
\label{regression_story1}
\end{table*}

\begin{table*}[]
\resizebox{\textwidth}{!}{
\centering
\begin{tabular}{|c|c|c|c|c|c|c|c|c|c|c|}
\hline
                                      & DV10                            & DV11                               & DV12                              & DV13                                                   & DV14                                                   & DV15                                                   & DV16                                                   & DV17                            & DV18                                            & DV19                                                 \\
                                      \hline
Human-Primary                         & \begin{tabular}[c]{@{}c@{}}0.06\\ (0.15)\end{tabular}    & \begin{tabular}[c]{@{}c@{}}-0.05\\ (0.15)\end{tabular}    & \begin{tabular}[c]{@{}c@{}}0.08\\ (0.11)\end{tabular}    & \begin{tabular}[c]{@{}c@{}}-0.39*\\ (0.18)\end{tabular}  & \begin{tabular}[c]{@{}c@{}}-0.38\\ (0.22)\end{tabular}   & \begin{tabular}[c]{@{}c@{}}-0.51*\\ (0.22)\end{tabular}  & \begin{tabular}[c]{@{}c@{}}-0.35\\ (0.22)\end{tabular}   & \begin{tabular}[c]{@{}c@{}}-0.80\\ (1.59)\end{tabular}  & \begin{tabular}[c]{@{}c@{}}-2.17***\\ (0.63)\end{tabular} & \begin{tabular}[c]{@{}c@{}}-0.01\\ (0.02)\end{tabular}    \\
\hline
AI-Primary                            & \begin{tabular}[c]{@{}c@{}}0.50***\\ (0.13)\end{tabular} & \begin{tabular}[c]{@{}c@{}}0.12\\ (0.13)\end{tabular}     & \begin{tabular}[c]{@{}c@{}}0.47***\\ (0.10)\end{tabular} & \begin{tabular}[c]{@{}c@{}}-0.45**\\ (0.16)\end{tabular} & \begin{tabular}[c]{@{}c@{}}-0.32\\ (0.20)\end{tabular}   & \begin{tabular}[c]{@{}c@{}}-0.33\\ (0.20)\end{tabular}   & \begin{tabular}[c]{@{}c@{}}-0.17\\ (0.19)\end{tabular}   & \begin{tabular}[c]{@{}c@{}}-1.54\\ (1.42)\end{tabular}  & \begin{tabular}[c]{@{}c@{}}-1.60**\\ (0.56)\end{tabular}  & \begin{tabular}[c]{@{}c@{}}-0.01\\ (0.02)\end{tabular}    \\
\hline
ChatGPT usage frequency (High)        & \begin{tabular}[c]{@{}c@{}}0.12\\ (0.13)\end{tabular}    & \begin{tabular}[c]{@{}c@{}}0.24\\ (0.12)\end{tabular}     & \begin{tabular}[c]{@{}c@{}}0.14\\ (0.09)\end{tabular}    & \begin{tabular}[c]{@{}c@{}}0.05\\ (0.15)\end{tabular}    & \begin{tabular}[c]{@{}c@{}}-0.23\\ (0.18)\end{tabular}   & \begin{tabular}[c]{@{}c@{}}-0.13\\ (0.19)\end{tabular}   & \begin{tabular}[c]{@{}c@{}}0.01\\ (0.18)\end{tabular}    & \begin{tabular}[c]{@{}c@{}}1.35\\ (1.34)\end{tabular}   & \begin{tabular}[c]{@{}c@{}}-1.14*\\ (0.53)\end{tabular}   & \begin{tabular}[c]{@{}c@{}}0.01\\ (0.01)\end{tabular}     \\
\hline
Gender (Female)                        & \begin{tabular}[c]{@{}c@{}}0.15\\ (0.11)\end{tabular}    & \begin{tabular}[c]{@{}c@{}}0.02\\ (0.11)\end{tabular}     & \begin{tabular}[c]{@{}c@{}}-0.02\\ (0.08)\end{tabular}   & \begin{tabular}[c]{@{}c@{}}-0.05\\ (0.14)\end{tabular}   & \begin{tabular}[c]{@{}c@{}}-0.01\\ (0.17)\end{tabular}   & \begin{tabular}[c]{@{}c@{}}-0.02\\ (0.17)\end{tabular}   & \begin{tabular}[c]{@{}c@{}}-0.10\\ (0.16)\end{tabular}   & \begin{tabular}[c]{@{}c@{}}-1.45\\ (1.22)\end{tabular}  & \begin{tabular}[c]{@{}c@{}}-0.92\\ (0.48)\end{tabular}    & \begin{tabular}[c]{@{}c@{}}0.01\\ (0.01)\end{tabular}     \\
\hline
Gender (Others)                       & \begin{tabular}[c]{@{}c@{}}0.21\\ (0.56)\end{tabular}    & \begin{tabular}[c]{@{}c@{}}-0.06\\ (0.54)\end{tabular}    & \begin{tabular}[c]{@{}c@{}}0.09\\ (0.42)\end{tabular}    & \begin{tabular}[c]{@{}c@{}}-1.22\\ (0.68)\end{tabular}   & \begin{tabular}[c]{@{}c@{}}-1.06\\ (0.81)\end{tabular}   & \begin{tabular}[c]{@{}c@{}}-1.10\\ (0.82)\end{tabular}   & \begin{tabular}[c]{@{}c@{}}-1.21\\ (0.79)\end{tabular}   & \begin{tabular}[c]{@{}c@{}}-2.95\\ (5.76)\end{tabular}  & \begin{tabular}[c]{@{}c@{}}-1.06\\ (2.29)\end{tabular}    & \begin{tabular}[c]{@{}c@{}}-0.02\\ (0.01)\end{tabular}    \\
\hline
Age (35-44)                         & \begin{tabular}[c]{@{}c@{}}0.01\\ (0.14)\end{tabular}    & \begin{tabular}[c]{@{}c@{}}-0.07\\ (0.14)\end{tabular}    & \begin{tabular}[c]{@{}c@{}}-0.12\\ (0.11)\end{tabular}   & \begin{tabular}[c]{@{}c@{}}0.21\\ (0.17)\end{tabular}    & \begin{tabular}[c]{@{}c@{}}0.30\\ (0.21)\end{tabular}    & \begin{tabular}[c]{@{}c@{}}0.26\\ (0.21)\end{tabular}    & \begin{tabular}[c]{@{}c@{}}0.19\\ (0.20)\end{tabular}    & \begin{tabular}[c]{@{}c@{}}2.69\\ (1.50)\end{tabular}   & \begin{tabular}[c]{@{}c@{}}-0.83\\ (0.60)\end{tabular}    & \begin{tabular}[c]{@{}c@{}}0.02\\ (0.01)\end{tabular}     \\
\hline
Age (above 45)                        & \begin{tabular}[c]{@{}c@{}}-0.06\\ (0.14)\end{tabular}   & \begin{tabular}[c]{@{}c@{}}0.04\\ (0.14)\end{tabular}     & \begin{tabular}[c]{@{}c@{}}-0.05\\ (0.11)\end{tabular}   & \begin{tabular}[c]{@{}c@{}}0.32\\ (0.17)\end{tabular}    & \begin{tabular}[c]{@{}c@{}}0.33\\ (0.21)\end{tabular}    & \begin{tabular}[c]{@{}c@{}}0.46*\\ (0.21)\end{tabular}   & \begin{tabular}[c]{@{}c@{}}0.44*\\ (0.20)\end{tabular}   & \begin{tabular}[c]{@{}c@{}}4.83**\\ (1.49)\end{tabular} & \begin{tabular}[c]{@{}c@{}}-0.36\\ (0.59)\end{tabular}    & \begin{tabular}[c]{@{}c@{}}-0.05\\ (0.02)\end{tabular}    \\
\hline
Race (Black or African American)       & \begin{tabular}[c]{@{}c@{}}-0.10\\ (0.20)\end{tabular}   & \begin{tabular}[c]{@{}c@{}}-0.06\\ (0.19)\end{tabular}    & \begin{tabular}[c]{@{}c@{}}-0.04\\ (0.15)\end{tabular}   & \begin{tabular}[c]{@{}c@{}}-0.40\\ (0.24)\end{tabular}   & \begin{tabular}[c]{@{}c@{}}0.07\\ (0.29)\end{tabular}    & \begin{tabular}[c]{@{}c@{}}-0.27\\ (0.29)\end{tabular}   & \begin{tabular}[c]{@{}c@{}}-0.47\\ (0.29)\end{tabular}   & \begin{tabular}[c]{@{}c@{}}1.08\\ (2.10)\end{tabular}   & \begin{tabular}[c]{@{}c@{}}1.73*\\ (0.83)\end{tabular}    & \begin{tabular}[c]{@{}c@{}}-0.01\\ (0.02)\end{tabular}    \\
\hline
Race (Hispanic or Latino)           & \begin{tabular}[c]{@{}c@{}}-0.03\\ (0.23)\end{tabular}   & \begin{tabular}[c]{@{}c@{}}0.02\\ (0.22)\end{tabular}     & \begin{tabular}[c]{@{}c@{}}-0.03\\ (0.17)\end{tabular}   & \begin{tabular}[c]{@{}c@{}}-0.10\\ (0.28)\end{tabular}   & \begin{tabular}[c]{@{}c@{}}0.18\\ (0.34)\end{tabular}    & \begin{tabular}[c]{@{}c@{}}-0.01\\ (0.34)\end{tabular}   & \begin{tabular}[c]{@{}c@{}}-0.18\\ (0.33)\end{tabular}   & \begin{tabular}[c]{@{}c@{}}1.73\\ (2.41)\end{tabular}   & \begin{tabular}[c]{@{}c@{}}-1.06\\ (0.96)\end{tabular}    & \begin{tabular}[c]{@{}c@{}}0.02\\ (0.01)\end{tabular}     \\
\hline
Race (Others)                       & \begin{tabular}[c]{@{}c@{}}-0.18\\ (0.22)\end{tabular}   & \begin{tabular}[c]{@{}c@{}}0.12\\ (0.21)\end{tabular}     & \begin{tabular}[c]{@{}c@{}}0.12\\ (0.17)\end{tabular}    & \begin{tabular}[c]{@{}c@{}}0.13\\ (0.27)\end{tabular}    & \begin{tabular}[c]{@{}c@{}}0.34\\ (0.32)\end{tabular}    & \begin{tabular}[c]{@{}c@{}}0.10\\ (0.32)\end{tabular}    & \begin{tabular}[c]{@{}c@{}}0.19\\ (0.31)\end{tabular}    & \begin{tabular}[c]{@{}c@{}}2.52\\ (2.30)\end{tabular}   & \begin{tabular}[c]{@{}c@{}}-0.75\\ (0.91)\end{tabular}    & \begin{tabular}[c]{@{}c@{}}-0.01\\ (0.03)\end{tabular}    \\
\hline
Education (Some college)              & \begin{tabular}[c]{@{}c@{}}0.05\\ (0.17)\end{tabular}    & \begin{tabular}[c]{@{}c@{}}-0.15\\ (0.17)\end{tabular}    & \begin{tabular}[c]{@{}c@{}}-0.09\\ (0.13)\end{tabular}   & \begin{tabular}[c]{@{}c@{}}0.37\\ (0.21)\end{tabular}    & \begin{tabular}[c]{@{}c@{}}0.52\\ (0.26)\end{tabular}    & \begin{tabular}[c]{@{}c@{}}0.57*\\ (0.26)\end{tabular}   & \begin{tabular}[c]{@{}c@{}}0.45\\ (0.25)\end{tabular}    & \begin{tabular}[c]{@{}c@{}}2.23\\ (1.84)\end{tabular}   & \begin{tabular}[c]{@{}c@{}}-1.01\\ (0.73)\end{tabular}    & \begin{tabular}[c]{@{}c@{}}0.04\\ (0.02)\end{tabular}     \\
\hline
Education (Bachelor degree)           & \begin{tabular}[c]{@{}c@{}}0,17\\ (0.19)\end{tabular}    & \begin{tabular}[c]{@{}c@{}}-0.15\\ (0.17)\end{tabular}    & \begin{tabular}[c]{@{}c@{}}0.01\\ (0.13)\end{tabular}    & \begin{tabular}[c]{@{}c@{}}0.09\\ (0.21)\end{tabular}    & \begin{tabular}[c]{@{}c@{}}0.26\\ (0.25)\end{tabular}    & \begin{tabular}[c]{@{}c@{}}0.25\\ (0.25)\end{tabular}    & \begin{tabular}[c]{@{}c@{}}-0.04\\ (0.25)\end{tabular}   & \begin{tabular}[c]{@{}c@{}}1.33\\ (1.82)\end{tabular}   & \begin{tabular}[c]{@{}c@{}}-1.25\\ (0.72)\end{tabular}    & \begin{tabular}[c]{@{}c@{}}-0.05\\ (0.02)\end{tabular}    \\
\hline
Education (Graduate school or higher)  & \begin{tabular}[c]{@{}c@{}}-0.22\\ (0.19)\end{tabular}   & \begin{tabular}[c]{@{}c@{}}-0.21\\ (0.18)\end{tabular}    & \begin{tabular}[c]{@{}c@{}}-0.27\\ (0.15)\end{tabular}   & \begin{tabular}[c]{@{}c@{}}0.10\\ (0.23)\end{tabular}    & \begin{tabular}[c]{@{}c@{}}0.20\\ (0.28)\end{tabular}    & \begin{tabular}[c]{@{}c@{}}0.12\\ (0.28)\end{tabular}    & \begin{tabular}[c]{@{}c@{}}-0.32\\ (0.27)\end{tabular}   & \begin{tabular}[c]{@{}c@{}}0.51\\ (1.98)\end{tabular}   & \begin{tabular}[c]{@{}c@{}}-1.33\\ (0.79)\end{tabular}    & \begin{tabular}[c]{@{}c@{}}-0.03\\ (0.03)\end{tabular}    \\
\hline
Payment                               & \begin{tabular}[c]{@{}c@{}}0.24**\\ (0.08)\end{tabular}  & \begin{tabular}[c]{@{}c@{}}0.03\\ (0.08)\end{tabular}     & \begin{tabular}[c]{@{}c@{}}0.10\\ (0.06)\end{tabular}    & \begin{tabular}[c]{@{}c@{}}0.14\\ (0.10)\end{tabular}    & \begin{tabular}[c]{@{}c@{}}0.04\\ (0.12)\end{tabular}    & \begin{tabular}[c]{@{}c@{}}0.19\\ (0.12)\end{tabular}    & \begin{tabular}[c]{@{}c@{}}0.17\\ (0.12)\end{tabular}    & \begin{tabular}[c]{@{}c@{}}1.52\\ (0.89)\end{tabular}   & \begin{tabular}[c]{@{}c@{}}-0.07\\ (0.35)\end{tabular}    & \begin{tabular}[c]{@{}c@{}}0.04\\ (0.01)\end{tabular}     \\
\hline
Writing confidence in creative story  & \begin{tabular}[c]{@{}c@{}}-0.11*\\ (0.05)\end{tabular}  & \begin{tabular}[c]{@{}c@{}}-0.29***\\ (0.05)\end{tabular} &                                                          & \begin{tabular}[c]{@{}c@{}}0.14*\\ (0.06)\end{tabular}   & \begin{tabular}[c]{@{}c@{}}0.10\\ (0.07)\end{tabular}    & \begin{tabular}[c]{@{}c@{}}0.13\\ (0.08)\end{tabular}    & \begin{tabular}[c]{@{}c@{}}0.16*\\ (0.12)\end{tabular}   & \begin{tabular}[c]{@{}c@{}}0.32\\ (0.56)\end{tabular}   & \begin{tabular}[c]{@{}c@{}}0.61**\\ (0.22)\end{tabular}   & \begin{tabular}[c]{@{}c@{}}0.02\\ (0.01)\end{tabular}     \\
\hline
Average writing confidence in others          &                                                          &                                                           & \begin{tabular}[c]{@{}c@{}}-0.18**\\ (0.06)\end{tabular} &                                                          &                                                          &                                                          &                                                          &                                                         &                                                           &                                                           \\
\hline
familiarity with ChatGPT               & \begin{tabular}[c]{@{}c@{}}0.08\\ (0.06)\end{tabular}    & \begin{tabular}[c]{@{}c@{}}0.10\\ (0.06)\end{tabular}     & \begin{tabular}[c]{@{}c@{}}0.07\\ (0.05)\end{tabular}    & \begin{tabular}[c]{@{}c@{}}0.09\\ (0.07)\end{tabular}    & \begin{tabular}[c]{@{}c@{}}0.06\\ (0.09)\end{tabular}    & \begin{tabular}[c]{@{}c@{}}0.08\\ (0.09)\end{tabular}    & \begin{tabular}[c]{@{}c@{}}0.10\\ (0.09)\end{tabular}    & \begin{tabular}[c]{@{}c@{}}0.001\\ (0.66)\end{tabular}  & \begin{tabular}[c]{@{}c@{}}-0.35\\ (0.26)\end{tabular}    & \begin{tabular}[c]{@{}c@{}}0.01\\ (0.01)\end{tabular}     \\
\hline
Constant                              & \begin{tabular}[c]{@{}c@{}}-0.82\\ (0.46)\end{tabular}   & \begin{tabular}[c]{@{}c@{}}0.86\\ (0.45)\end{tabular}     & \begin{tabular}[c]{@{}c@{}}0.09\\ (0.35)\end{tabular}    & \begin{tabular}[c]{@{}c@{}}2.87***\\ (0.56)\end{tabular} & \begin{tabular}[c]{@{}c@{}}3.04***\\ (0.67)\end{tabular} & \begin{tabular}[c]{@{}c@{}}2.47***\\ (0.68)\end{tabular} & \begin{tabular}[c]{@{}c@{}}2.60***\\ (0.66)\end{tabular} & \begin{tabular}[c]{@{}c@{}}3.77\\ (4.79)\end{tabular}   & \begin{tabular}[c]{@{}c@{}}5.88**\\ (1.90)\end{tabular}   & \begin{tabular}[c]{@{}c@{}}0.29***\\ (0.05)\end{tabular} \\
\hline
\end{tabular}
}
\caption{Regression models predicting experiences and perceptions of writers who completed creative story writing tasks, considering the writing mode they are in, their demographic information, the payment they could earn in their writing modes, their own writing confidence, and their familiarity with and usage frequency of ChatGPT. Coefficients and standard errors (in parentheses) are reported. $*$,$**$, and $***$ represent significance levels of 0.05, 0.01, and 0.001, respectively ($p$-values are unadjusted).}
\label{regression_story2}
\end{table*}

\section{Additional Results}

\subsection{Detailed results on people's 
perceptions of the final writing outcome}
\label{app:more_results}
\noindent \textbf{\em The provision of content generation assistance by AI significantly decreases people's perceived ownership of the final writing outcome.} As shown in Figure~\ref{ownership}, when examining participants' perceived ownership over the final article produced as the outcome of the writing processes, we found that participants who received content generation assistance from ChatGPT consistently decreased their ratings compared to other participants, regardless of the types of writing tasks they worked on. Specifically, when writing argumentative essays, participants in the {\em AI-primary} writing mode reported significantly lower levels of ownership over the final article compared to participants in the {\em independent} or {\em human-primary} modes  ({\em adjusted}-$p<0.001$ for both comparisons). Similarly, when writing creative stories, participants in the {\em AI-primary} writing mode also perceived significantly lower levels of ownership over the final article compared to participants in the {\em independent} writing mode ({\em adjusted}-$p<0.001$). 

\vspace{2pt}
\noindent \textbf{\em In argumentative essay writing, the provision of content generation assistance by AI significantly decreases how much people take pride in the final writing outcome and how much people consider it as unique.} As illustrated in Figure~\ref{pride}, 
when participants received content generation assistance from ChatGPT, they expressed a significantly reduced level of pride in the final article relative to participants who wrote them independently ({\em adjusted}-$p=0.002$), and those who only received editing assistance from ChatGPT ({\em adjusted}-$p=0.007$). Additionally, when comparing the perceived sense of uniqueness in the final article in Figure~\ref{unique}, we also found that participants in the {\em AI-primary} mode reported a noticeable decrease compared to participants who took the {\em independent} writing mode ({\em adjusted}-$p<0.001$), and those in the {\em human-primary} writing mode ({\em adjusted}-$p=0.005$).

\subsection{Detailed results on people's writing performance}
\label{app:grammar}

\noindent \textbf{\em AI assistance leads to significantly decreased number of grammar and spelling mistakes in people's writing.}
Our results demonstrate that 
participants in the {\em human-primary} and {\em AI-primary} modes significantly reduced the number of grammar and spelling errors in their final articles compared to participants in the {\em independent} mode, both in argumentative essay writing (independent vs. human-primary: $p=0.003$; independent vs. AI-primary: $p=0.003$), and in creative story writing (independent vs. human-primary: $p<0.001$; independent vs. AI-primary: $p=0.005$).

\end{document}